\documentclass[amsmath,10pt,twocolumn,footinbib,pre,aps]{revtex4-1}
\usepackage[utf8]{inputenc}
\usepackage{hyperref}
\usepackage{graphicx}
\usepackage{subcaption}
\usepackage{tabularx}
\usepackage{amsmath,bm, amssymb}
\usepackage{stackengine}
\usepackage{float}

\usepackage{uri}

\usepackage[x11names, rgb]{xcolor}

\begin{document}
\title{\texorpdfstring{Mechanically Interlocked Polymers in Dilute Solution\\ under Shear and Extensional Flows: A Brownian Dynamics Study}{Mechanically Interlocked Polymers in Dilute Solution — under Shear and Extensional Flows: A Brownian Dynamics Study}}
\author{Ali Seyedi, Alex Albaugh*}
\affiliation{Department of Chemical Engineering and Materials Science,\\
Wayne State University, 5050 Anthony Wayne Drive, Detroit, Michigan 48202, USA}

\begin{abstract}

Mechanically interlocked polymers (MIPs) are a class of polymer structures in which the components are connected by mechanical bonds instead of covalent bonds. 
We measure the single-molecule rheological properties of polyrotaxanes, daisy chains, and polycatenanes under steady shear and steady uniaxial extension using coarse-grained Brownian dynamics simulations with hydrodynamic interactions. 
We obtain key rheological features, including tumbling dynamics, molecular extension, stress, and viscosity. 
By systematically varying structural features, we demonstrate how MIP topology governs flow response. 
Compared to linear polymers, all three MIP architectures exhibit enhanced tumbling in shear flow, weaker shear thinning, and lower normal stress differences in extensional flow. 
While polyrotaxanes show higher shear and extensional viscosities, polycatenanes and daisy chains have lower viscosities. 
In extensional and shear flows, MIPs typically extend more in the flow direction and at weaker flow strength than linear polymers. 
These effects arise from the mechanically bonded rings in MIPs, which expand the polymer profile in gradient direction and increase backbone rigidity due to ring-backbone repulsions. 
This study provides key insights into MIP flow properties, providing the foundation for their systematic development in engineering applications.
\end{abstract}

\maketitle

\section{Introduction}
\label{sec:intro}

\begin{figure}
  \centering
  \begin{subfigure}[t]{0.49\textwidth}
    \centering
    \includegraphics[width=\textwidth]{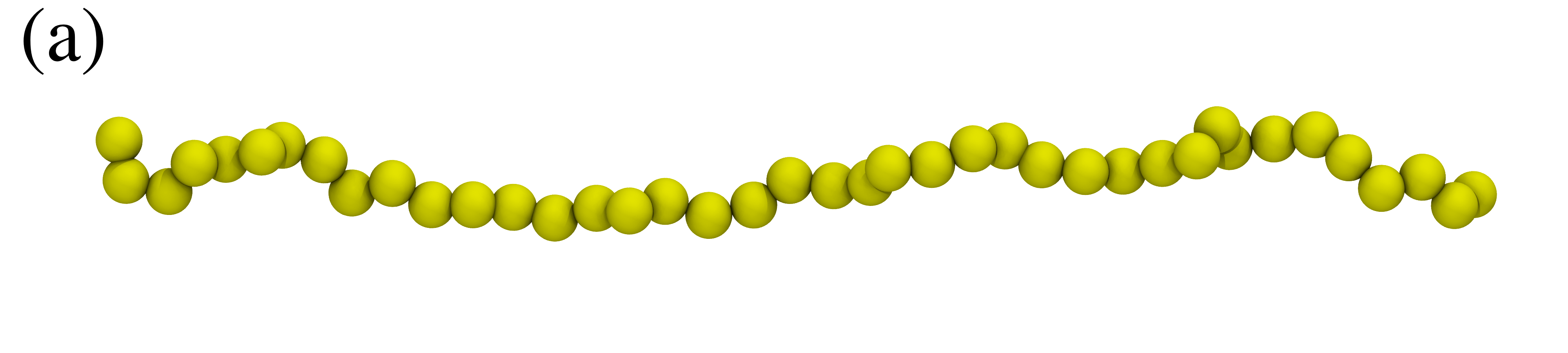}
    \label{fig:linear_polymer}
  \end{subfigure}
  \vspace{0.5em}

    \begin{subfigure}[t]{0.49\textwidth}
    \centering
    \includegraphics[width=\textwidth]{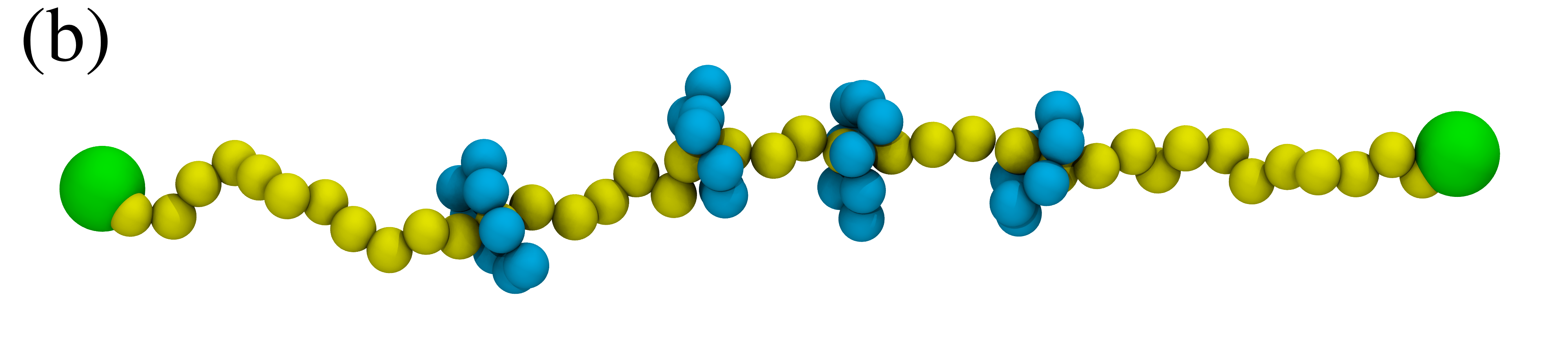}
    \label{fig:polyrotaxane}
  \end{subfigure}
  \vspace{0.5em}

  \begin{subfigure}[t]{0.49\textwidth}
    \centering
    \includegraphics[width=\textwidth]{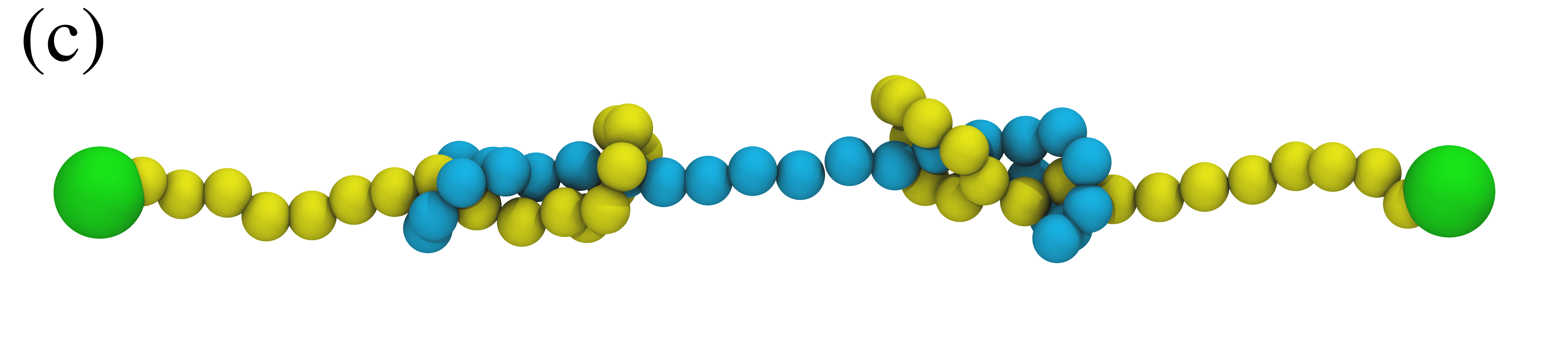}
    \label{fig:daisy_chain}
  \end{subfigure}
  \vspace{0.5em}
  
  \begin{subfigure}[t]{0.49\textwidth}
    \centering
    \includegraphics[width=\textwidth]{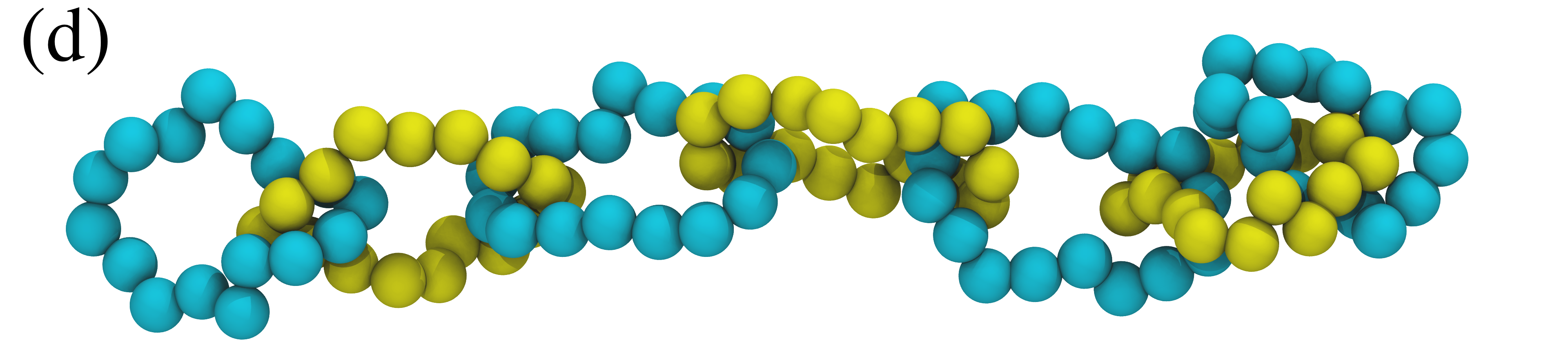}
    \label{fig:polycatenane}
  \end{subfigure}
  \vspace{0.5em}
   \captionsetup{justification=raggedright,singlelinecheck=false}
  \caption{Schematic representation of polymers: (a) linear polymer, (b) polyrotaxane, (c) daisy chain, and (d) polycatenane. Yellow and blue beads are equivalent, the colors are chosen to highlight the interlocked architecture.  Green beads are end caps to prevent dethreading in polyrotaxanes and daisy chains.}
  \label{fig:Schematic_polymers}
\end{figure}

Mechanically interlocked polymers (MIPs) are a class of polymer architectures distinguished by components linked with mechanical entanglements rather than covalent bonds.
The schematic representation in Fig. \ref{fig:Schematic_polymers} shows different classes of MIPs in comparison with a typical linear polymer (Fig. \ref{fig:Schematic_polymers}a). 
Polyrotaxanes (Fig. \ref{fig:Schematic_polymers}b) consist of multiple ring-shaped molecules threaded onto a linear backbone, held in place by bulky end caps called stoppers\cite{harrison1967synthesis,wenz1992threading}.
Daisy chains (Fig. \ref{fig:Schematic_polymers}c) are made from covalently-bonded rings and linear sections interlocked in series\cite{ashton1998supramolecular,wenz2006cyclodextrin}.
Polycatenanes (Fig. \ref{fig:Schematic_polymers}d) are polymers made of interlocked rings, like a chain\cite{schill1964preparation,frisch1961chemical,weidmann1999poly,wu2017poly}. 
These mechanical bonds impart MIPs with increased conformational freedom, setting them apart from typical polymers \cite{hart2021material} and leading to unique properties\cite{sluysmans2019burgeoning,liu2022polycatenanes} such as shape memory, ductile glassy states, large strain-at-break, scratch resistance, and stimuli-responsive mechanical properties \cite{hart2021material,zhou2024mechanically,ando2025recent}.
From theoretical studies it is anticipated that MIPs will show large loss modulus, low activation energy for viscous flow, and rapid stress relaxation \cite{weidmann1999poly}, potentially acting as superior energy-damping materials with excellent toughness \cite{wu2017poly}.
In addition, equilibrium and strain simulations of rotaxanes and rotaxane-based materials, like slide-ring gels and tendomer gels, have demonstrated exceptional elasticity and yield stresses surpassing conventional cross-linked polymer gels~\cite{li2021sliding,yasuda2019sliding,chen2023elasticity,tanahashi2021molecular,uehara2022molecular,yasuda2020molecular,muller2019tendomers,muller2021swelling,muller2022elasticity}.
These examples demonstrate the unique properties imparted by mechanical bonds.

Although mechanically interlinking molecules is a synthetic challenge~\cite{segawa2019topological,leigh2014star}, recent advances have enabled the production of MIPs with moderate molecular weights. The largest polyrotaxane synthesized to date has weight-average molecular weight $M_w$ = 119,000 g/mol with 86 threaded rings \cite{araki2005efficient} while the largest daisy chain polymer has 11 repeat units and number-average molecular weight $M_n\approx$ 25,000 g/mol \cite{cai2020molecular}. 
Ultra-high-molecular-weight polycatenanes have recently been synthesized, consisting of up to 2,800 rings with a molecular weight of 4.5 MDa \cite{wang2025fully}.
Nano-scale polycatenanes comprising up to 22 interlocked rings have been synthesized through the self-assembly of supramolecular toroidal building blocks \cite{datta2020self}. 

Despite advances in synthesis, the relationship between mechanical bonds and flow properties is largely unexplored. 
Computational techniques present an effective method to link the molecular structure, dynamics, and macroscopic behavior of polymers\cite{gartner2019modeling,li2013challenges, Gedde2021,seyedi2020initiator,kremer1990dynamics}.
In the realm of MIPs, simulations have quantified the structural and dynamic properties of doubly interlocked rings (2-catenanes) in equilibrium conditions \cite{rauscher2020hydrodynamic,bohn2010influence}, in shear \cite{li2025effects}, and in glassy states \cite{li2023glass}. 
Simulation studies have explored the equilibrium behavior of longer polycatenanes (up to 300 rings), isolated \cite{tubiana2022circular,chiarantoni2022effect,rauscher2018topological,dehaghani2020effects,pakula1999simulation,li2021double} and in a melt state \cite{rauscher2020thermodynamics,rauscher2020dynamics,hagita2021mathematical}, uncovering characteristics that blend features of ring and linear polymers. 
In addition, computational investigations have characterized the linear and nonlinear elasticity of single polycatenanes~\cite{chen2023topological,chen2024nonlinear} and probed the structure and dynamics of catenanes under confinement \cite{chiarantoni2023linear} and during pore translocation \cite{caraglio2017driven,suma2017pore}.
Simulations have also been used to analyze rotaxanes in equilibrium \cite{li2021sliding,yasuda2019sliding} and daisy chains formed in ring polymer melts under extensional flow \cite{o2020topological}.
These simulations have highlighted the unique behavior of MIPs compared to conventional polymers, demonstrating how structural features, such as ring size and the number of mechanical bonds, influence properties like chain stiffness \cite{pakula1999simulation,rauscher2020thermodynamics}, elasticity \cite{chen2023topological,chen2024nonlinear}, $\theta$-temperature \cite{dehaghani2020effects, guo2023theta}, and relaxation dynamics \cite{pakula1999simulation,rauscher2018topological}. Importantly, these studies show that MIPs exhibit emergent properties not observed in typical linear and cyclic polymer architectures\cite{pakula1999simulation}, highlighting the crucial role of topology in governing their behavior.

Here we simulate dilute solutions of MIPs with coarse-grained Brownian dynamics, an approach that has been effective at characterizing the rheology of many polymer morphologies, like linear \cite{lyulin1999brownian}, combs \cite{mai2018stretching}, bottlebrushes \cite{dutta2024brownian}, and ring polymers \cite{young2019ring}.
We report rheological properties for polyrotaxanes, daisy chains, and polycatenanes in both steady shear and steady extensional flow, a critically uncharacterized domain for MIPs.
We show that all three MIP types show enhanced tumbling in shear flow, weaker shear thinning, and lower normal stress differences in extensional flow.
Also, while polyrotaxanes show higher viscosity both in shear and extension, polycatenanes and daisy chains exhibit lower viscosities.
Polymer extension is typically greater in both shear and extensional flows, and approaches linear values as the density of mechanical bonds in MIPs decreases.

\section{Methods}
We simulated single polymers in the dilute solution limit with a coarse-grained bead-based model. 
Using a previously established model for MIPs~\cite{rauscher2020thermodynamics,becerra2024single}, each polymer consists of \(N\) beads that interact through bonded and non-bonded potentials.
The pairwise non-bonded interaction potential is the Weeks–Chandler–Anderson potential~\cite{weeks1971role}

\begin{equation}
U_{\mathrm{WCA}} = \sum_{\substack{i,j \\ j>i}} \left\{ 
\begin{array}{ll}
\displaystyle 4\epsilon \left[ \left(\frac{\sigma}{r_{ij}} \right)^{12} - \left(\frac{\sigma}{r_{ij}} \right)^{6} \right] - U_c, & r_{ij} < r_c \\[20pt]
0, & r_{ij} \geq r_c
\end{array} 
\right\}
\label{eq:Lennard-Jones}
\end{equation}
where \( r_{ij} = \left| \mathbf{r}_j - \mathbf{r}_i \right| \) is the distance between particles \(i\) and \(j\), \(r_c=2.2449241\) is the minimum energy cutoff distance, \( U_c = 4 \epsilon \left[ \left(\frac{\sigma}{r_c} \right)^{12} - \left(\frac{\sigma}{r_c} \right)^{6} \right] \) is the Lennard-Jones energy at the cutoff distance,  and \( \sigma \) and \( \epsilon \) are the Lennard-Jones length and energy parameters, respectively. 
To prevent polyrotaxane and daisy chain rings from dethreading, we introduced non-bonded interactions between capping beads and ring beads that are modeled without the attractive term, shift and cutoff by using \( U_{\mathrm{cap}} = 4 \epsilon_{\mathrm{cap}} \left(\frac{\sigma}{r_{ij}} \right)^{12} \), where $\epsilon_{\mathrm{cap}} = 10^{5} \epsilon$  in daisy chains and $\epsilon_{\mathrm{cap}} = 10^{10} \epsilon$ in polyrotaxanes.
Due to the absence of an attractive term, these \(\epsilon_{\mathrm{cap}}\) values are equivalent to 2.6 and 6.8 fold increase in $\sigma$ size parameter, respectively.
Bonds between beads are modeled with finitely extensible nonlinear elastic (FENE) springs~\cite{kremer1990dynamics}:
\begin{equation}
U_{\mathrm{FENE}} = -\sum_{i,j} \frac{1}{2} k_s r_{\text{max}}^2 \log \left[1 - \left(\frac{r_{ij}}{r_{\text{max}}} \right)^2 \right], \quad r_{ij} < r_{\text{max}},
\label{eq:FENE}
\end{equation}
where $k_{s}=30\varepsilon/\sigma^2$ is the spring strength and $r_{\mathrm{max}}=1.5\sigma$ is the maximum bond extension.

The bead positions evolve according to the following stochastic differential equation incorporating hydrodynamic interactions and implicit solvent~\cite{ermak1978brownian,ottinger1996stochastic}:
\begin{equation}
d\mathbf{r} = \left[\mathbf{v} + \frac{1}{k_{B}T} \mathbf{D} \cdot \mathbf{f}\right] dt + \sqrt{2} \, \mathbf{B} \cdot d\mathbf{w},
\label{eq:equation of moton}
\end{equation}
\noindent
where $\mathbf{r}$ is a vector of particle positions with \(3N\) elements, $\mathbf{v}$ is the solvent flow velocity at the particle positions, $\mathbf{f}$ is a vector of forces acting on the particles, $\mathbf{D}$ is the diffusivity matrix, $\mathbf{B} = \sqrt{\mathbf{D}}$, and $\mathbf{w}$ is a vector containing independent Wiener processes with zero mean and variance $dt$.
In Eq. \ref{eq:equation of moton}, the term in the brackets is the deterministic drift, while the other term accounts for the Brownian motion of the particles due to random collisions with solvent molecules. 
The force vector is $\mathbf{f} = -\nabla U$, where $U = U_{\text{WCA}} + U_{\text{FENE}} + U_{\mathrm{cap}}$ is the total intramolecular potential. 
The matrix $\mathbf{D}$ incorporates hydrodynamic interactions between the particles. 
Each $3 \times 3$ block of the matrix $\mathbf{D}$ represents hydrodynamic interactions between particles \(i\) and \(j\) and is given by
\begin{equation}
\displaystyle
\mathbf{D}_{ij} = \frac{k_B T}{\zeta} 
\left[ (1 - \delta_{ij}) \bm{\Omega}_{ij} + \delta_{ij} \mathbf{I} \right], \quad i, j = 1, \ldots, N, 
\label{eq:diffusion matrix}
\end{equation}
where $\delta_{ij}$ is the Kronecker delta, $\mathbf{I}$ is the $3 \times 3$ identity matrix, \(\zeta\) is the friction coefficient, \(a=\sigma/2\) is the hydrodynamic radius, and
\begin{equation}
\bm{\Omega}_{ij} = 
\begin{cases} 
\frac{3a}{4r_{ij}} 
\left[ 
\left( 1 + \frac{2a^2}{3r_{ij}^2} \right) \mathbf{I} + 
\left( 1 - \frac{2a^2}{r_{ij}^2} \right) \frac{\mathbf{r}_{ij} \mathbf{r}_{ij}}{r_{ij}^2}
\right], & r_{ij} \geq 2a, \\[15pt]
\left( 1 - \frac{9r_{ij}}{32a} \right) \mathbf{I} + 
\left( \frac{3r_{ij}}{32a} \right)\frac{\mathbf{r}_{ij} \mathbf{r}_{ij}}{r_{ij}^2} , & r_{ij} < 2a,
\end{cases}
\label{eq:RPY tensor}
\end{equation}
is the Rotne–Prager–Yamakawa tensor~\cite{rotne1969variational,yamakawa1970transport}.
To remove the overall translational motion of the molecule, the solvent velocity is centered around the center of mass of the polymer $\mathbf{r}_{\mathrm{COM}}=[x_{\mathrm{COM}}, y_{\mathrm{COM}}, z_{\mathrm{COM}}]^\top$ \cite{miao2017iterative}. 
The solvent velocity $\mathbf{v}_i$ for bead $i$ with spatial coordinate $\mathbf{r}_{i}=[x_i, y_i, z_i]^\top$ is zero at equilibrium; $[\dot{\gamma} (y_i-y_{\mathrm{COM}}), 0, 0]^\top$ in shear flow, where $\dot{\gamma}$ is the shear rate; and  $[\dot{\varepsilon} (x_i-x_{\mathrm{COM}}), -\dot{\varepsilon} (y_i-y_{\mathrm{COM}}) / 2, -\dot{\varepsilon} (z_i-z_{\mathrm{COM}}) / 2]^\top$ in uniaxial extensional flow, where $\dot{\varepsilon}$ is the extension rate. 
Furthermore, the Weissenberg number (Wi) measures the flow strength and is defined as $\dot{\gamma} \tau_R$ for shear flow and $\dot{\varepsilon} \tau_R$ for uniaxial extensional flow, where $\tau_R$ is the longest relaxation time of the polymer.
Relaxation time is $\tau_R =R_g^2/6D_{\mathrm{COM}}$, where the center-of-mass diffusivity was calculated from the equilibrium conformations using the Kirkwood formalism \cite{dunweg2002corrections}. 
Relaxation time results are summarized in SI Tables A.1 to A.4. 
The variables are nondimensionalized using $a$ as the unit of length, $k_B T$ as the unit of energy, and the diffusion time of a bead $\tau=\zeta a^2/{k_B T}$ as the unit of time. 
Eq. \ref{eq:equation of moton} is integrated numerically using a semi-implicit Euler scheme~\cite{jendrejack2000hydrodynamic,jendrejack2002stochastic} to allow the use of larger time steps compared to explicit methods.
The diffusivity matrix $\mathbf{D}$ and the Brownian term are treated explicitly and the remaining terms are treated implicitly.
The following set of nonlinear algebraic equations in
$\mathbf{r}^{*}$ are obtained using the semi-implicit algorithm:
\begin{align}
\mathbf{r}^*(t^*{+}\Delta t^*) =\, \mathbf{r}^*(t^*) 
+ \sqrt{2\Delta t^*}\, \mathbf{B}^*(t^*) \cdot \Delta \mathbf{w}^* \notag \\
+ \Big[\mathbf{v}^*(t^*{+}\Delta t^*) + \mathbf{D}^*(t^*) \cdot \mathbf{f}^*(t^*{+}\Delta t^*)\Big] \Delta t^*
\label{eq:semi-implicit}
\end{align}
where $\Delta t^{*}$ is the time step, $\Delta \mathbf{w}^*$ is a vector of normally-distributed random numbers with zero mean and unit variance, and the star represents non-dimensional quantities. $\mathbf{B}^{*}$ is calculated at time $t^{*}$ using Cholesky decomposition of $\mathbf{D}^{*}$.

Equation \ref{eq:semi-implicit} is solved iteratively by a Jacobian-free Newton–Krylov method~\cite{knoll2004jacobian} through the NITSOL software package~\cite{pernice1998nitsol} without using a preconditioner. 
This approach has recently been used to simulate bottlebrush polymers~\cite{dutta2024brownian}.
We use OpenMP shared memory parallelism with 1-4 CPUs per simulation (depending on system size) to more efficiently calculate forces, perform the Cholesky decomposition of the $\mathbf{D^*}$ matrix, and integrate the equation of motion. Initial configurations were generated with equilibrium bond lengths and energy minimization was carried out prior to production simulations. 
All properties were calculated by averaging over 10 to 20 trajectories of at least 100 relaxation times long ($10^6-10^8$ time steps) with time steps ranging from $10^{-2}$ to $10^{-4}$ depending on the strain rate. Error bars in the graphs represent the standard error of the mean. 
We calculated power-law exponents and their error using parametric bootstrap (Monte Carlo resampling) \cite{efron1992bootstrap}. 
The fractional extension in the flow direction is $\langle X \rangle / L$, where $X$ is the difference between minimum and maximum coordinates in the $x$ direction and $L$ is the maximum contour length.
The maximum contour length is $L = 3a (N- 1)$ and $L = 3a (N_{bb}- 1)$ for linear polymers and polyrotaxanes respectively, where $N$ is the total number of beads and $N_{bb}$ is the number of backbone beads. 
For daisy chains and polycatenanes, a pair of equal and opposite forces is applied to the end beads of the polymers in the $x$-direction and gradually increased under zero-temperature conditions. 
The polymer extension just before a mechanical bond fails, corresponding to a maximum in stress ($\tau^{p}_{xx}$) is recorded as the maximum contour length, $L$.

In dilute polymer solutions, the polymer contribution to the stress tensor ($\boldsymbol{\tau}^p$) is calculated using the following equation \cite{bird1977dynamics, Graham_2018, bosko2008effect}
\[
\frac{\boldsymbol{\tau}^p}{n_p k_B T} = -(N - 1) \mathbf{I} - \sum_{i=1}^{N-1} \sum_{j=i+1}^{N} {\mathbf{f}{^*_{ij}}} {\mathbf{r}{^*_{ij}}}.
\]
where $n_p$ is the polymer number density,
${\mathbf{r}{^*_{ij}}} = {\mathbf{r}{^*_j}} - {\mathbf{r}{^*_i}}
$ is the non-dimensionalized distance vector, and ${\mathbf{f}}^*_{ij}$ is the non-dimensionalized force on bead $i$ due to bead $j$. 
From this, we can obtain the shear stress $\tau_{xy}$ and the first normal stress difference, $N_1 = \tau^p_{xx} - \tau^p_{yy}$. 
The uniaxial extensional viscosity is $\eta_E = N_1 / \dot{\varepsilon}$, shear viscosity is $\eta =\tau^p_{xy} / \dot{\gamma}$, and the first normal stress coefficient is $\psi_1 = N_1 / \dot{\gamma}^2$.

Under steady uniaxial extension, we calculate the fractional extension \( \langle X \rangle / L \), extensional viscosity $\eta_e$ and first normal stress difference \(N_1\) across a range of Wi values. 
Under steady shear flow, we calculate the fractional extension in $x$ (flow), $y$ (gradient), and $z$ (vorticity) directions (\( \langle X \rangle / L \), \( \langle Y \rangle / L \), \( \langle Z \rangle / L \)), first normal stress coefficient ($\psi_1$), shear viscosity $\eta$, orientation angle with respect to the flow $\theta$, and tumbling frequency ($\omega \tau_R$) over a range of Wi values. 
In shear, we track and analyze the evolution of the angle between the end-to-end vector of the polymer chains and the $x$-axis to calculate the tumbling time, similar to the work of Dalal et. al. \cite{saha2012tumbling}. 
We calculate tumbling frequency as inverse tumbling time, normalized by the relaxation time ($\tau_{R}$).

We note that the parameters of this model allow the mechanical bonds to maintain integrity up to moderate flow strength in extensional flow and high flow strength in shear flow.
For strong enough flows, however, the hydrodynamic forces cause bond crossing events that break up the mechanical interlocks.
We report data for flow strengths up until the point that the mechanical bonds fail.
Mechanical bond failure in MIPs is expected from experiments~\cite{muramatsu2021rotaxane, zhang2019mechanical,lee2016mechanical}, and we address this point in greater detail in our discussion.

\section{Results}
\begin{figure*}[htbp]
  \centering
  \begin{minipage}[t]{0.48\textwidth}
    \centering
    \begin{subfigure}[t]{\textwidth}
      \includegraphics[width=0.99\textwidth]{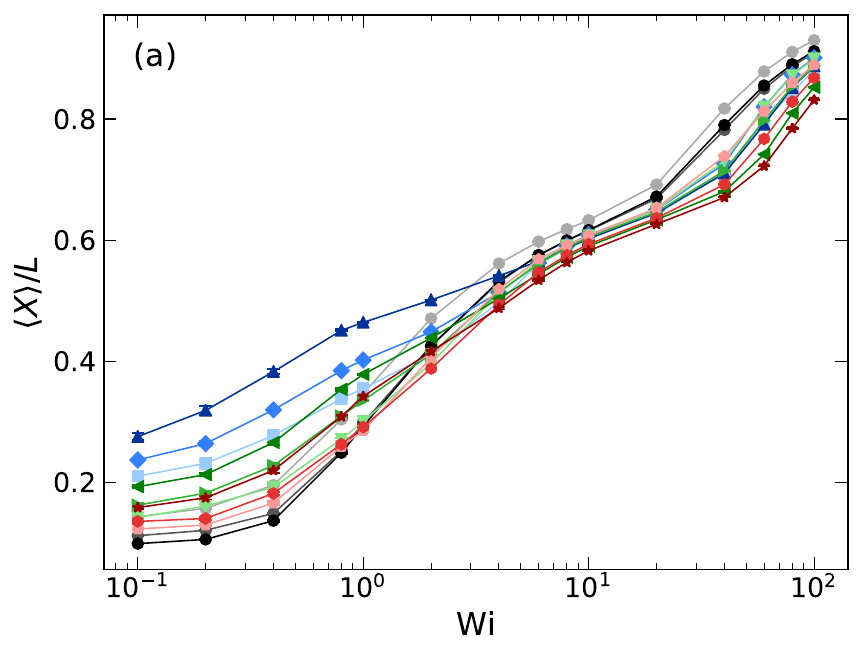}
    \end{subfigure}
  \end{minipage}
  \hfill
  \begin{minipage}[t]{0.48\textwidth}
    \centering
    \begin{subfigure}[t]{\textwidth}
      \includegraphics[width=0.99\textwidth]{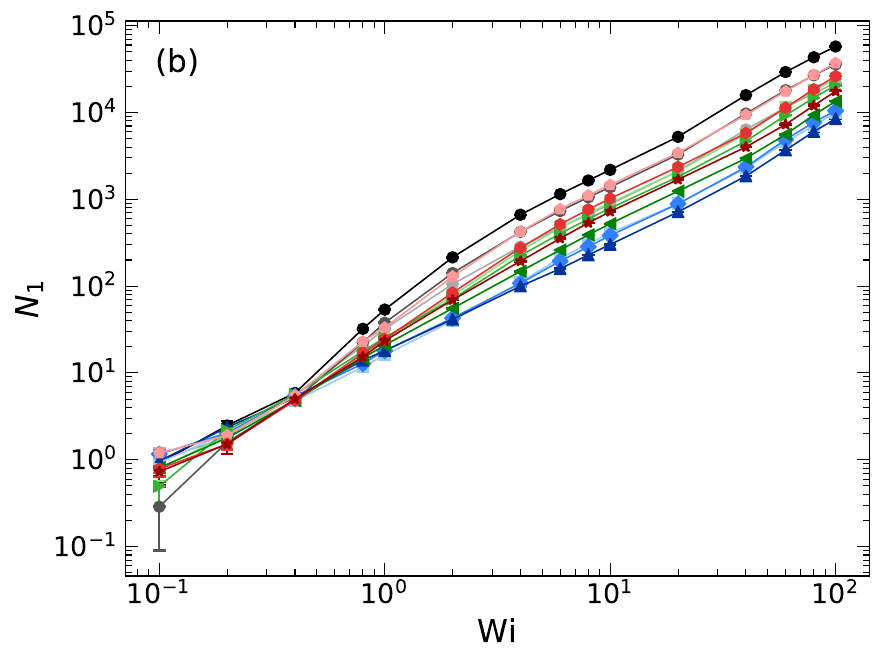}
    \end{subfigure}
  \end{minipage}

  \vspace{0.5em}
  \begin{subfigure}[b]{\textwidth}
    \centering
    \includegraphics[width=0.7\textwidth]{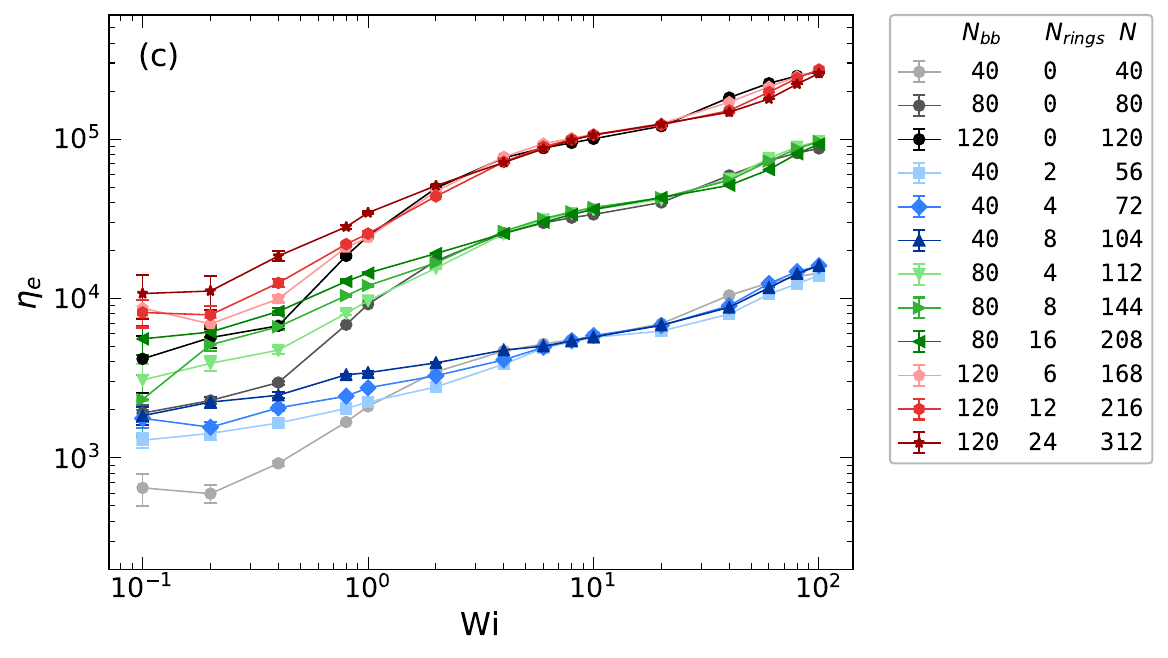}
  \end{subfigure}
     \captionsetup{justification=raggedright,singlelinecheck=false}

  \caption{Polyrotaxanes properties in extensional flow as a function of Weissenberg number (Wi): (a) fractional extension in flow direction $( \langle X \rangle / L)$,
  (b) first normal stress difference $(N_{1})$,
  (c) extensional viscosity $(\eta_{e})$. Linear polymers are shown in shades of gray with darker shades indicating more beads. Polyrotaxanes with 40, 80, and 120 backbone beads ($N_{bb}$) are shown in blue, green, and red, respectively; darker shades indicate higher ring density ($N_\mathrm{rings}/N_{bb}$). The rings in the polyrotaxanes are composed of 8 beads, regardless of backbone length.}
  \label{fig:combined_PR_extensional_plots}
\end{figure*}

\begin{figure*}[htbp]
  \centering
  % Left block: 2x2 plots
  \begin{minipage}[c]{0.85\textwidth}
    \centering
    % Row 1
    \begin{subfigure}[b]{0.48\textwidth}
      \centering
      \includegraphics[width=\textwidth]{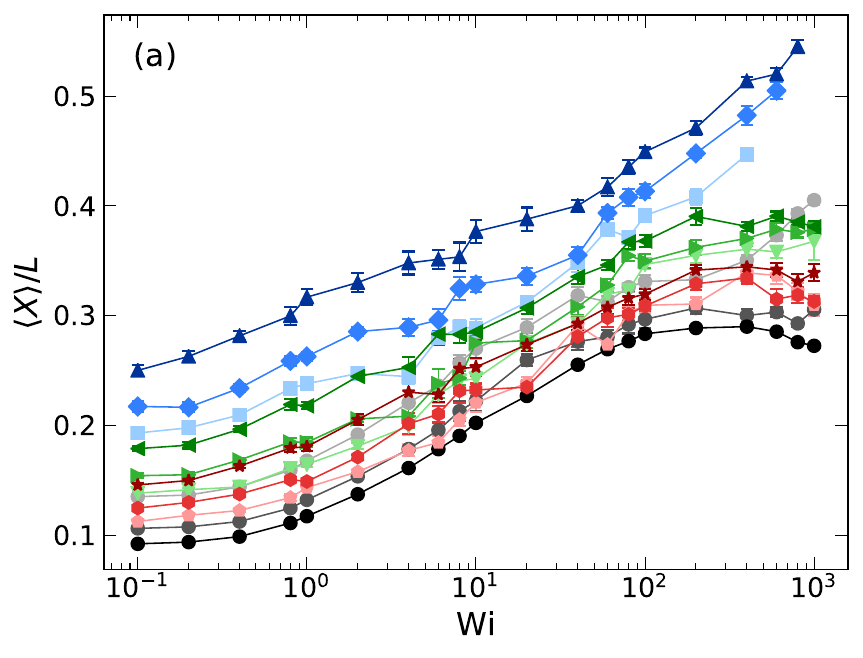}
    \end{subfigure}\hfill
    \begin{subfigure}[b]{0.48\textwidth}
      \centering
      \includegraphics[width=\textwidth]{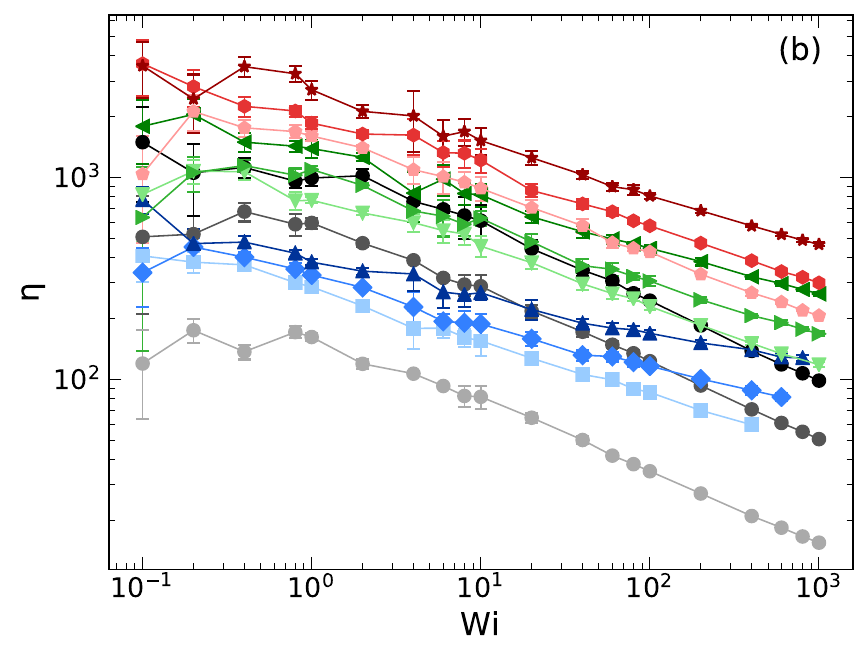}
    \end{subfigure}

    \vspace{0.5em}

    % Row 2
    \begin{subfigure}[b]{0.48\textwidth}
      \centering
      \includegraphics[width=\textwidth]{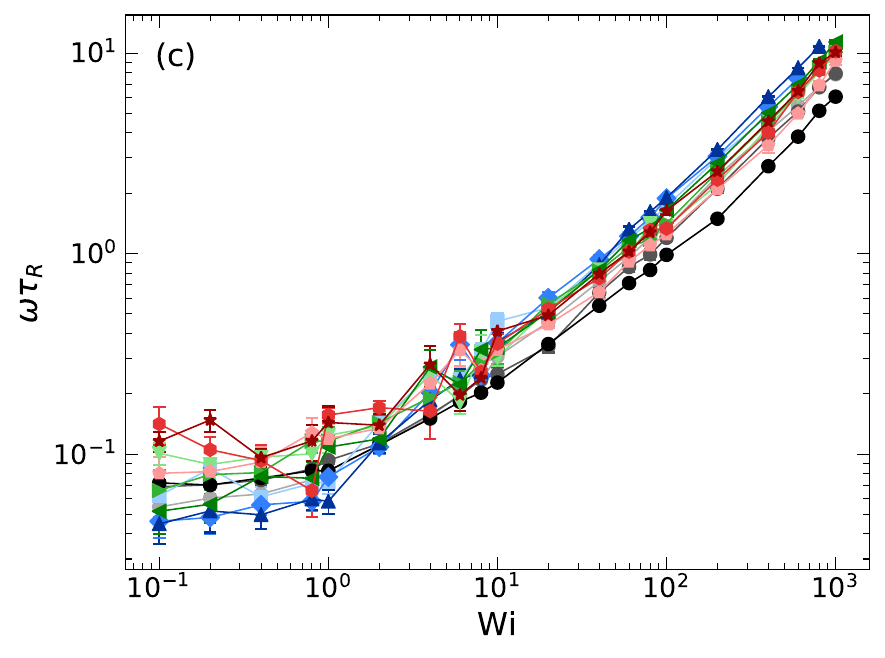}
    \end{subfigure}\hfill
    \begin{subfigure}[b]{0.48\textwidth}
      \centering
      \includegraphics[width=\textwidth]{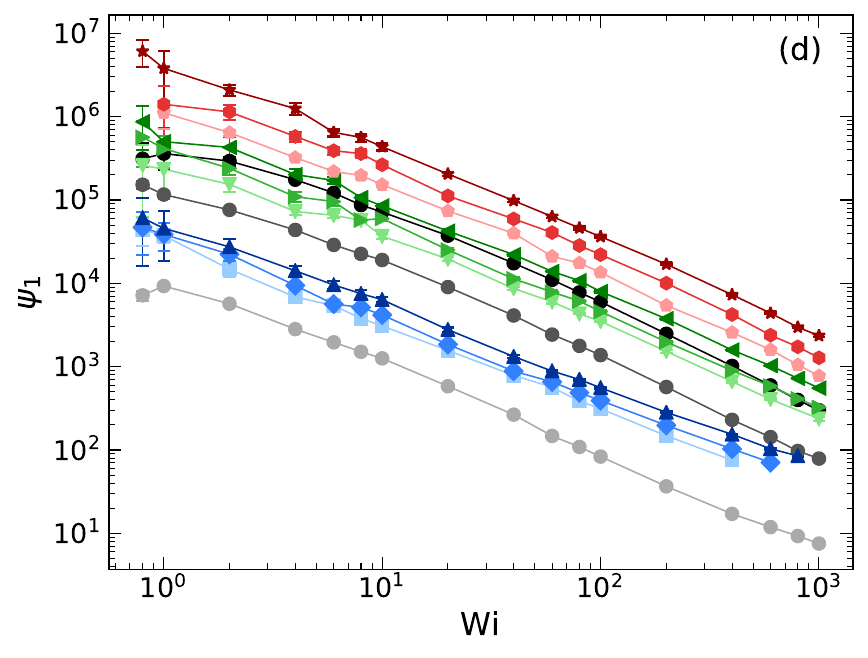}
    \end{subfigure}
  \end{minipage}%
  \hfill
  % Legend on the right
  \begin{minipage}[c]{0.13\textwidth}
    \centering
    \includegraphics[width=\textwidth]{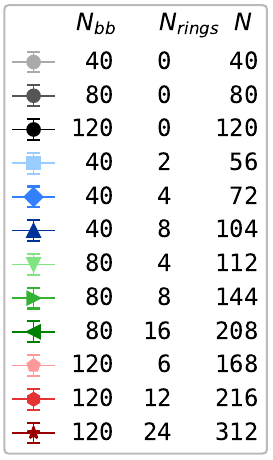}
  \end{minipage}
     \captionsetup{justification=raggedright,singlelinecheck=false}

  \caption{Polyrotaxane properties in shear flow as a function of Weissenberg number (Wi): (a) fractional extension in flow direction $( \langle X \rangle / L)$, (b) shear viscosity ($\eta$), (c) tumbling frequency ($\omega\tau_{R}$), and (d) first normal stress coefficient (\(\psi_1\)). 
  Linear polymers are shown in shades of gray with darker shades indicating more beads. 
  Polyrotaxanes with 40, 80, and 120 backbone beads ($N_{bb}$) are shown in blue, green, and red, respectively; darker shades indicate higher ring density ($N_\mathrm{rings}/N_{bb}$). 
  The rings in the polyrotaxanes are composed of 8 beads, regardless of backbone length.
  Fitted power law exponents for \(\mathrm{Wi}>10^{2}\) for \(\eta\) and \(\psi_1\) are reported in SI Tables E.1 and E.2, respectively.
  }
  \label{fig:PR_all_plots}
\end{figure*}

\subsection{Polyrotaxanes: Extensional Flow}
At the lower end of Wi range, the behavior of polyrotaxanes diverges significantly from linear polymers. 
Fig. \ref{fig:combined_PR_extensional_plots}a shows that adding rings to linear polymers increases the fractional extension in the flow direction.
This low Wi effect is largely equilibrium in nature, as demonstrated in SI Tables A.1 and A.2, which give the equilibrium radius of gyration \(R_g\), asphericity \cite{theodorou1985shape}, and prolateness \cite{vatin2024conformation} of linear polymers and polyrotaxanes.  
Shorter polyrotaxanes with a high density of mechanical bonds are more extended (higher \(R_g\)), less spherical (higher asphericity), and more rod-like (higher prolateness) than linear polymers with a similar backbone length.
This is in agreement with experimental works on PEG-based polyrotaxanes where more rings elongate and rigidify the backbone \cite{qiu2020precise}.
In general, both rings and backbones are rigidified after threading as rotaxanes, even when the starting components are flexible because the intermolecular and steric interactions reduce the conformational space and increase barriers \cite{fadler2022rigidity}.
This leads to a much less pronounced transition between a coiled state and stretched state in polycatenanes, where the stiffening effect of the rings is similar to the reduced coil-stretch transition seen in linear polymers as the backbone becomes stiffer~\cite{andrews1998configuration}.
The differences in extension become less pronounced for longer backbones or fewer mechanical bonds, which is also reflected in Fig.~\ref{fig:combined_PR_extensional_plots}a.  
The low Wi differences, then, can be understood as a difference in equilibrium shape that will align with weak flows.
At higher Wi values, differences due to interlocked rings become negligible as both linear and polyrotaxane polymers reach more extended backbone configurations.
Examples of linear polymers and polyrotaxanes in these steady-state stretched configurations are shown in SI Fig. C.1.
We omit extensional flow data for $\mathrm{Wi} > 100$ because mechanical bonds in polyrotaxanes fail in this regime as hydrodynamic forces become strong enough to de-thread rings past the end caps.

The first normal stress difference $(N_{1})$ is lower in polyrotaxanes than linear polymers across the range of simulated Wi values, as shown in Fig. \ref{fig:combined_PR_extensional_plots}b. 
Polymers with more rings exhibit lower $N_{1}$ values compared with linear polymers, demonstrating that mechanical bonds lead to weaker elastic behavior.
This weaker elastic behavior is notably similar to bottlebrush polymers~\cite{dutta2024brownian}, with increasing bristle length playing a similar role to backbone ring density in polyrotaxanes.

As shown in Fig. \ref{fig:combined_PR_extensional_plots}c, at low Wi, polyrotaxanes show greater extensional viscosity $(\eta_e)$ compared to linear polymers at the same backbone length, with higher number of rings resulting in higher $\eta_e$. 
The higher $\eta_e$ can be attributed to the more extended configuration of chains with threaded rings than chains without rings. 
As a result, polyrotaxanes have higher resistance to further elongation. 
At higher Wi values, the $\eta_e$ difference between polyrotaxanes and linear polymers lessens because all chains reach a relatively extended configuration. 
As shown in SI Fig. D.1, in this regime, the rings are pulled to the ends of the backbone and therefore, do not play a significant role in the backbone conformation and resistance to extension.

\subsection{Polyrotaxanes: Shear Flow}
Polymers in shear flow undergo continuous stretch-tumble cycles \cite{de1974coil, smith1999single, teixeira2005shear} and the fractional extension values reported in this work are averaged over those cycles.
Example configurations of a polyrotaxane (and other MIP types) during these cycles are shown in SI Fig. C.1.
According to Fig. \ref{fig:PR_all_plots}a, fractional extension in flow direction ($\langle X \rangle / L$) in polyrotaxanes is higher than the linear polymers with the same number of backbone beads across the Wi range tested. 
A greater number of rings result in higher extension and more deviation from linear polymers. 
Similar to low Wi extensional flow, the ring-backbone interactions increase rigidity and cause the molecules to exhibit more elongated configurations on average, resulting in comparatively higher extension in flow direction.
In real rotaxanes, this increased rigidity can be caused by hydrogen bonding between the ring and backbone, $\pi$-$\pi$ stacking between rings, metal coordination, steric interactions between rings, or electrostatic repulsions \cite{fadler2022rigidity}.

Polyrotaxanes exhibit higher shear viscosity $(\eta)$ than their linear counterparts, with larger number of threaded rings leading to further rise in viscosity, as shown in Fig. \ref{fig:PR_all_plots}b. 
The size of a polymer in solution has a direct relationship with the viscosity of the solution~\cite{krigbaum1953molecular,hiemenz2007polymer,rubinstein2003polymer}. 
As evident from SI Tables A.1 and A.2, the radius of gyration of polyrotaxanes is higher than linear polymers with the same backbone length, with additional threaded rings leading to larger polymer size.
For example, a 120 bead linear polymer has a radius of gyration of 15.9, compared to a polyrotaxane with 24 rings on a 120 bead backbone, which has a radius of gyration of 23.8. 

Similar to linear polymers, polyrotaxanes show shear thinning, although to a lesser degree, as evident from the power-law scaling exponent $\beta$ values shown in SI Table E.1. 
Additionally, polyrotaxanes with more rings have lower $\beta$ values. 
For example, a linear polymer with 80 beads has $\beta=-0.380$ in agreement with the previously reported value for this model~\cite{dutta2024brownian} and similar to other dilute linear polymers with excluded volume and hydrodynamic interactions (\(\beta=-0.28\))~\cite{moghani2017computationally}.
Threading 4 or 16 rings onto the backbone decreases the power law scaling value to \(-0.290\) and \(-0.225\), respectively, showing that more rings further suppress shear thinning.
The weaker shear thinning is due to the polyrotaxanes' decreased alignment with the flow, demonstrated by the average orientation angle in SI Fig. B.1c.
Increasing the number of rings results in less alignment with the flow on average (an angle \(\theta\) with greater deviation from \(0^{\circ}\)) because the rings cannot fully align with the flow due to repulsive interactions within the rings and between the rings and the backbone.
This misalignment leads to less relaxation and weaker shear thinning~\cite{teixeira2005shear,ryder2006shear}.

In weak flow ($\mathrm{Wi} \le 1$), shear forces are not sufficient to induce significant tumbling in either polyrotaxanes or linear polymers, as shown in Fig. \ref{fig:PR_all_plots}c. Consequently, both polymers have flat tumbling frequency profiles. As the flow strength increases ($1<\mathrm{Wi}<10$), polyrotaxanes start to show enhanced tumbling compared to the linear counterparts. 
At $\mathrm{Wi}\geq 10$, the polyrotaxanes show significantly higher tumbling frequency compared with linear polymers.
At $\mathrm{Wi}=800$, polyrotaxanes with 40 and 80 backbone beads (with 8 and 16 threaded rings, respectively) exhibit tumbling frequencies 38\% and 35\% higher than their linear polymer counterparts. 
Furthermore, increasing the number of threaded rings increases the tumbling frequency. For instance, at $\mathrm{Wi}=10^3$, a polyrotaxane with 80 beads in the backbone and 16 threaded rings tumbles 14\% more frequently than one with only 4 rings. 
Threading rings along the backbone also enhances the scaling of the tumbling frequency with Wi. 
As shown in SI Table E.3, adding 8 rings to a linear polymer with $N=40$ increases the scaling exponent from 0.741 to 0.826. 
The exponent obtained for linear polymers is consistent with the previously reported 3/4 power-law scaling of tumbling time with shear rate \cite{saha2012tumbling}.
The enhanced tumbling in polyrotaxanes can be explained by the threaded rings increasing the polymer’s hydrodynamic cross-section along the gradient direction, as shown is SI Fig. B.1a.
The expanded profile exposes a larger surface area to shear forces of the solvent, leading to more frequent tumbling events.
The polyrotaxanes also have a greater extent in the vorticity direction, demonstrated in SI Fig. B.1b.

Fig. \ref{fig:PR_all_plots}d shows the first normal stress coefficient ($\psi_1$) as a function of Wi. 
At a comparable backbone length, polyrotaxanes have higher $\psi_1$ than linear polymers, with more rings further increasing the value. 
As shown in SI Table E.2, the power-law exponent ($\beta$) in polyrotaxanes is only slightly higher than linear polymers. 
For example, for a polyrotaxane with 120 beads in the backbone and 24 threaded rings, $\beta=-1.24$ while for a linear polymer with $N=120$, we report $\beta=-1.32$. %\pm0.01$
The latter value is in close agreement with the previous reported value of -1.30 for this model~\cite{dutta2024brownian} and other excluded volume linear polymer models with hydrodynamic interactions (-1.13)~\cite{moghani2017computationally}.
Generally, the power law scaling of $\psi_1$ is reduced in polyrotaxanes relative to that of linear polymers with the same backbone size.
A similar effect was seen in bottlebrush polymers where side chains are grafted onto a linear backbone~\cite{dutta2024brownian}.

\begin{figure*}[htbp]
  \centering
  \begin{minipage}[t]{0.48\textwidth}
    \centering
    \begin{subfigure}[t]{\textwidth}
      \includegraphics[width=0.99\textwidth]{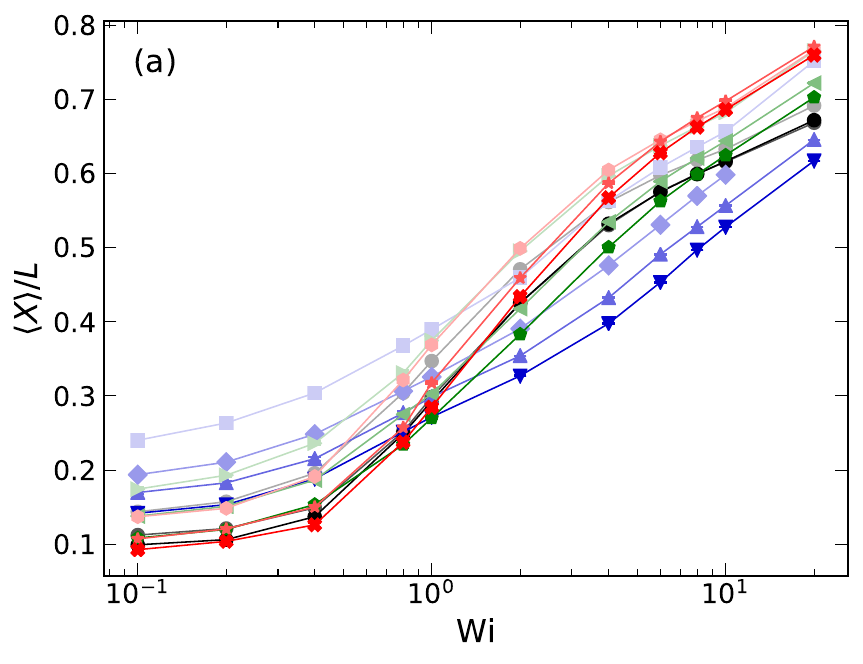}
    \end{subfigure}
  \end{minipage}
  \hfill
  \begin{minipage}[t]{0.48\textwidth}
    \centering
    \begin{subfigure}[t]{\textwidth}
      \includegraphics[width=0.99\textwidth]{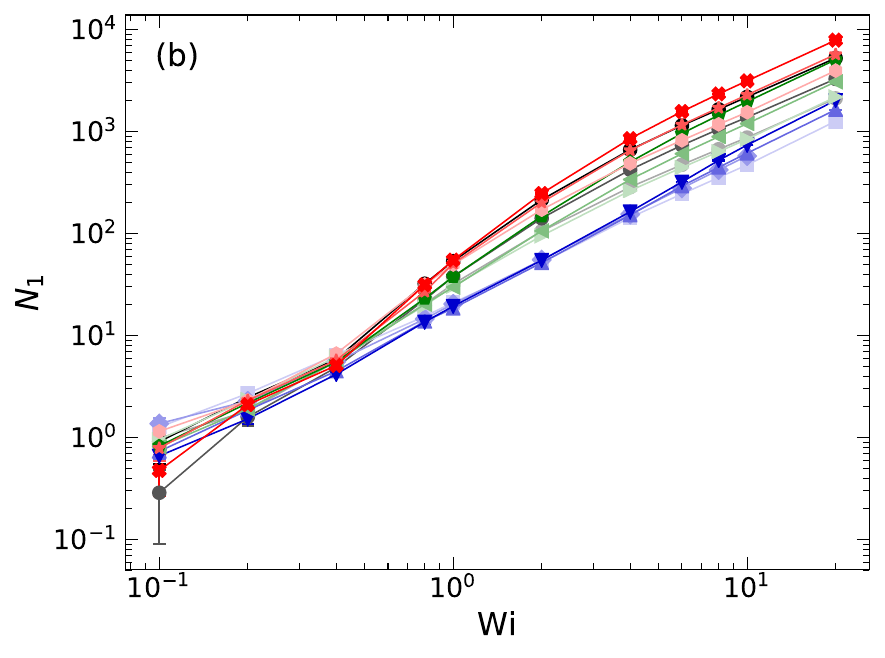}
    \end{subfigure}
  \end{minipage}
  
  \vspace{0.5em}
  \begin{subfigure}[b]{\textwidth}
    \centering
    \includegraphics[width=0.67\textwidth]{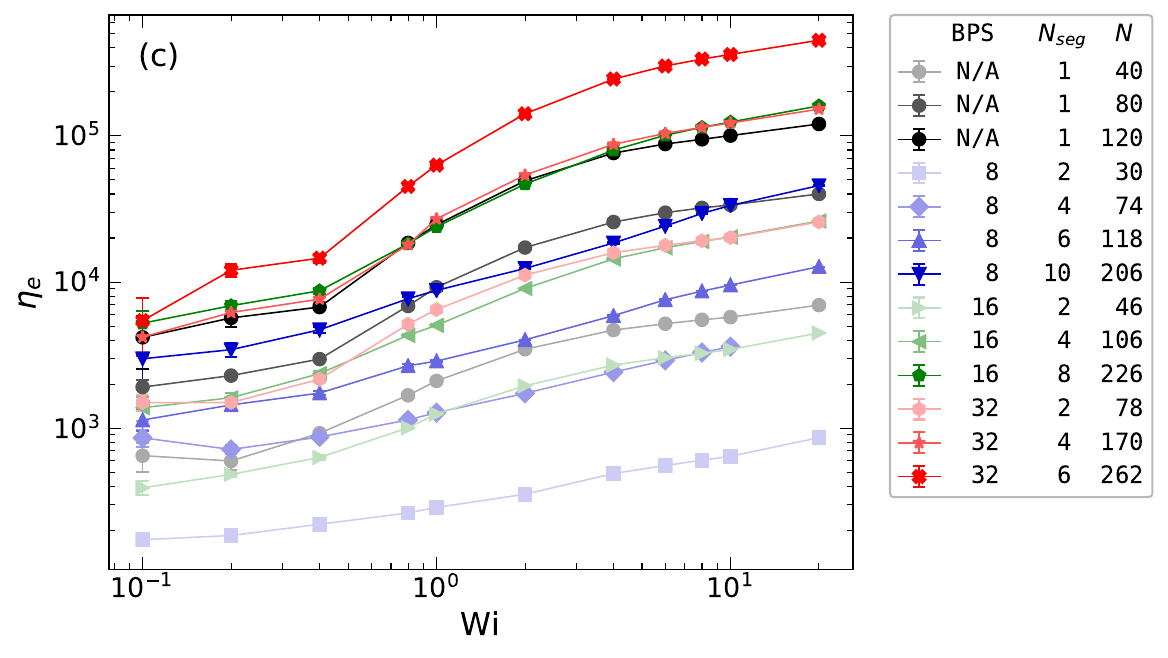}
  \end{subfigure}
     \captionsetup{justification=raggedright,singlelinecheck=false}

  \caption{Daisy chain properties in extensional flow as a function of Weissenberg number (Wi): (a) fractional extension in flow direction $( \langle X \rangle / L)$,
  (b) first normal stress difference $(N_{1})$,
  (c) extensional viscosity $(\eta_{e})$. Linear polymers are shown in
shades of gray with darker shades indicating more beads. Daisy chains with 8, 16, and 32 beads per segment (BPS) are shown in blue, green, and red, respectively; darker shades indicate more segments ($N_\mathrm{seg}$). The rings in all daisy chains are composed of 8 beads.}
  \label{fig:combined_DC_extensional_plots}
\end{figure*}

\begin{figure*}[htbp]
  \centering
  % Left block: 2x2 plots
  \begin{minipage}[c]{0.85\textwidth}
    \centering
    % Row 1
    \begin{subfigure}[b]{0.48\textwidth}
      \centering
      \includegraphics[width=\textwidth]{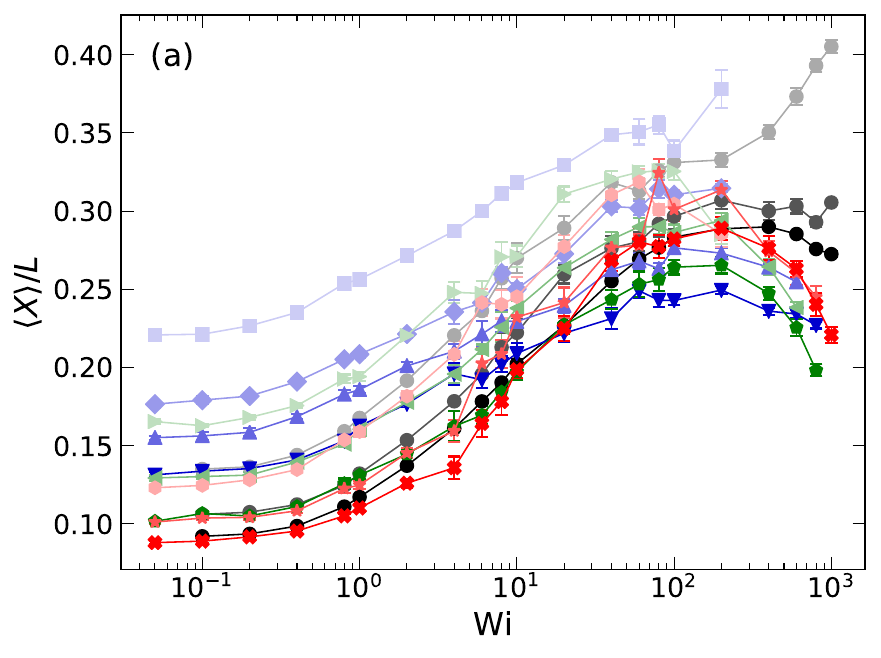}
    \end{subfigure}\hfill
    \begin{subfigure}[b]{0.48\textwidth}
      \centering
      \includegraphics[width=\textwidth]{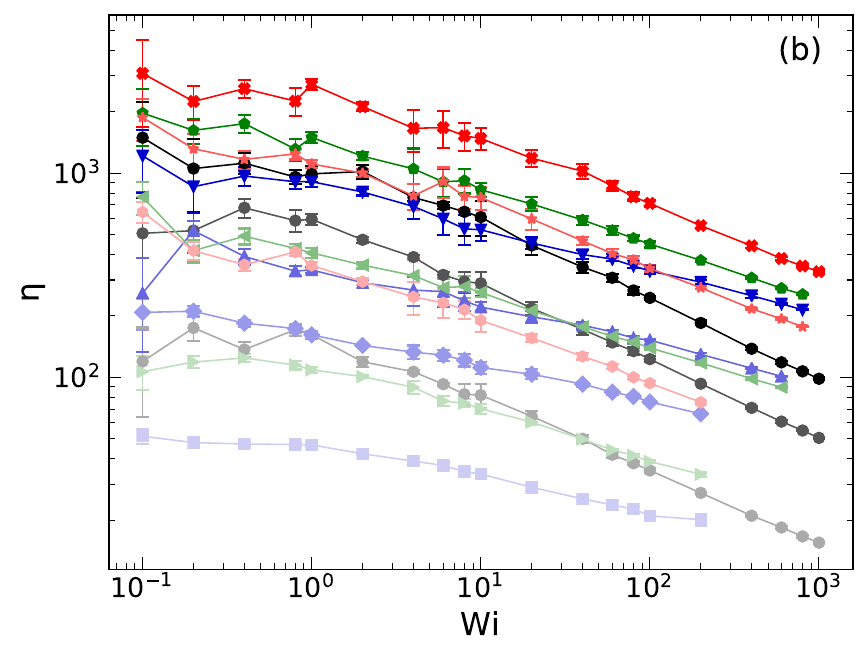}
    \end{subfigure}

    \vspace{0.5em}

    % Row 2
    \begin{subfigure}[b]{0.48\textwidth}
      \centering
      \includegraphics[width=\textwidth]{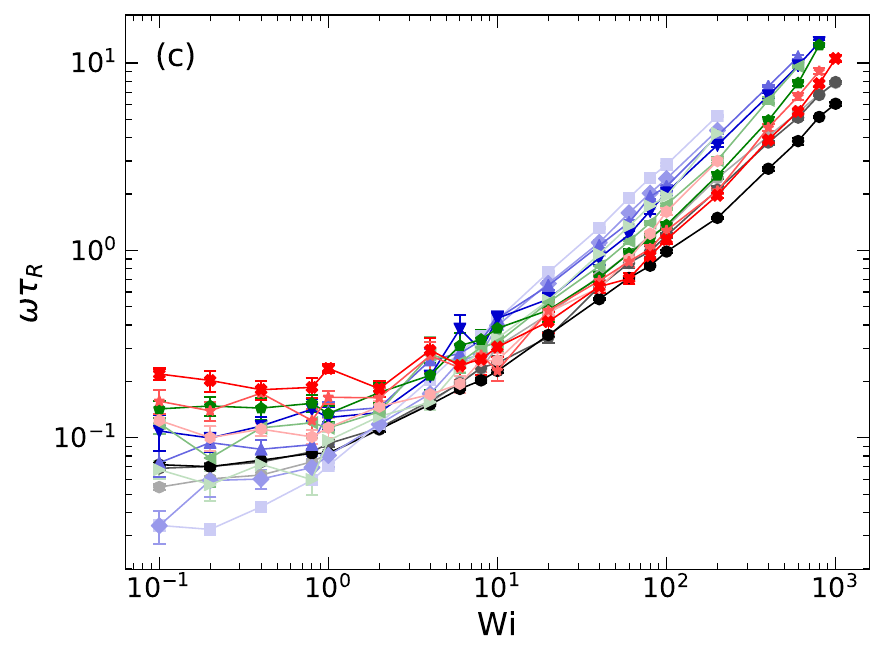}
    \end{subfigure}\hfill
    \begin{subfigure}[b]{0.48\textwidth}
      \centering
      \includegraphics[width=\textwidth]{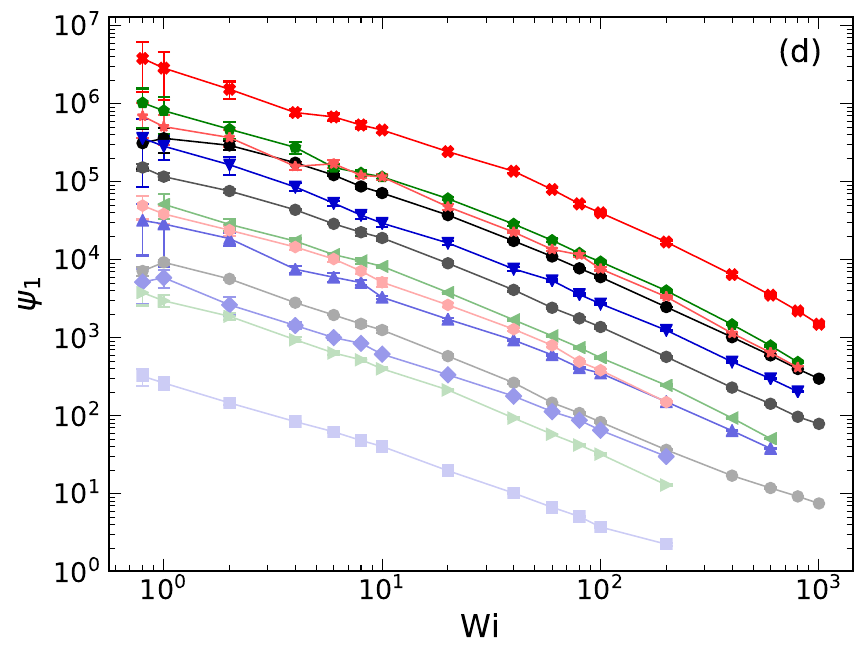}
    \end{subfigure}
  \end{minipage}%
  \hfill
  % Legend on the right
  \begin{minipage}[c]{0.13\textwidth}
    \centering
    \includegraphics[width=\textwidth]{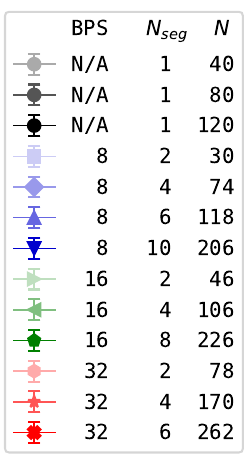}
  \end{minipage}
   \captionsetup{justification=raggedright,singlelinecheck=false}

  \caption{Daisy chain properties in shear flow as a function of Weissenberg number (Wi): (a) fractional extension in flow direction $( \langle X \rangle / L)$, (b) shear viscosity ($\eta$), (c) tumbling frequency ($\omega\tau_{R}$), and (d) first normal stress coefficient (\(\psi_1\)). 
  Linear polymers are shown in shades of gray with darker shades indicating more beads. 
  Daisy chains with 8, 16, and 32 beads per segment (BPS) are shown in blue, green, and red, respectively; darker shades indicate higher number of segments ($N_\mathrm{seg}$). 
  The rings in all daisy chains are composed of 8 beads. 
  Fitted power law exponents for \(\mathrm{Wi}>10^{2}\) for \(\eta\) and \(\psi_1\) are reported in SI Tables E.1 and E.2, respectively.}
  \label{fig:DC_all_shear_plots}
\end{figure*}

\subsection{Daisy Chains: Extensional Flow}
From Fig. \ref{fig:combined_DC_extensional_plots}a, at low Wi in extensional flow, daisy chains are more extended than linear polymers at similar $M_w$ with higher density of mechanical bonds leading to greater fractional extension. 
As an example, at \(\mathrm{Wi}=0.1\), a daisy chain with 8 beads per segment and 4 segments ($M_w=74$) is 72\% more extended than a comparable linear polymer ($M_w=80$) whereas a daisy chain with 32 beads per segment and 2 segments ($M_w=78$), which has lower density of mechanical bonds, is only 22\% more extended.
This is reflected in the shape parameters of these two polymers in equilibrium, where the chain with more short segments is more prolate (0.74 vs. 0.49) and has higher asphericity (0.70 vs. 0.61), demonstrated in SI Tables A.1 and A.3.

This behavior is similar to the backbone rigidifying effect of the rings as seen in the polyrotaxanes. In chains with longer segments, this effect diminishes and the chains become more coiled like a linear polymer.
Interestingly, shorter linear segments lead to greater extension at low Wi, but the trend reverses at high Wi where daisy chains with longer linear segments are more extended.
In the strong flow regime where the chains are far from equilibrium and more extended, the repulsions between rings become important and a stronger flow is be needed to stretch daisy chains with more segments. 
Making segments longer diminishes this effect and the chains become more like a linear polymer.
Overall, a higher density of mechanical bonds ``stiffens'' the polymer, leading to lower variability in the stretch and a weaker coil-stretch transition.
We report extensional flow data up to $\mathrm{Wi}=20$ where the hydrodynamic forces in cause bond crossing between the interlocked rings.

As Fig. \ref{fig:combined_DC_extensional_plots}b shows, at low Wi, first normal stress difference ($N_1$) values are close in linear and daisy chains. 
At higher Wi values, daisy chains have lower \(N_1\) values compared with linear polymers with the same molecular weight.
Also, comparing a daisy chain with 8 beads per segment and 4 segments to a chain with 32 beads per segment and 2 segments, which have comparable molecular weights of 74 and 78 respectively, shows that higher density of mechanical bonds leads to lower $N_1$ values.
At $\mathrm{Wi} = 10$, the former yields $N_1 = 577$, whereas the latter reaches a higher value of $N_1 = 1548$.

Extensional viscosity ($\eta_e$) follows a similar trend as \(N_1\), as shown in Fig. \ref{fig:combined_DC_extensional_plots}c. 
Daisy chains exhibit lower $\eta_e$ than linear polymers of equivalent molecular weight. 
For example, at $\mathrm{Wi}=10$, a daisy chain with 32 beads per segment and 2 segments ($M_w=78$) has 67\% lower $\eta_e$ compared to a linear polymer with 80 beads. Higher mechanical bond density increases the difference.
A daisy chain with 8 beads per segment and 6 segments ($M_w=118$) has an extensional viscosity 10 times lower than a linear polymer with 120 beads.
The same trend is evident at lower Wi values, but to a lesser extent. 
Overall, mechanical bonds reduce stress and $\eta_e$, but increase extension at low Wi while decreasing extension at high Wi.

\subsection{Daisy Chains: Shear Flow}
As shown in Fig. \ref{fig:DC_all_shear_plots}a, fraction extension in the flow direction has different regimes depending on flow strength. 
Previous studies on semi-flexible linear polymers show a similar shear response where polymers stretch at low shear rates and collapse at high shear rates \cite{lyulin1999brownian}. 
At $\mathrm{Wi} <0.8$, extension remains nearly constant in linear polymers and daisy chains, with daisy chains exhibiting greater extension than linear polymers of similar molecular weight.
For example, at \(\mathrm{Wi}=10^{-1}\), a daisy chain of 6 linear segments with 8 beads each (\(M_w=118\)) has a fractional extension of 0.16 compared to a linear polymer of \(M_w=120\) with an extension of 0.09.
At \(10^0 < \mathrm{Wi} < 10^2\), extension increases and then decreases around $\mathrm{Wi} \approx 10^2$. 
While the conformational change is seen in both linear and daisy chains, the variation is less significant in daisy chains with more mechanical bonds. 
For example, a daisy chain with 8 beads per segment and 6 segments (\(M_w = 118\)) has a fractional extension between 0.16 and 0.27 across the Wi range compared to a linear polymer of \(M_w=120\), which has extensions between 0.08 and 0.029 across the same Wi range.
Decreasing the number of segments or increasing the length of the linear segments increases the variation in extension.
Also, in daisy chains with the same number of segments, longer segments result in behavior similar to a linear polymer.
A terminal regime with increasing extension is seen at very high Wi in linear polymers.
However, this behavior could not be confirmed in daisy chains because with our current model the chains disentangle under very strong flow conditions due to bond crossing. 
We plan to address this shortcoming in future work. 

Fig. \ref{fig:DC_all_shear_plots}b shows that the  shear thinning degree is lower in daisy chains compared to linear polymers and the shear viscosity power-law scaling exponent \(\beta\) (SI Table~E.1) is lower in daisy chains with higher mechanical bond density. 
For example, a daisy chain with 8 beads per segment and 6 segments ($M_w=118$) has $\beta=-0.225$ while the corresponding linear polymer ($M_w=120$) has $\beta=-0.390$, which shows the less pronounced shear-thinning character of the daisy chains. 
Shear-thinning behavior becomes more linear-like as the daisy chains' linear segment length increases.
Similar to polyrotaxanes, this decreased shear-thinning in daisy chains is caused by their misalignment with the flow direction.
SI Fig. B.2c shows that the orientation angle deviates from the flow direction more for daisy chains with shorter linear segments or fewer segments.
At a comparable molecular weight, daisy chains also have lower zero-shear viscosity due to their smaller equilibrium size, as shown in SI Table A.3, highlighting another significant deviation from linear polymers.
For example, at $\mathrm{Wi}=20$, a daisy chain with 32 beads per segment and 2 segments ($M_w=78$) has 42\% lower viscosity than the corresponding linear polymer ($M_w=80$).
It is worth noting that because of the difference in shear-thinning degree between linear and daisy chain polymers, the viscosity difference is more pronounced in weaker flows and diminishes in stronger flows. 
As an example, a daisy chain with 8 beads per segment and 6 segments ($M_w=118$) has a factor of 5.2 lower viscosity than the comparable linear polymer ($M_w=120$) at $\mathrm{Wi}=0.1$ but the difference decreases at $\mathrm{Wi}=600$ to a less significant value of ~27\%.

Based on Fig.~\ref{fig:DC_all_shear_plots}c, at low Wi, the flow is not strong enough to cause significant tumbling in either the daisy chains or linear polymers.
At higher Wi (\(>7\)), daisy chains show significantly more frequent tumbling in shear flow. 
At $\mathrm{Wi}=200$, a daisy chain with 8 beads per segment and 6 segments ($M_w=118$) shows 87\% higher $\omega\tau_R$ than a linear polymer with 120 beads. 
Similarly, a daisy chain with 32 beads per segment and 2 segments ($M_w=78$) has 60\% higher $\omega\tau_R$ than the linear counterpart with 80 beads. 
Tumbling frequency also scales more strongly with Wi in daisy chains than linear polymers. 
As shown in SI table E.3, the scaling exponent of a daisy chain with $M_w=118$ is 0.838 compared to 0.732 of the equivalent linear polymer.
In fact, a daisy chain of two 16-bead linear segments exhibits a scaling exponent of 0.875, compared to the expected 0.75 exponent of a linear polymer~\cite{saha2012tumbling}.
The rings in the daisy chains increase the gradient-direction extent (shown in SI Fig.~B.2a) causes larger shear forces from the flow to be applied to the chains, resulting in more frequent tumbles.
This effect is less present in configurations with lower density of mechanical bonds, making them  more linear-like in terms of aspect ratio, showing less tumbling. 
On the other hand, the role of the rings in the aspect ratio is more pronounced in daisy chains with higher mechanical bond density, causing the chains to experience larger shear forces from the flow, resulting in higher tumbling frequency. 

According  to Fig. \ref{fig:DC_all_shear_plots}d, daisy chains show lower $\psi_1$ compared with linear polymers at the same molecular weight, with higher density of mechanical bonds further decreasing   $\psi_1$ value. 
As shown in SI Table E.2, the power-law scaling exponent ($\beta$) for daisy chains can be lower or higher than linear polymers with comparable molecular weight depending on the density of mechanical bonds. 
For example, a daisy chain with 32 beads per segment and 2 segments ($M_w=78$), $\beta=-1.349$ which is lower than the value of -1.239 for a linear polymer with 80 beads. On the other hand, a daisy chain with 8 beads per segment and 4 segments ($M_w=74$) has $\beta=-1.105$ due to higher density of mechanical bonds.

\begin{figure*}[htbp]
  \centering
  % Row 1: fractional extension in flow direction
  \begin{subfigure}[b]{0.48\textwidth}
    \centering
    \includegraphics[width=\textwidth]{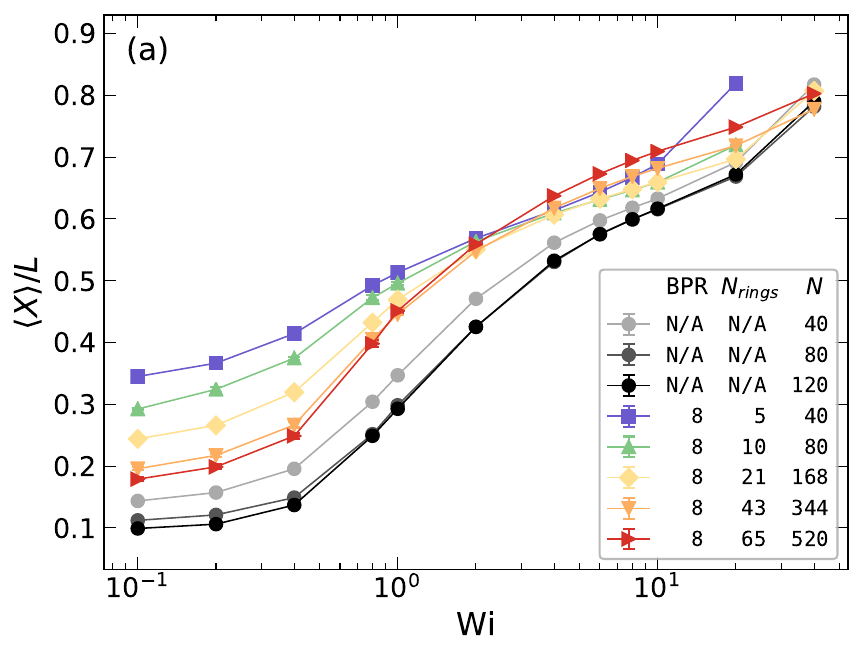}
  \end{subfigure}
  \hfill
  \begin{subfigure}[b]{0.48\textwidth}
    \centering
    \includegraphics[width=\textwidth]{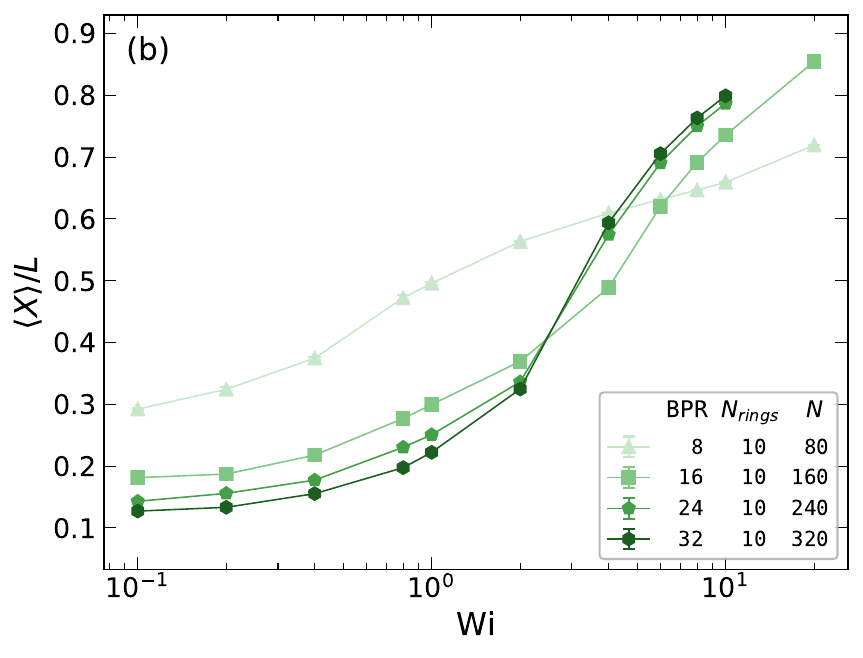}
  \end{subfigure}

  \vspace{0.5em}

  % Row 2: first normal stress difference
  \begin{subfigure}[b]{0.48\textwidth}
    \centering
    \includegraphics[width=\textwidth]{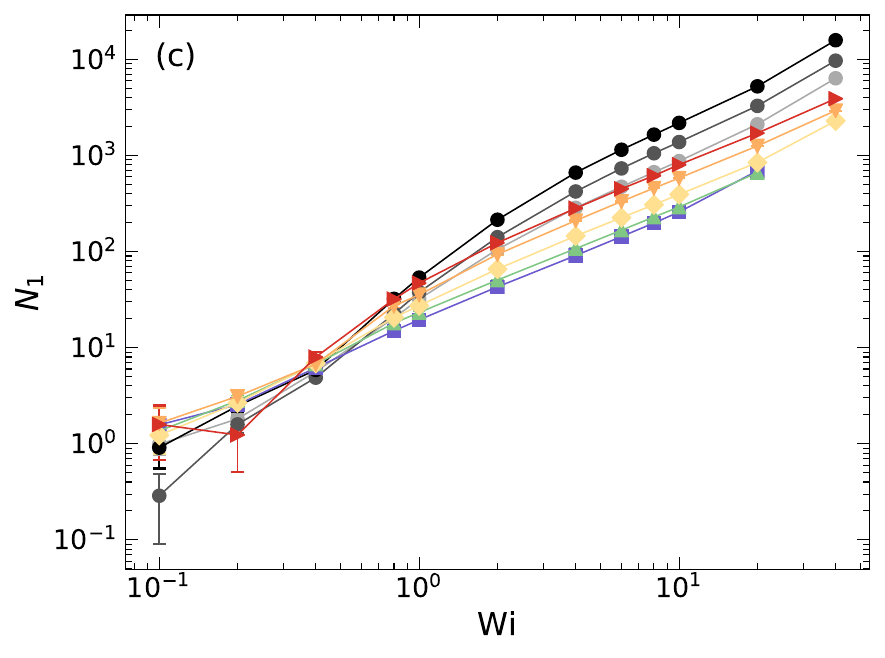}
  \end{subfigure}
  \hfill
  \begin{subfigure}[b]{0.48\textwidth}
    \centering
    \includegraphics[width=\textwidth]{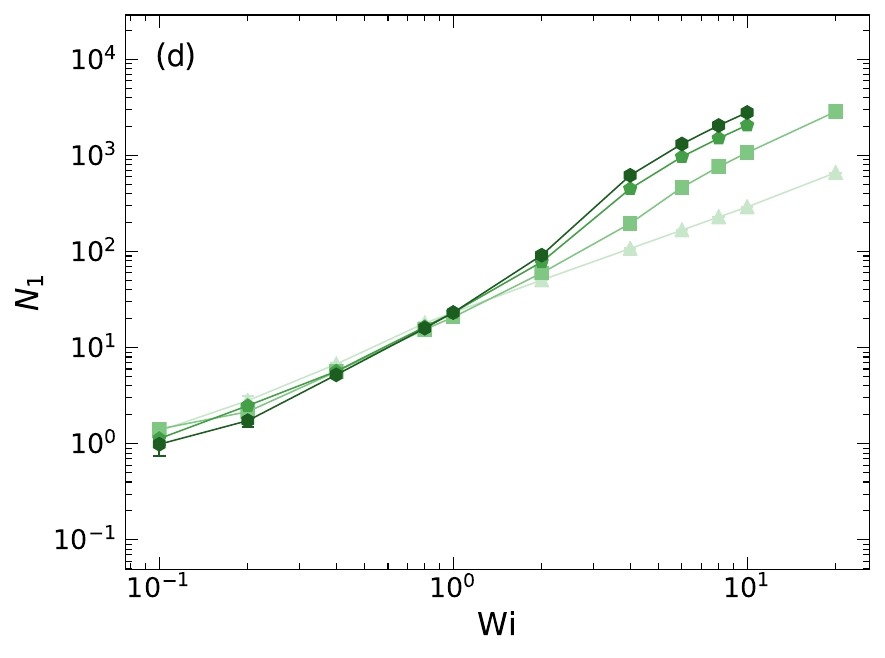}
  \end{subfigure}

  \vspace{0.5em}

  % Row 3: extensional viscosity
  \begin{subfigure}[b]{0.48\textwidth}
    \centering
    \includegraphics[width=\textwidth]{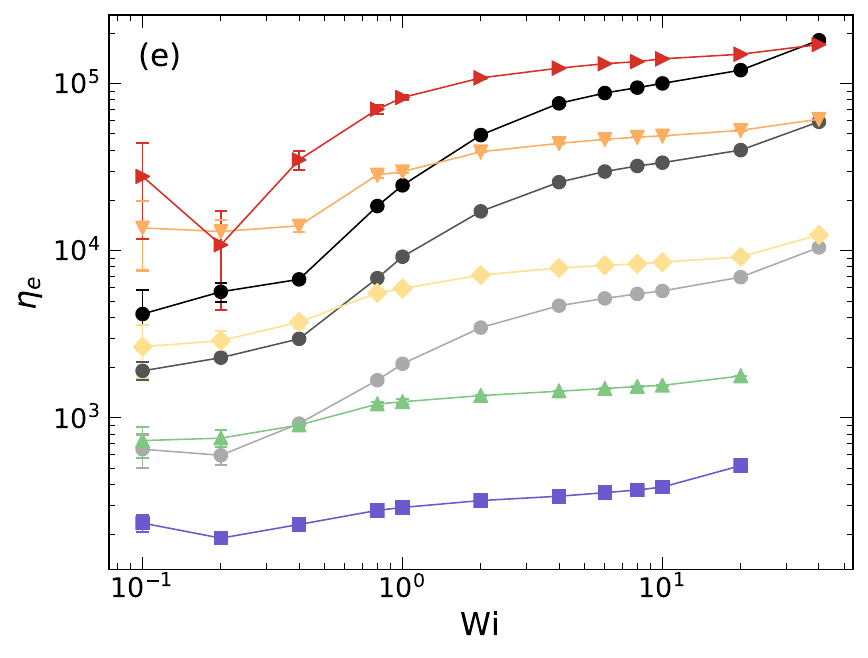}
  \end{subfigure}
  \hfill
  \begin{subfigure}[b]{0.48\textwidth}
    \centering
    \includegraphics[width=\textwidth]{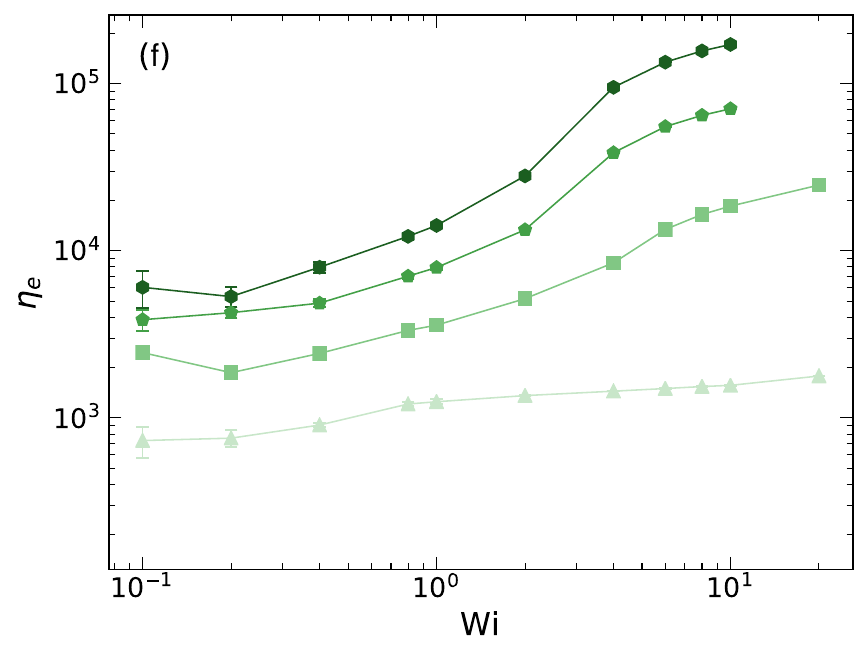}
  \end{subfigure}

   \captionsetup{justification=raggedright,singlelinecheck=false}
  \caption{Polycatenane properties in extensional flow as a function of Weissenberg number (Wi): (a, b) fractional extension in flow direction $( \langle X \rangle / L)$,
  (c, d) first normal stress difference $(N_{1})$,
  (e, f) extensional viscosity $(\eta_{e})$. Linear polymers are shown in shades of gray with darker shades indicating more beads. Colors from cold to hot indicate higher number of rings ($N_\mathrm{rings}$) with fixed beads per ring ($\mathrm{BPR}=8$). Darker shades of green indicate increasing beads per ring (BPR) in polycatenanes with fixed number of rings ($N_\mathrm{rings}=10)$.}
  \label{fig:combined_PC_extensional_plots}
\end{figure*}

\begin{figure*}[htbp]
  \centering
  % Row 1: eta
  \begin{subfigure}[b]{0.48\textwidth}
    \centering
    \includegraphics[width=\textwidth]{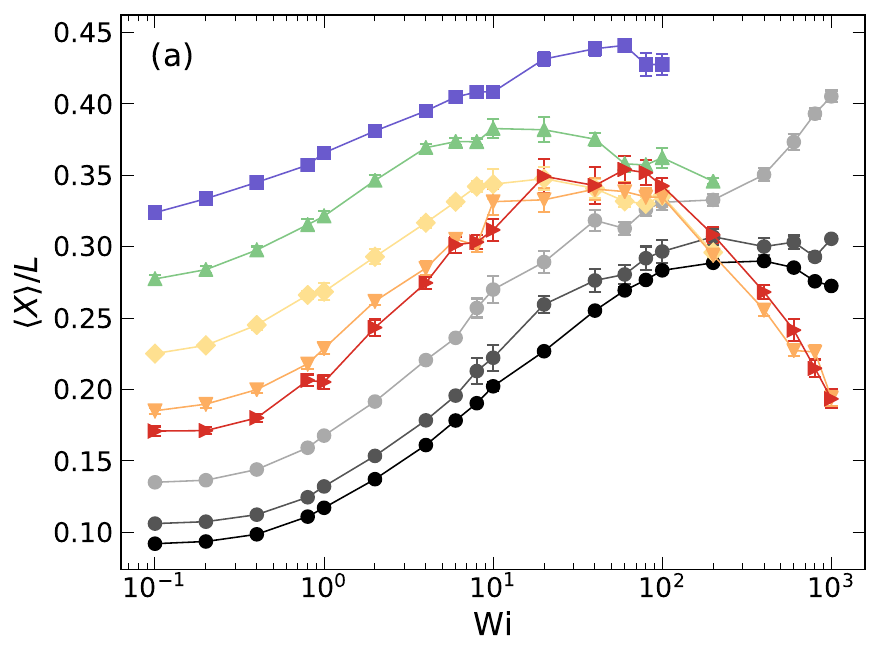}
  \end{subfigure}
  \hfill
  \begin{subfigure}[b]{0.48\textwidth}
    \centering
    \includegraphics[width=\textwidth]{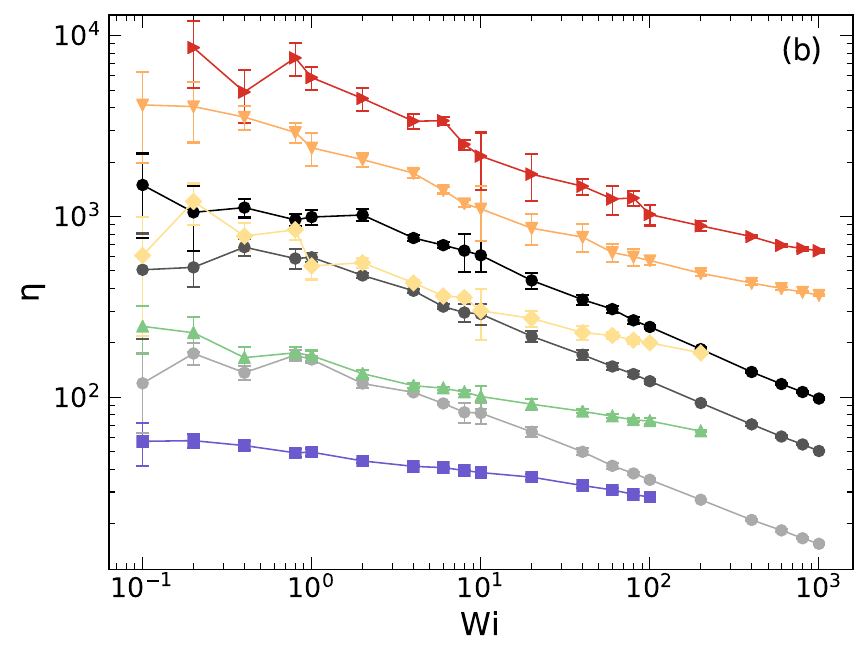}
  \end{subfigure}

  \vspace{0.5em}

  % Row 2: xbb
  \begin{subfigure}[b]{0.48\textwidth}
    \centering
    \includegraphics[width=\textwidth]{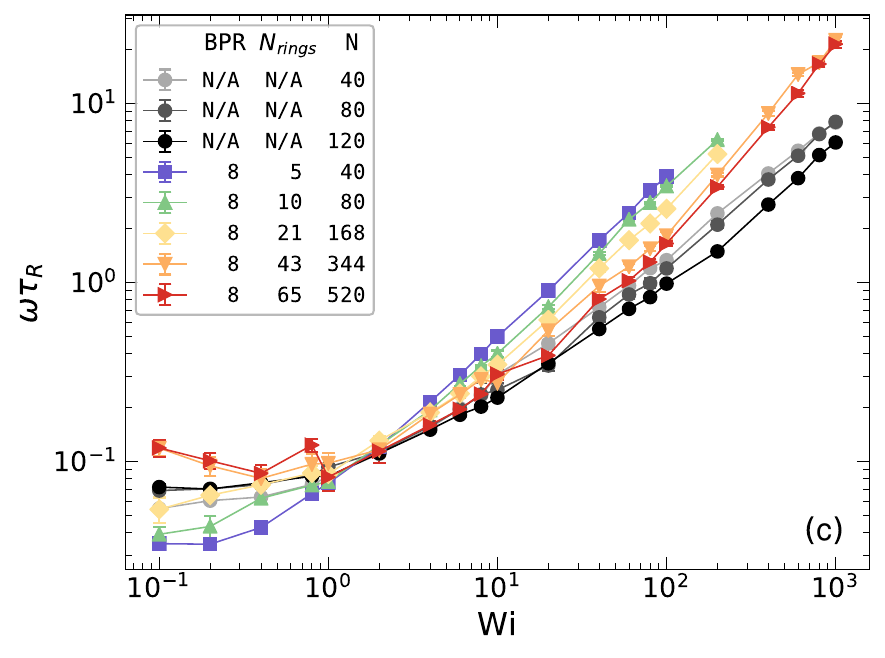}
  \end{subfigure}
  \hfill
  \begin{subfigure}[b]{0.48\textwidth}
    \centering
    \includegraphics[width=\textwidth]{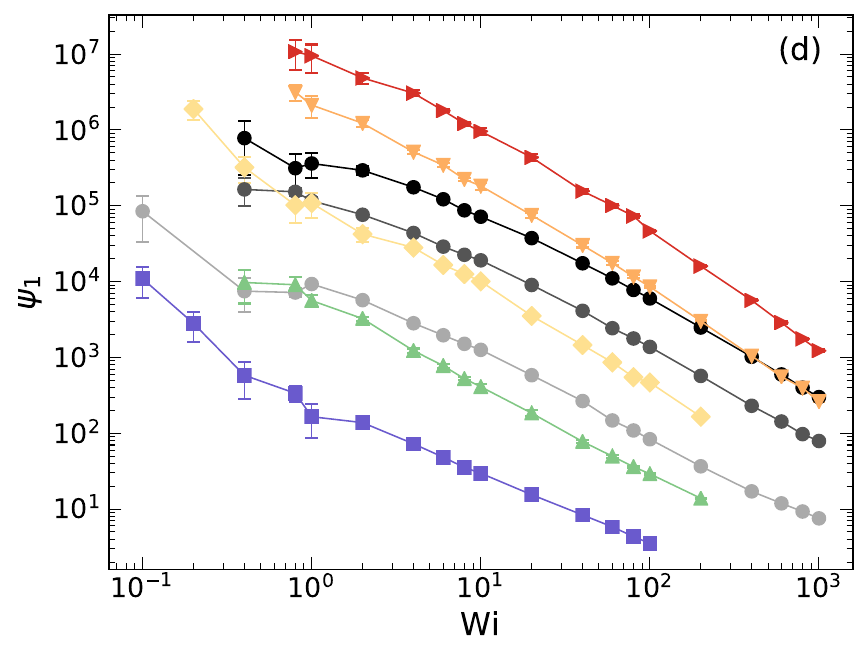}
  \end{subfigure}

  \vspace{0.5em}
  
\captionsetup{justification=raggedright,singlelinecheck=false}
  \caption{Poylcatenane properties in shear flow as a function of Weissenberg number (Wi): (a) fractional extension in flow direction $( \langle X \rangle / L)$, (b) shear viscosity ($\eta$), (c) tumbling frequency ($\omega\tau_{R}$), and (d) first normal stress coefficient ($\psi_1$). 
  Linear polymers are shown in shades of gray with darker shades indicating more beads. 
  Colors from cold to hot indicate higher number of rings ($N_\mathrm{rings}$) with fixed beads per ring ($\mathrm{BPR}=8$).
  Fitted power law exponents for \(\mathrm{Wi}>10^{2}\) for \(\eta\) and \(\psi_1\) are reported in SI Tables E.1 and E.2, respectively.}
  \label{fig:PC_numrings_shear_plots}
\end{figure*}

\begin{figure*}[htbp]
  \centering
  % Row 1
  \begin{subfigure}[b]{0.48\textwidth}
    \centering
    \includegraphics[width=\textwidth]{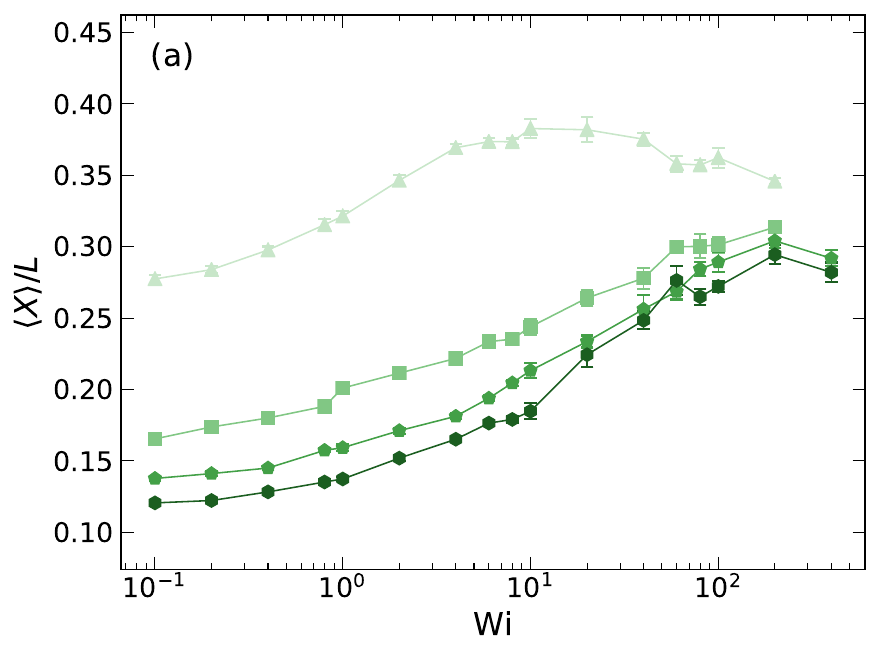}
  \end{subfigure}
  \hfill
  \begin{subfigure}[b]{0.48\textwidth}
    \centering
    \includegraphics[width=\textwidth]{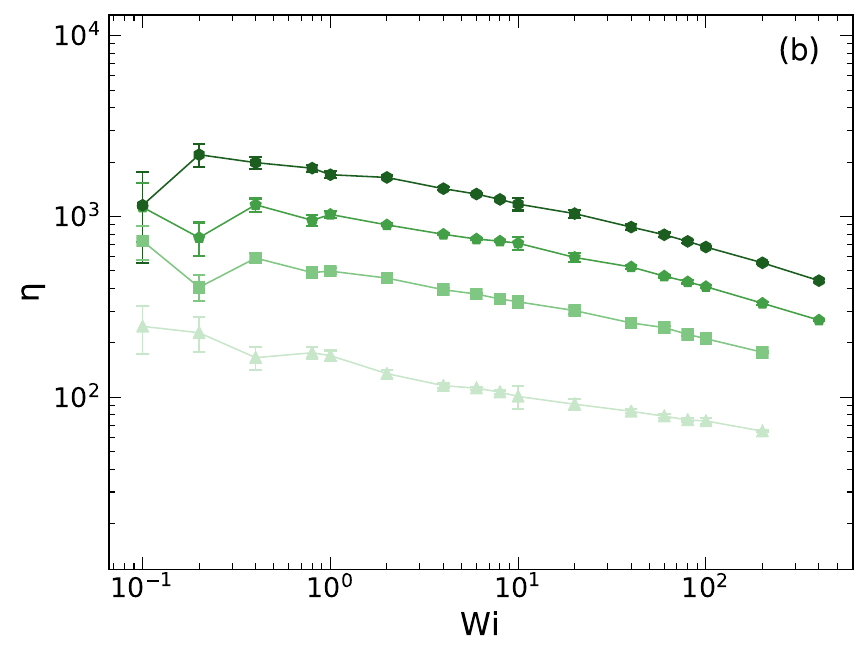}
  \end{subfigure}

  \vspace{0.5em}

  % Row 2: xbb
  \begin{subfigure}[b]{0.48\textwidth}
    \centering
    \includegraphics[width=\textwidth]{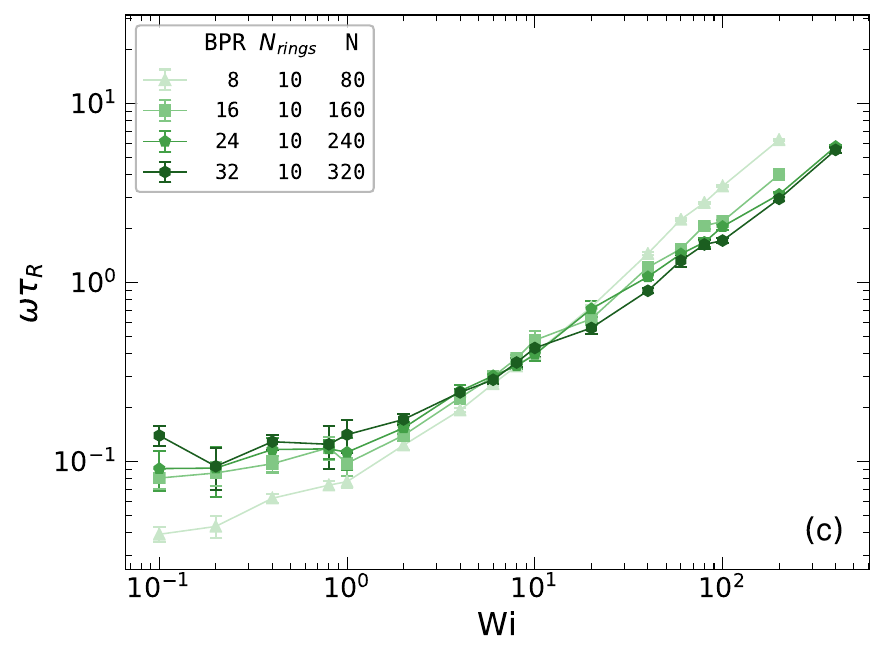}
  \end{subfigure}
  \hfill
  \begin{subfigure}[b]{0.48\textwidth}
    \centering
    \includegraphics[width=\textwidth]{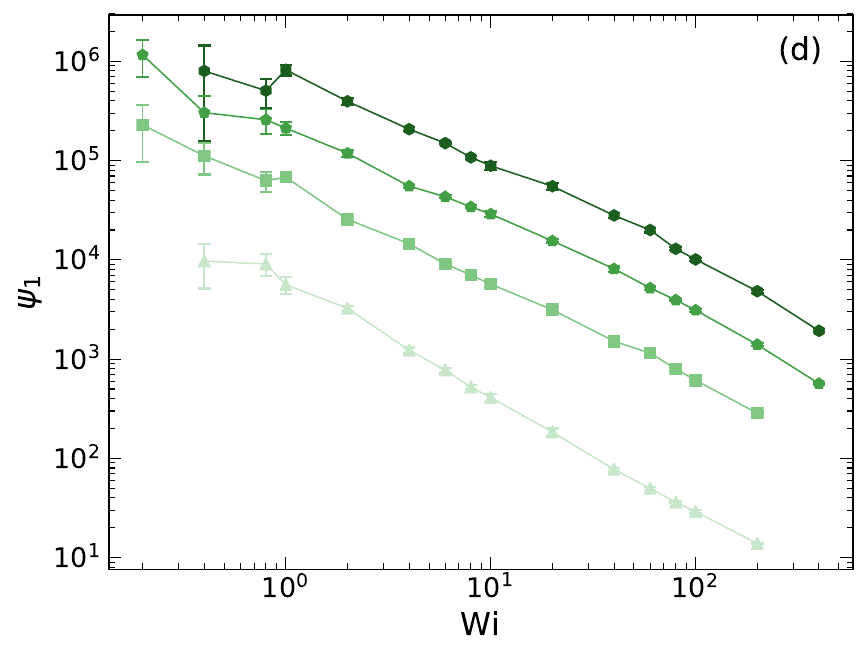}
  \end{subfigure}

  \vspace{0.5em}
  
\captionsetup{justification=raggedright,singlelinecheck=false}
  \caption{Poylcatenane properties in shear flow as a function of Weissenberg number (Wi): (a) fractional extension in flow direction $( \langle X \rangle / L)$, (b) shear viscosity ($\eta$), (c) tumbling frequency ($\omega\tau_{R}$), and (d) first normal stress coefficient ($\psi_1$). 
  Darker shades of green indicate increasing beads per ring (BPR) in polycatenanes with fixed number of rings ($N_\mathrm{rings}=10)$.
  Fitted power law exponents for \(\mathrm{Wi}>10^{2}\) for \(\eta\) and \(\psi_1\) are reported in SI Tables E.1 and E.2, respectively.}  \label{fig:PC_ringsize_shear_plots}
\end{figure*}

\subsection{Polycatenanes: Extensional Flow}

Fig. \ref{fig:combined_PC_extensional_plots}a shows that the variation of fractional extension of polycatenanes in the flow direction is less than linear polymers, and it is lower in polycatenanes that have fewer rings. For instance, a polycatenane  with 8 beads per ring and 5 rings extends only 2.4 times beyond equilibrium, while a polycatenane chain with the same ring size and 65 rings stretches 4.2 times. 
More rings give the chain more degrees of freedom, leading to greater relative fluctuations in size and shape. In the limiting case, if we continue increasing the number of rings indefinitely, the chain approaches a linear polymer \cite{brereton2001statistical}.
At high Wi, the difference becomes smaller as both polycatenanes and linear polymers approach fully extended configurations.
As shown in Fig. \ref{fig:combined_PC_extensional_plots}b, increasing the ring size while keeping the number of rings constant also leads to a more pronounced coil-stretch transition due to the higher conformational degrees of freedom from the larger rings.
The chains with larger rings have more flexibility \cite{rauscher2020thermodynamics} and can collapse and extend more readily. As a result, in weak flows, chains with smaller rings exhibit higher extension, whereas in strong flows, chains with larger rings show greater extension. 
No extension data is presented at $\mathrm{Wi}>40$ because strong hydrodynamic forces cause bond crossing between interlocked rings in this regime.

According to Fig. \ref{fig:combined_PC_extensional_plots}c, at low Wi, polycatenanes have higher $N_{1}$ than linear polymers with the same molecular weight because the interlocks prevent chain collapse and increase stress. 
At medium to high Wi, linear polymers have higher $N_{1}$. 
Furthermore, $N_{1}$ increases with increasing number of rings because the polymer becomes subjected to larger forces from the extensional flow since there are more segments farther away from the stagnation point. Increasing ring size has a similar effect, as evident from Fig. \ref{fig:combined_PC_extensional_plots}d.

Extensional viscosity ($\eta_{e}$) is significantly lower in polycatenanes compared to linear polymers with the same molecular weight, as shown in Fig. \ref{fig:combined_PC_extensional_plots}e. 
For instance, a polycatenane with 8 beads per ring and 10 rings ($M_w=80$) has 22 times lower ($\eta_{e}$) than a linear polymer with 80 beads at $\mathrm{Wi}=10$. 
Furthermore, adding more rings to a polycatenane increases the extensional viscosity due to increased molecular weight. 
The same effect is achieved by making the rings larger, as seen in Fig. \ref{fig:combined_PC_extensional_plots}f. 
Comparing a polycatenane with 16 beads per ring and 10 rings ($M_w=160$) to a polycatenane with 8 beads per ring and 21 rings ($M_w=168$) allows us to decouple the effects of molecular weight from the impact of chain topology on viscosity. 
Based on Figs. \ref{fig:combined_PC_extensional_plots}e and \ref{fig:combined_PC_extensional_plots}f, at low Wi, the chain with more small rings has higher viscosity. 
This reflects the more extended conformation of the smaller rings, leading to higher stress and resistance to extension. 
At high Wi, the chain with fewer large rings show significantly higher extensional viscosity, up to 270\% higher at the highest Wi. 
Under strong extensional flow, the rings in the chain with fewer large rings are more effectively stretched, leading to an increased resistance to further deformation. 
The extended conformation under high strain produces a larger stress response, which is reflected in the higher measured extensional viscosity. 
In contrast, the greater number of flexible mechanical interlocks in the chain with more small rings facilitates lower stress response when subjected to high strain rates which 
%This relaxation mechanism 
ultimately reduces $\eta_{e}$ relative to the chain with fewer large rings.
Previous polycatenane stretching simulations identified a stress-softening regime driven by rotational sliding of rings under high force \cite{chen2024nonlinear}. We observed no equivalent decrease in extensional viscosity; this likely stems from differences in topology and deformation mode (constant force stretching versus extensional flow). However, we cannot rule out this behavior at high Wi number regime, where bond crossing limits our data availability.
\subsection{Polycatenanes: Shear Flow}
As shown in Fig.~\ref{fig:PC_numrings_shear_plots}a, at low Wi, polycatenanes exhibit a range of extensional behaviors based on the number of rings in their structure.
Chains with fewer rings have limited variation of fractional extension throughout the Wi range, while chains with higher number of rings can be deformed more readily in the flow, leading to larger variation of fractional extension in flow direction. 
While it is somewhat trivial that smaller polymers stretch less than longer polymers, the scaling of catenanes is different than that of linear polymers.
For a linear polymer in good solvent in equilibrium, the radius of gyration scales at \(N^{0.6}\)~\cite{de1972exponents} and the contour length as \(N\).
This leads to a ratio of maximum stretch to equilibrium extent that scales as \(N^{1.67}\).
We find that from fitting equilibrium data for polycatenanes in SI Table A.4, the radius of gyration scales as \(N^{0.65}\) and from our pulling simulations to determine contour length, the maximum stretch scales as \(N^{0.95}\).
This leads to a ratio of stretch to coil of \(N^{1.46}\) for polycatenanes, demonstrating that they will stretch less than linear polymers.
The extensional data in Fig.~\ref{fig:PC_numrings_shear_plots}a reflects this resistance to stretching.
For example, a polycatenane of 5 rings stretches between 0.32 and 0.43 of its maximum, compared to a 65 ring polycatenane, which extends between 0.17 and 0.35.
Shorter, less flexible polycatenanes begin to stretch in the flow direction at lower flow strength than longer polycatenanes.
For example, a 5-ring polycatenane shows increased stretch at \(\mathrm{Wi}=0.1\), but is delayed to \(\mathrm{Wi}=0.2\) for a 65-ring polycatenane.
For $10^0 < \mathrm{Wi} < \ 10^2$, there is a regime with increasing extension, which corresponds to chains being extended in flow direction with corresponding thinning in the gradient and vorticity directions, shown in SI Figs.~B.3a and B.3b, respectively. 
In chains with 21 or more rings, there is a decreasing regime in the $100 \le \mathrm{Wi}$ range. 
This corresponds to expansion in vorticity direction, shown in SI Fig.~B3.b. 
This behavior could not be verified in chains with 5 and 10 rings because of failure of the mechanical bonds due to bond crossing, a subject of future work.
At low to moderate Wi, $\langle X \rangle / L$ is higher in polycatenanes than linear polymers with the same molecular weight due to entangled ring topologies leading to strong repulsive interactions, keeping the chains more extended.
or example, at \(\mathrm{Wi}=1\), a 10-ring polycatenane has a fractional extension of 0.32 compared to a linear polymer of the same molecular weight (\(M_w=80\)) with an extension of 0.13.
At high Wi, linear polymers are more extended, due to the more pronounced decreasing regime in polycatenanes.
Based on \ref{fig:PC_ringsize_shear_plots}a, increasing the ring size has the same effect on extension at low to medium Wi as number of rings.
The greater flexibility of larger rings leads to more coiling behavior and less extended configurations.

Polycatenanes show significantly lower shear viscosity than linear polymers of similar molecular weight, as shown in Fig. \ref{fig:PC_numrings_shear_plots}b. 
For example at $\mathrm{Wi}=10$, the viscosity of a polycatenane with 8 beads per ring and 10 rings ($M_w=80$) is 35\% of the linear polymer with 80 beads.
This decreased viscosity comes from the decreased size.
Compared with linear polymers of the same molecular weight, polycatenanes have smaller size \cite{lei2021dimensional,pakula1999simulation}, also demonstrated by equilibrium $R_g$ data in SI Table~A.4.
For example, a polycatenane with \(M_w=80\) has $R_g=7.36$, about 57\% of linear polymer with the same molecular weight, $R_g=12.85$.
While their viscosity is lower, polycatenanes also have weakened shear-thinning.
The viscosity power-law scaling exponent can be up to 50\% less than the linear polymers, shown in SI Table~E.1.
For example, compared to the scaling exponent of -0.38 for a linear polymer of \(M_w=80\), the 10-ring, 8-bead-per-ring polycatenane has a much reduced scaling of -0.154.
Generally, the scaling of polycatenanes is reduced from linear polymers with shorter polycatenanes or smaller rings leading to weakened shear thinning.
Consistent with both polyrotaxanes and daisy chains, polycatenanes also do not align as well with the flow as linear polymers, showcased by their larger orientation angles at high Wi in SI Fig.~B.3c.
Flow alignment is better for longer polycatenanes, leading to more shear thinning for these larger molecules.
From Fig.~\ref{fig:PC_ringsize_shear_plots}b, increasing the ring size in polycatenanes leads to a higher viscosity, primarily because of the increase in polymer size. 
The shear thinning exponent (SI Table~E.1) also increases slightly for larger rings. 
This is due to the additional molecular degrees of freedom, leading to improved alignment with the flow, demonstrated with decreased orientation angles in SI Fig.~B.4c.

At very low Wi number, short, stiff polycatenanes have decreased tumbling frequency compared to linear polymers, as shown in Fig. \ref{fig:PC_numrings_shear_plots}c. 
However, because of the interlocked rings, polycatenanes are thicker in gradient direction (as seen in SI Fig. B.3a), making them more responsive to the flow strength. 
As a result, a slight increase in Wi causes more tumbling in short polycatenanes compared with long polycatenanes and linear polymers, where tumbling frequency remains fairly constant in low Wi regime.
For example, a polycatenane with 5 rings has increased tumbling frequency at \(\mathrm{Wi}\approx0.2\), compared to linear polymers and polycatenanes with more than 10 rings where the increase occurs around \(\mathrm{Wi}\approx1\).
At moderate to high Wi number, polycatenanes tumble more frequently than linear polymers. 
For instance, at $\mathrm{Wi}=200$, a polycatenane with 8 beads per ring and 10 rings ($M_w=80$) tumbles 3 times more frequently than the linear polymer with 80 beads. 
Additionally, more rings in a polycatenane decreases tumbling frequency at moderate to high Wi number. 
This can be explained by the resulting increase in aspect ratio of the polymer, making the polymer thickness less effective, causing the chains to behave more like linear polymers. 
Increasing the ring size has a similar effect on tumbling frequency as increasing the number of rings,  as shown in Fig.~\ref{fig:PC_ringsize_shear_plots}c. 
The larger more flexible rings cause the polycatenanes to behave more like linear polymers with decreased tumbling frequency. 
SI Table E.3 shows that polycatenanes with 8-bead rings have a much stronger scaling of $\omega \tau_R$ with Wi than linear polymers.
This scaling increases with more rings.
For example, a 5-ring polycatenane with 8-bead rings exhibits a scaling exponent of 0.917 with tumbling frequency and a 65-ring polycatenane with 8-bead rings has an exponent of 1.029.
In both cases the scaling is significantly increased from the expected linear polymer scaling of 0.75.
Conversely, increasing the ring size reduces the exponent toward the linear polymer value.
For example, while a 10-ring polycatenane with 8-bead rings has a scaling exponent of 0.935, a 32-ring polycatenane with the same size rings has an exponent of 0.753, identical to that expected from a linear polymer.

According  to Fig. \ref{fig:PC_numrings_shear_plots}d, polycatenanes have a $\psi_1$ value about two orders of magnitude lower than that of linear polymers at the same molecular weight, with the number of rings having only a minimal effect. 
As shown in SI Table E.2, the power-law scaling exponent ($\beta$) for the polycatenanes is slightly higher than linear polymers of comparable molecular weight, with more rings further increasing the exponent. 
For example, for a polycatenane with 8 beads per ring and 10 rings ($M_w=80$), $\beta=-1.067$ compared with a value of -1.239 corresponding to a linear polymer with 80 beads.

\section{Discussion}

\begin{figure*}[htbp]
  \centering
  % Row 1: fractional extension and viscosity in extension
  \begin{subfigure}[b]{0.48\textwidth}
    \centering
    \includegraphics[width=\textwidth]{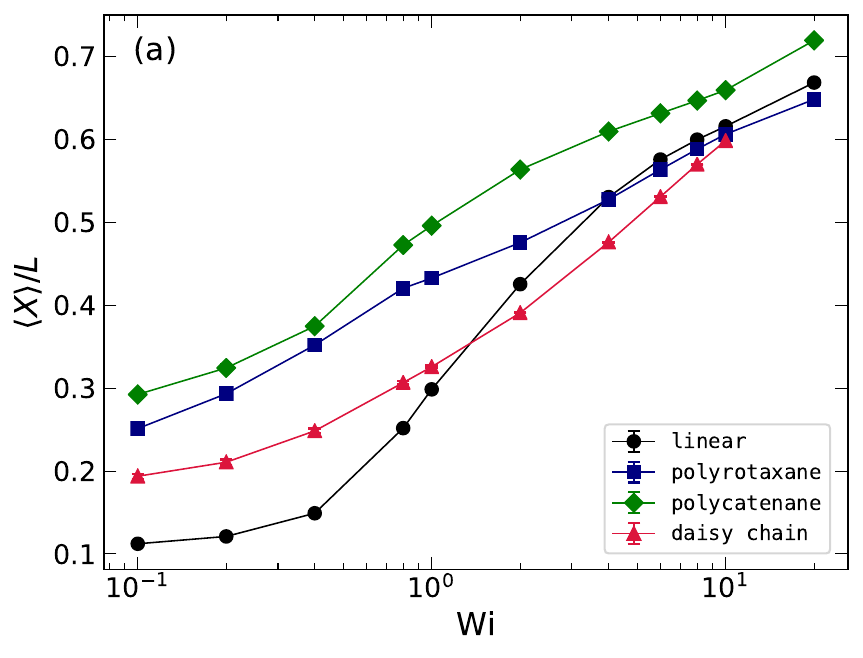}
  \end{subfigure}
  \hfill
  \begin{subfigure}[b]{0.48\textwidth}
    \centering
    \includegraphics[width=\textwidth]{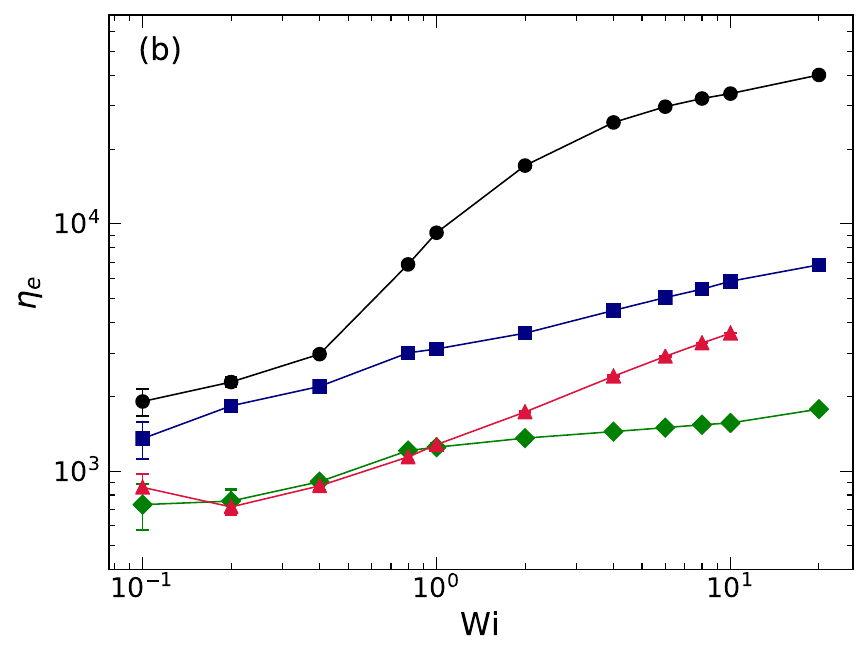}
  \end{subfigure}

  \vspace{0.5em}

  % Row 2: fractional extension and viscosity in shear
  \begin{subfigure}[b]{0.48\textwidth}
    \centering
    \includegraphics[width=\textwidth]{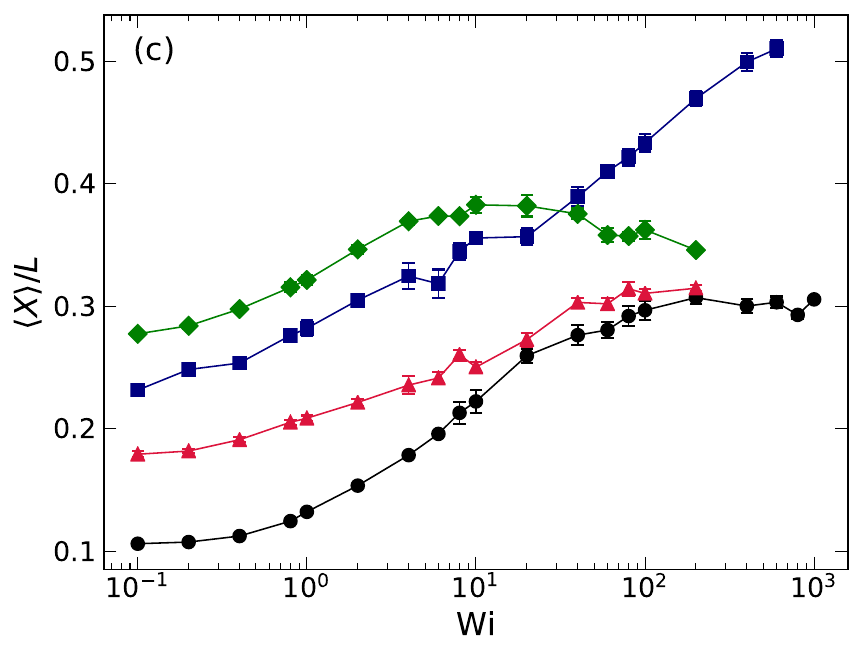}
  \end{subfigure}
  \hfill
  \begin{subfigure}[b]{0.48\textwidth}
    \centering
    \includegraphics[width=\textwidth]{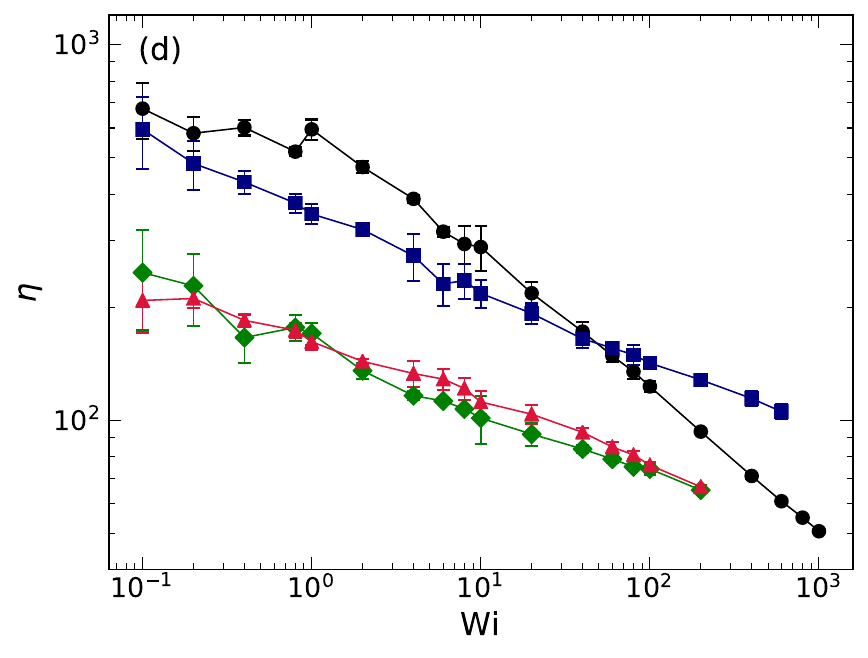}
  \end{subfigure}

  \vspace{0.5em}

   \captionsetup{justification=raggedright,singlelinecheck=false}
  \caption{Comparison of linear and MIP polymers of similar molecular weight in extensional (a, b) and shear (c, d) flow.
    Fractional extension in the flow direction $( \langle X \rangle / L)$ for extension (a) and shear (c) along with extensional viscosity $(\eta_{e})$ (b) and shear viscosity $(\eta)$ are given as a function of Weissenberg number (Wi).  The linear polymer (black) has 80 backbone beads ($M_w=80$).  The polyrotaxane (blue) has 40 backbone beads including 2 capping beads and 6 rings ($M_w=88$).  The daisy chain (red) has 4 linear segments of 8 beads each interlocked with 6 8 bead rings ($M_w=74$) and includes 2 capping beads. The polycatenane (green) has 10 interconnected rings of 8 beads each ($M_w=80$).}
  \label{fig:MIP_comparison_mw}
\end{figure*}

The presence of mechanical bonds gives rise to notable quantitative differences compared with linear polymers. Unlike linear polymers that coil under weak flow, in polyrotaxanes, ring-backbone repulsion causes more extended equilibrium configurations and earlier chain extension in flow. Polyrotaxanes show lower normal stress difference and weaker elasticity.
Extensional viscosity is higher in polyrotaxanes especially in shorter chains with more rings due to increased rigidity and resistance to elongation. 
Under shear flow, polyrotaxanes extend more in all directions and have higher viscosity, due their larger size, with reduced shear thinning originating from less alignment with flow.
In addition, tumbling frequency is higher and it increases with number of rings due to enhanced hydrodynamic cross-section from threaded rings.

Daisy chains extend more than linear polymers especially in weak to moderate flows and have lower normal stress difference, extensional and shear viscosity, as well as reduced shear thinning.
They tumble more frequently at high Wi, due to higher gradient-direction extent and aspect ratio effects with higher density of mechanical bonds enhancing this effect, causing greater deviation from linear polymers. 

Polycatenanes are more extended and have lower first normal stress difference than linear chains except at very low Wi. 
In line with previous experimental work on kinetoplasts (catenated DNA networks) \cite{soh2020deformation}, these polymers do not show an abrupt transition from coiled to stretched states typical of linear polymers.
Polycatenanes show significantly lower extensional and shear viscosities, as well as weaker shear thinning. More or larger rings in polycatenanes increases viscosity, primarily due to increased size.
Polycatenanes tumble more frequently, with fewer or smaller rings enhancing this effect.

Overall, the interlocked topology of each MIP leads to a distinct set of rheological properties, as illustrated in Fig.~\ref{fig:MIP_comparison_mw}.  
Here we give an example of each polymer type, all with a similar molecular weight.
For direct comparison with a linear polymer, properties of MIPs are normalized by those of a linear polymer, as shown in SI Fig.~B.6.
We compare polymer types with similar maximum contour lengths ($L$) in SI Fig.~B.5.
For a given molecular weight, the polycatenane exhibits the greatest relative stretch in extensional flow (Fig.~\ref{fig:MIP_comparison_mw}a), while the polyrotaxane shows the greatest stretch in shear flow (Fig.~\ref{fig:MIP_comparison_mw}c). 
In shear, the daisy chain and the polycatenane have fairly consistent stretch across a wide range of Wi, varying only between 0.18-0.31 and 0.28–0.34, respectively, much less than the variability observed in linear polymers and polyrotaxanes, showing a weaker coil–stretch response.
Additionally, similar to the effect of knots on relaxation dynamics and coil-stretch transition of DNA molecules \cite{soh2018knots}, mechanical bonds decrease the relaxation time of the MIPs (see SI Tables A.1 to A.4) by lowering the contour length of the polymer, and shift the coil-stretch transition to higher strain rates.
For a given molecular weight, the polymers also have distinct viscosity profiles.  
In extensional flow, the linear polymer has the greatest extensional viscosity, as seen in Fig.~\ref{fig:MIP_comparison_mw}b.  
The mechanical bonds in all MIP types lead to lower extensional viscosity, with the daisy chain behaving similar to the linear polymer and the polyrotaxane and polycatenane showing weaker scaling of their extensional viscosity than the linear polymer and daisy chain.  
In shear flow (Fig.~\ref{fig:MIP_comparison_mw}d), the linear polymer has higher shear viscosity than the MIPs except the polyrotaxane, which overtakes the linear polymer at a cross-over around $\mathrm{Wi}\approx50$.
This demonstrates the reduced shear thinning of the polyrotaxane. 
Indeed all of the MIPs' shear viscosity scales less strongly with Wi than the linear polymer, demonstrating that the mechanical bonds suppress shear-thinning behavior. 
This suppression is further reflected in the increase of the normalized viscosities shown in SI Fig.~B.6d.
Similar to extensional viscosity, the shear viscosity of the polycatenane is the lowest of all polymer types.

In line with the varied rheological responses that we have reported, MIPs have already shown their potential in a range of applications. 
Slide-ring gels, a category of polyrotaxanes in which mechanically interlocked rings function as movable cross-links, could potentially be used as protective coatings in automobiles and smartphones due to their scratch-resistance and self-healing properties \cite{noda2014topological}.
Cyclodextrin-based polyrotaxanes have shown promise in stretchable electronics and energy storage materials. Their innovative use in stretchable conductive composite and binders for anodes in lithium-ion batteries helps to overcome critical challenges in these areas \cite{du2024cyclodextrins}.
Olympic gels (materials formed entirely from interlocked rings) \cite{raphael1997progressive}, recently synthesized from over 16,000 DNA rings, exhibit unique nonlinear elasticity and swelling behavior with potential applications in synthetic biology and bioengineering, including artificial micro-reactors for targeted delivery and artificial nuclei in synthetic cells \cite{speed2024assembling}. Moreover, mechanically interlocked [c2]daisy chain structures embedded in polymer networks enable advanced shape-memory materials for use as actuators in releasing and lifting objects, with further potential in soft robots via programmed stimuli-responsive motions \cite{zhou2024mechanically}.

Our results hint at additional potential uses of MIPs, as well.
High molecular weight polymers are known to reduce drag in pipe flow~\cite{larson2003analysis,perkins1997single,dimitropoulos1998direct,paterson1970turbulent}, at least partially due to the suppression of vortex formation caused by extended polymer configurations.  
The lower shear viscosity for MIPs compared to typical linear polymers and the greater extension of polyrotaxanes, in particular, suggest that MIPs could be developed into drag reducing agents.
Similarly, extended polymers in liquid petroleum-based fuels favor the formation of larger droplets that are less prone to ignition during accidental fuel releases~\cite{chao1984antimisting,wei2015megasupramolecules,lhota2024mist}.
The greater extension of MIPs also suggests their potential development as mist-suppressing fuel additives to improve liquid fuel safety.

As a first step toward characterizing the flow behavior of mechanically interlocked polymers, this work provides valuable insights into MIP dynamics in dilute solutions under weak to moderate flow conditions. 
However, strong flow regimes remain unexplored as our current model fails to maintain polymer integrity at high Wi.
Under such conditions, large hydrodynamic forces can cause the rings in all three MIP architectures to cross the polymer backbone (polyrotaxanes), or each other (daisy chains and polycatenanes). 
In extensional flow, these bond crossings occur at flow strengths less than those required to fully elongate the MIPs.
As a result, the extension curves do not saturate to 1 for this model (Figs.~\ref{fig:combined_PR_extensional_plots}a,~\ref{fig:combined_DC_extensional_plots}a,~\ref{fig:combined_PC_extensional_plots}a, b).
Moreover, shielding from hydrodynamic interactions (HI) substantially increases the Wi at which bond crossing occurs. 
To isolate this effect, we simulated the polymers shown in Fig. 9 without hydrodynamic interactions; the results are provided in SI Fig. B.7. 
In the absence of HI, bond crossing happens at significantly lower Wi, as expected when the shielding effect of HI is removed and interior beads are subject to full hydrodynamic forces.
Correspondingly, all polymer types stretch more in both extensional and shear flow and exhibit higher viscosities.
The shear thinning trends and scaling, however, are preserved in the absence of HI.

Although we could change the model parameters to prevent bond crossing at very high Wi, we chose to use the standard set of FENE and WCA parameters to ensure that our results remain directly comparable with previous studies in the low to moderate Wi regime. 
Despite the model deficiencies at high Wi, the observed degradation is nevertheless consistent with experimental findings, which show that rotaxanes degrade via dethreading and unstoppering under strong mechanical forces \cite{muramatsu2021rotaxane, zhang2019mechanical}.
Other experimental observations suggest that mechanical bonds in catenanes are comparable in strength to covalent bonds~\cite{lee2016mechanical}.
To fully capture degradation mechanisms in strong flow, future work will move beyond the FENE bond model and incorporate dissociable bonds, revealing combined chemical and mechanical modes of polymer failure.
This future work will lead to a better understanding of the durability of MIPs in flow for potential engineering applications.

\section{Conclusions}
In this study, we explored how mechanical bonds affect the dynamics and rheology of dilute solutions of mechanically-interlocked polymers under steady uniaxial extension and shear. 
We used a coarse-grained bead-spring model with excluded volume and hydrodynamic interactions in a Brownian dynamics simulation. 
To elucidate the effect of mechanical interlocks on properties, we ran a series of simulations across a range of flow strengths (Wi numbers) and polymer structures. 
In polyrotaxanes, the number of rings was varied. In daisy chains, the number and the length of the linear segments were varied. 
In polycatenanes, the number of rings and their sizes were varied. 
Our simulations reveal that mechanical interlocks significantly alter polymer behavior under various flow conditions, with polyrotaxanes, daisy chains, and polycatenanes each exhibiting distinctive rheological signatures compared to linear polymers.
In particular we note that MIPs generally have greater tumbling rates in shear flow and weaker shear thinning compared to linear polymers, while differences in extension depend on flow strength, type, and polymer architecture.
This work provides a foundation for understanding MIP behavior in dilute solutions under weak to moderate flows. 
Addressing strong flow regimes and mechanical degradation mechanisms in future studies will help fully realize the potential of these novel materials.

\section{acknowledgments}
The authors gratefully acknowledge the financial support of Wayne State University and the computational resources of Wayne State University's High Performance Computing.
The authors also thank Charles Manke for valuable feedback and discussions.

\section{Supplementary Information}
The supplementary information includes linear and MIP equilibrium properties (diffusion coefficient, relaxation time, radius of gyration, asphericity, prolateness); additional shear flow measurements as a function of Wi (gradient direction and vorticity direction extent, orientation angle); comparisons of linear and MIPs with similar contour length; comparisons of linear and MIPs with similar molecular weight with and without HI;trajectory snapshots; probability distribution of rings in polyrotaxanes; and power-law scaling exponents ($\beta$) of $\eta$, $\psi_1$, and $\omega\tau_R$ fitted from shear data.

\section{Data Availability}
Data will be available in a persistent Zenodo repository. Simulation code is available at \href{https://github.com/albaugh/polymer}{https://github.com/albaugh/polymer}.

\bibliography{biblio.bib}
\clearpage

\onecolumngrid
\section*{Appendix}
\appendix
\setcounter{figure}{0}
\setcounter{table}{0}  
\renewcommand{\thefigure}{\Alph{section}.\arabic{figure}}  
\renewcommand{\thetable}{\Alph{section}.\arabic{table}}   

\section{Equilibrium Properties}

\begin{table}[h!]
\centering
\caption{Equilibrium properties of linear polymers}
\setlength{\tabcolsep}{4pt}
\begin{tabular}{c c c c c c}
\hline
$M_{w}$ & $D_{COM}\zeta/k_BT$ & $\tau_R/\tau$ & $R_g/a$ & Asphericity & Prolateness \\
\hline
40  & 0.1685 & 65.74  & 8.16 & 0.64 & 0.57 \\
72  & 0.1196 & 206.35 & 12.17 & 0.64 & 0.59 \\
80  & 0.1130 & 243.59 & 12.85 & 0.65 & 0.62 \\
120 & 0.0913 & 459.08 & 15.86 & 0.60 & 0.49 \\
\hline
\end{tabular}
\label{table:linear}
\end{table}

\begin{table}[h!]
\centering
\caption{Equilibrium properties of polyrotaxanes}
\setlength{\tabcolsep}{4pt} 
\begin{tabular}{c c c c c c c c}
\hline
$M_{w}$ & $N_{bb}$ & $N_\mathrm{rings}$ & $D_{COM}\zeta/k_BT$ & $\tau_R/\tau$ & $R_g/a$ & Asphericity & Prolateness \\
\hline
56 & 40 & 2 & 0.1433 & 139.57 & 10.95 & 0.73 & 0.82 \\
72 & 40 & 4 & 0.1373 & 152.78 & 11.22 & 0.72 & 0.80 \\
88 & 40 & 6 & 0.1312 & 169.60 & 11.56 & 0.73 & 0.84 \\
104 & 40 & 8 & 0.1256 & 189.94 & 11.96 & 0.75 & 0.86 \\
112 & 80 & 4 & 0.0992 & 398.37 & 15.40 & 0.64 & 0.57 \\
144 & 80 & 8 & 0.0932 & 470.75 & 16.22 & 0.63 & 0.55 \\
176 & 80 & 12 & 0.0871 & 579.55 & 17.41 & 0.66 & 0.63 \\
208 & 80 & 16 & 0.0823 & 690.90 & 18.47 & 0.66 & 0.62 \\
168 & 120 & 6 & 0.0801 & 727.28 & 18.69 & 0.61 & 0.49 \\
216 & 120 & 12 & 0.0730 & 1040.23 & 21.34 & 0.67 & 0.65 \\
264 & 120 & 18 & 0.0686 & 1234.32 & 22.53 & 0.64 & 0.59 \\
312 & 120 & 24 & 0.0643 & 1470.45 & 23.80 & 0.62 & 0.53 \\
\hline

\end{tabular}
\label{table:polyrotaxane}
\end{table}

\begin{table}[h!]
\centering
\caption{Equilibrium properties of daisy chains}
\setlength{\tabcolsep}{4pt}
\begin{tabular}{c c c c c c c c}
\hline
$M_{w}$ & BPS & $N_\mathrm{seg}$  & $D_{COM}\zeta/k_BT$ & $\tau_R/\tau$ & $R_g/a$ & Asphericity & Prolateness \\
\hline
30 & 8 & 2 & 0.2504 & 13.64 & 4.53 & 0.61 & 0.47 \\
74 & 8 & 4 & 0.1624 & 62.61 & 7.81 & 0.70 & 0.74 \\
118 & 8 & 6 & 0.1254 & 155.26 & 10.79 & 0.71 & 0.76 \\
206 & 8 & 10 & 0.0914 & 451.62 & 15.75 & 0.68 & 0.70 \\
46 & 16 & 2 & 0.1874 & 41.01 & 6.79 & 0.62 & 0.52 \\
106 & 16 & 4 & 0.1228 & 169.84 & 11.18 & 0.66 & 0.63 \\
166 & 16 & 6 & 0.0972 & 347.90 & 14.26 & 0.62 & 0.53 \\
226 & 16 & 8 & 0.0811 & 631.57 & 17.57 & 0.63 & 0.55 \\
78 & 32 & 2 & 0.1331 & 130.08 & 10.19 & 0.61 & 0.49 \\
170 & 32 & 4 & 0.0872 & 533.57 & 16.68 & 0.65 & 0.61 \\
262 & 32 & 6 & 0.0687 & 1145.30 & 21.79 & 0.66 & 0.65 \\
\hline
\end{tabular}
\label{table:daisychains}
\end{table}

\begin{table}[h!]
\centering
\caption{Equilibrium properties of polycatenanes}
\setlength{\tabcolsep}{4pt}
\begin{tabular}{c c c c c c c c}
\hline
$M_{w}$ & BPR & $N_\mathrm{rings}$ & $D_{COM}\zeta/k_BT$ & $\tau_R/\tau$ & 
$R_g/a$ & Asphericity & Prolateness \\
\hline
40 & 8 & 5 & 0.2396 & 14.91 & 4.61 & 0.64 & 0.57 \\
80 & 8 & 10 & 0.1685 & 53.74 & 7.36 & 0.67 & 0.66 \\
168 & 8 & 21 & 0.1122 & 217.18 & 12.09 & 0.66 & 0.63 \\
344 & 8 & 43 & 0.0745 & 835.88 & 19.32 & 0.65 & 0.60 \\
520 & 8 & 65 & 0.0590 & 1753.85 & 24.89 & 0.64 & 0.58 \\
160 & 16 & 10 & 0.1133 & 172.67 & 10.84 & 0.65 & 0.60 \\
240 & 24 & 10 & 0.0898 & 341.97 & 13.57 & 0.62 & 0.52 \\
320 & 32 & 10 & 0.0751 & 611.21 & 16.60 & 0.65 & 0.60 \\
\hline
\end{tabular}
\label{table:polycatenanes}
\end{table}
\clearpage

\section{Additional Shear Flow Properties}
\begin{figure*}[h!]
  \centering
  % Top row: two subfigures
  \begin{minipage}[t]{0.48\textwidth}
    \centering
    \includegraphics[width=0.95\textwidth]{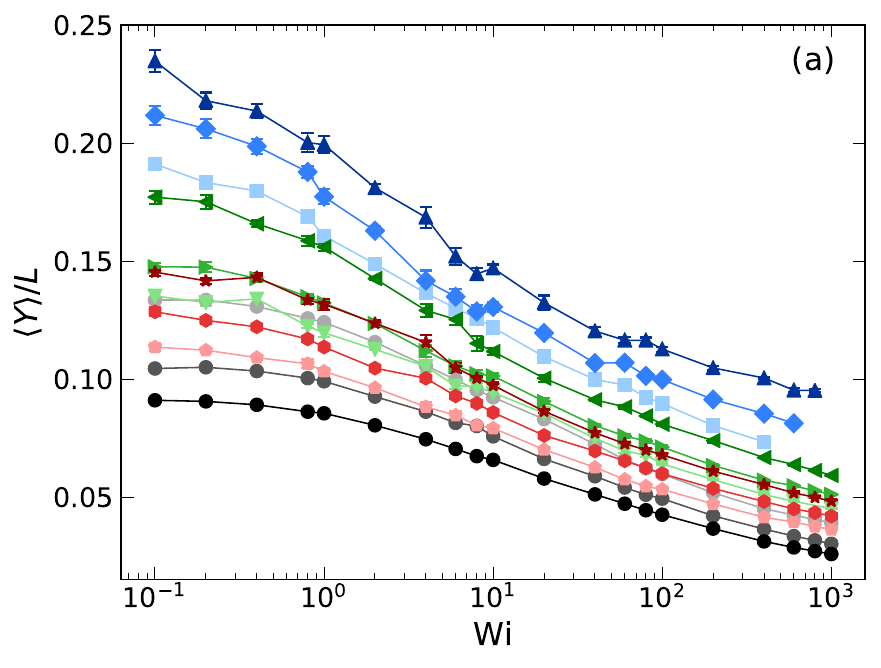}
  \end{minipage}
  \hfill
  \begin{minipage}[t]{0.48\textwidth}
    \centering
    \includegraphics[width=0.97\textwidth]{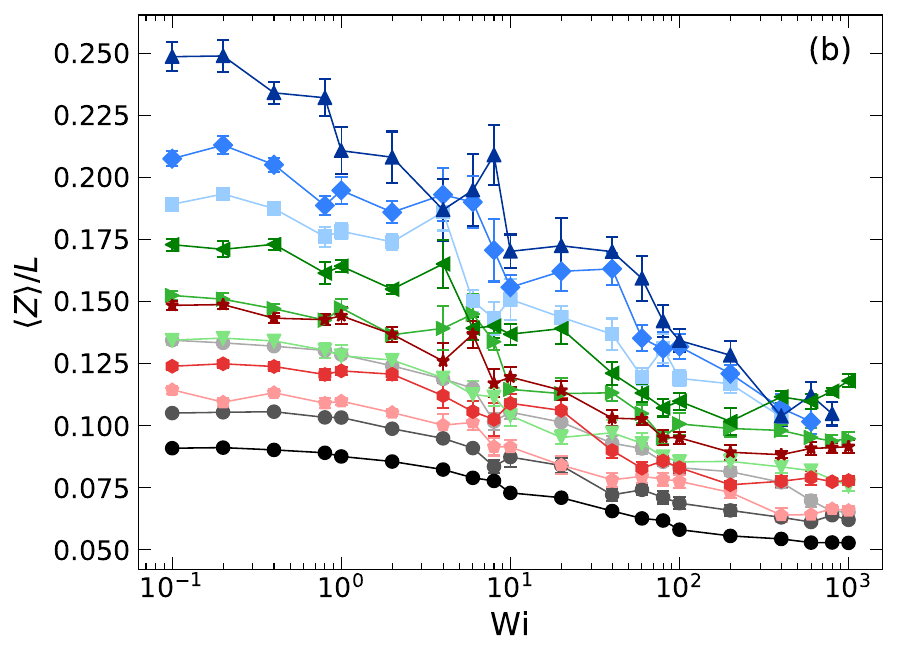}
  \end{minipage}
  
  \vspace{0.2em} % space between rows

  % Bottom row: single centered subfigure
  \begin{minipage}[t]{0.48\textwidth}
    \centering
    \includegraphics[width=0.95\textwidth]{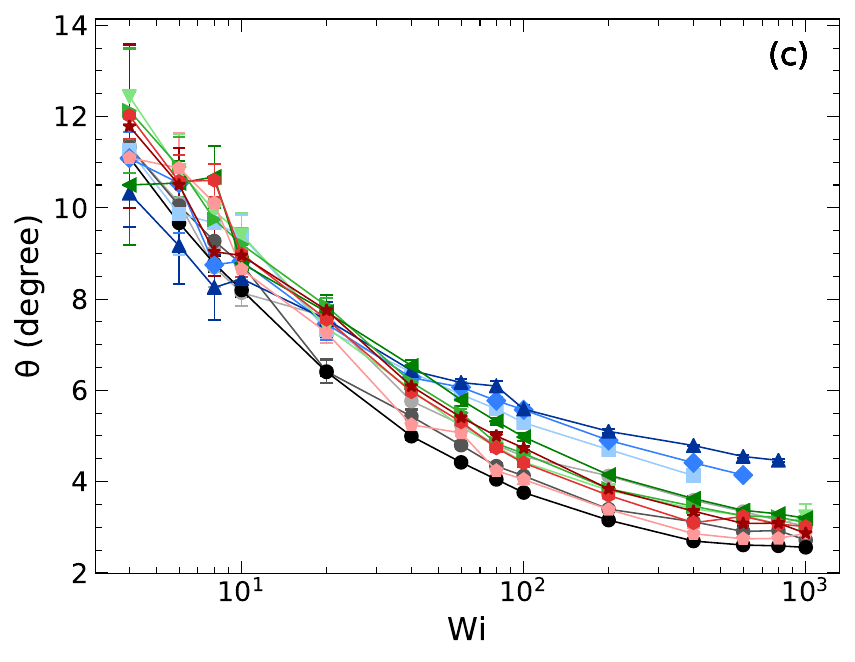}
  \end{minipage}
\captionsetup{justification=raggedright,singlelinecheck=false}

  \caption{Polyrotaxane properties in shear flow as a function of Weissenberg number (Wi): (a) fractional extension in gradient direction $(\langle Y \rangle / L)$, (b) fractional extension in vorticity direction $(\langle Z \rangle / L)$, (c) orientation angle ($\theta$). Linear polymers are shown in shades of gray with darker shades indicating more beads. Polyrotaxanes with 40, 80, and 120 backbone beads ($N_{bb}$) are shown in blue, green, and red, respectively; darker shades indicate higher ring density ($N_\mathrm{rings}/N_{bb}$). The rings in the polyrotaxanes are composed of 8 beads, regardless of backbone length.}
  \label{fig:PR_all_plots_app}
\end{figure*}
To calculate tumbling time, we monitor the angle between end-to-end vector of the polymers with the x-axis as it progresses through the four quadrants of a full $360^\circ$ rotation cycle and record the time taken for the chains to complete each full cycle (i.e., one complete rotation through all four quadrants) throughout the simulation and use the mean value as the tumbling time. We use the radius of gyration along the flow and gradient directions to calculate the orientation angle $\theta$ to measure the alignment of the chains in the $xy$ plane using the following equation \cite{schroeder2005dynamics}. 
\begin{equation*}
\tan(2\theta) = \frac{2R_{g,xy}}{R_{g,xx} - R_{g,yy}}
\label{eq:orientation}
\end{equation*}
According to Fig. B1c, the orientation angle ($\theta$) in polyrotaxanes follows a trend similar to linear polymers, showing that the threaded rings have minimal effect on polymer orientation angle.
\clearpage
\begin{figure*}[h!]
  \centering
  % Top row: two subfigures
  \begin{minipage}[t]{0.48\textwidth}
    \centering
    \includegraphics[width=0.95\textwidth]{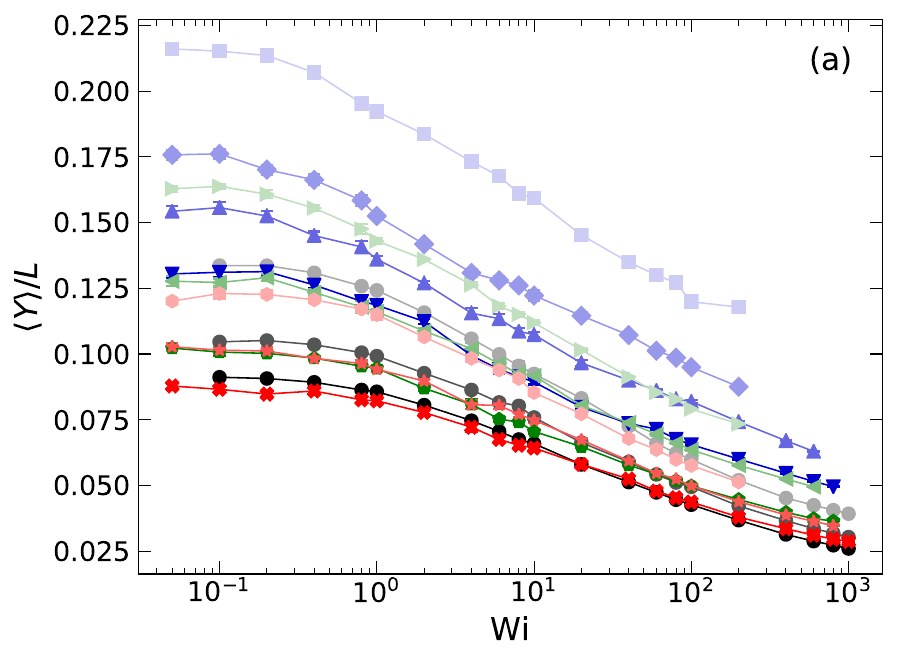}
  \end{minipage}
  \hfill
  \begin{minipage}[t]{0.48\textwidth}
    \centering
    \includegraphics[width=0.97\textwidth]{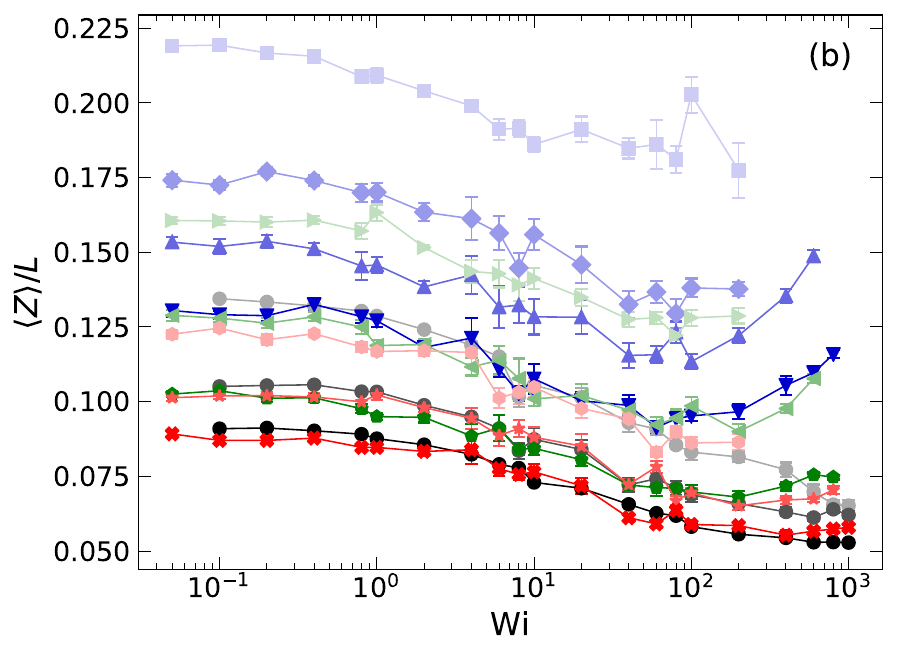}
  \end{minipage}
  
  \vspace{0.2em} % space between rows

  % Bottom row: single centered subfigure
  \begin{minipage}[t]{0.48\textwidth}
    \centering
    \includegraphics[width=0.95\textwidth]{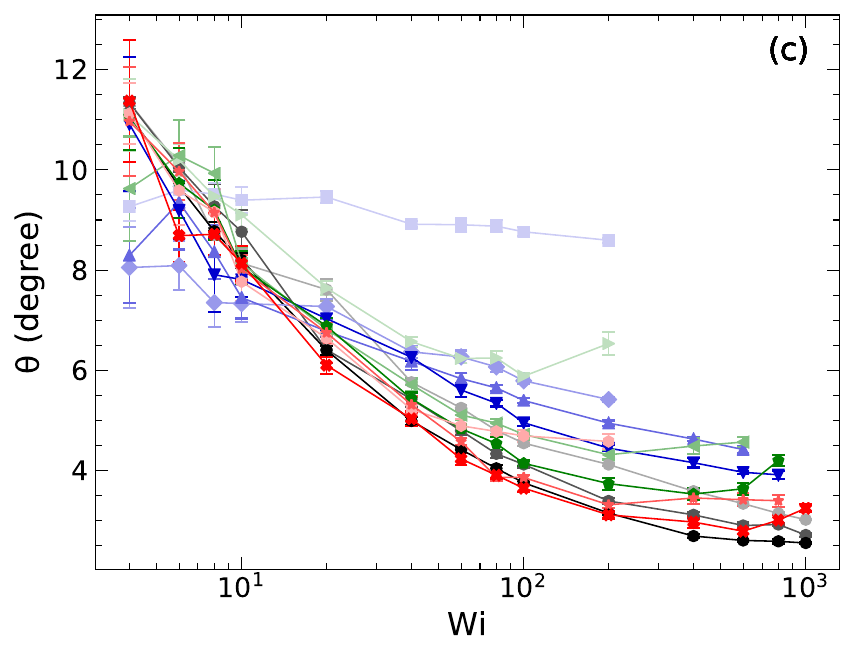}
  \end{minipage}\captionsetup{justification=raggedright,singlelinecheck=false}
  \caption{Daisy chain properties in shear flow as a function of Weissenberg number (Wi): (a) fractional extension in gradient direction $(\langle Y \rangle / L)$, (b) fractional extension in vorticity direction $(\langle Z \rangle / L)$, and (c) orientation angle ($\theta$). Linear polymers are shown in shades of gray with darker shades indicating more beads. Daisy chains with 8, 16, and 32 beads per segment (BPS) are shown in blue, green, and red, respectively; darker shades indicate more segments ($N_\mathrm{seg}$). The rings in all daisy chains are composed of 8 beads.}
  \label{fig:DC_all_shear_plots_app}
\end{figure*}

As shown by the orientation angle in Fig.~B2c, daisy chains exhibit little orientation in weak flows and reach their peak orientation angle at $\mathrm{Wi} \approx 7$, compared to $\mathrm{Wi} \approx 3$ for linear polymers, indicating that a stronger flow is needed to align daisy chains. 
This can be attributed to rearrangements at the mechanical bonds. 
At high $\mathrm{Wi}$, linear polymers reach a small angle of $\approx 2^\circ$, while daisy chains of equal molecular weight maintain a higher angle of $\approx 4.5^\circ$.
\clearpage
\begin{figure*}[h!]
  \centering
  % Top row: two subfigures
  \begin{minipage}[t]{0.48\textwidth}
    \centering
    \includegraphics[width=0.95\textwidth]{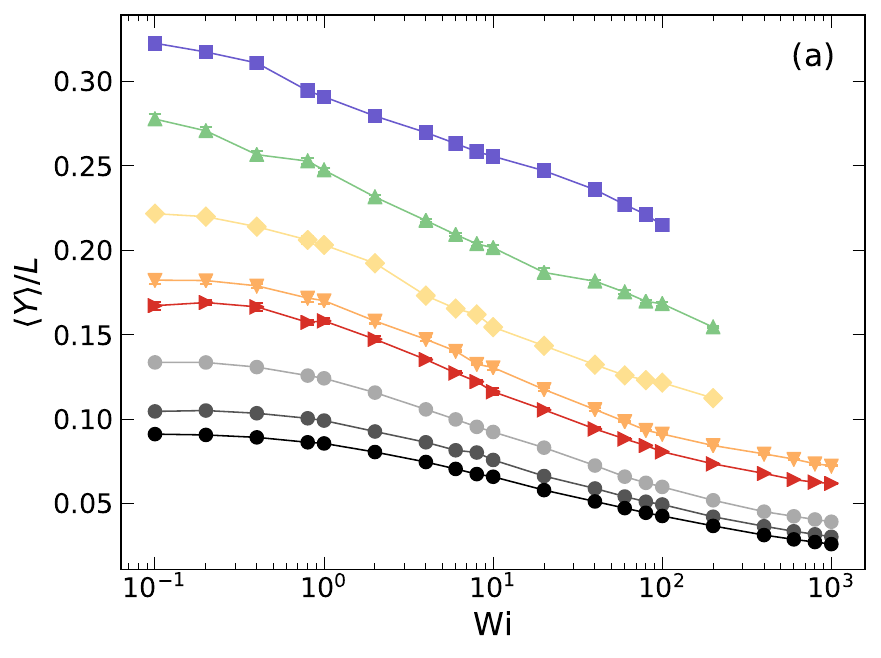}
  \end{minipage}
  \hfill
  \begin{minipage}[t]{0.48\textwidth}
    \centering
    \includegraphics[width=0.97\textwidth]{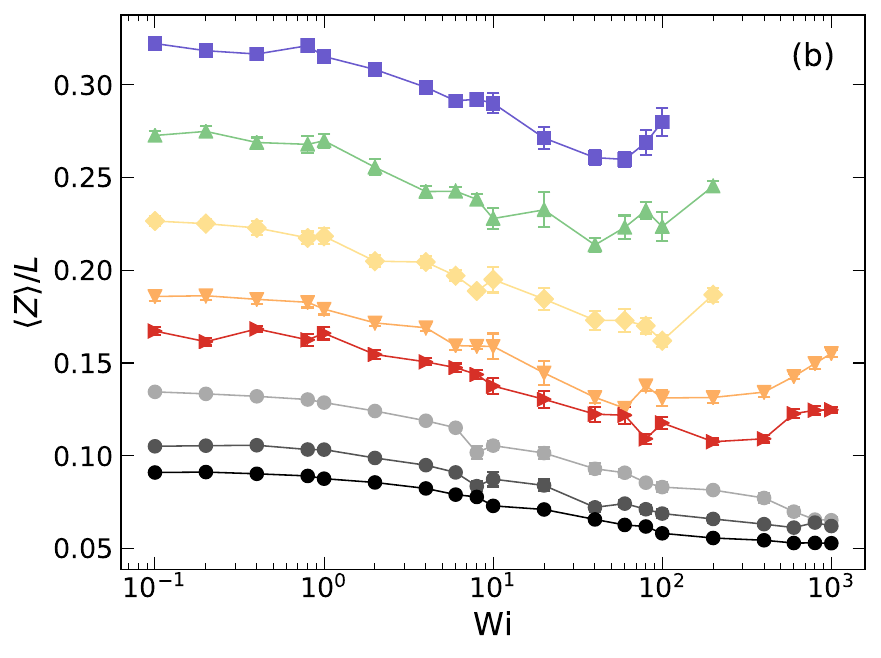}
  \end{minipage}
  
  \vspace{0.2em} % space between rows

  % Bottom row: single centered subfigure
  \begin{minipage}[t]{0.48\textwidth}
    \centering
    \includegraphics[width=0.95\textwidth]{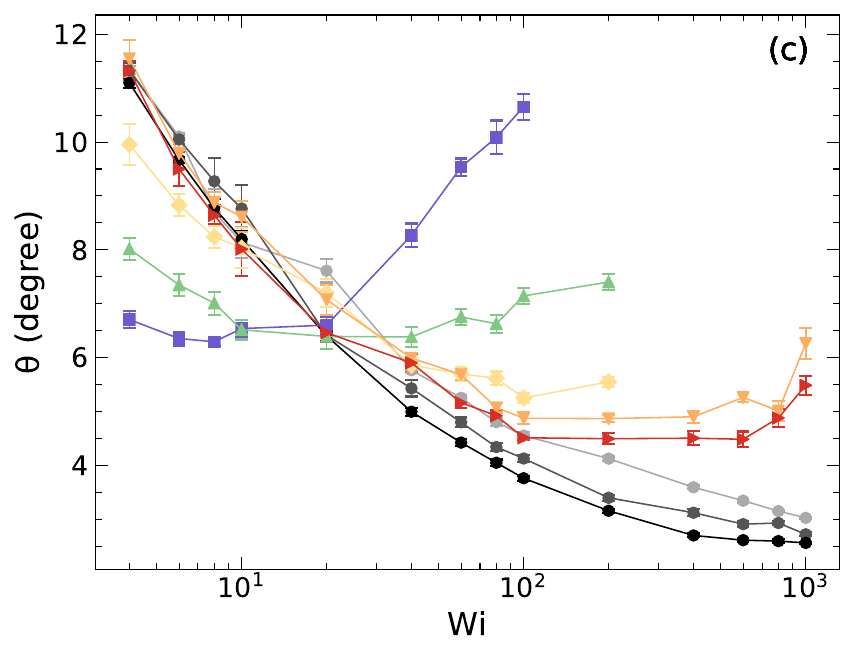}
  \end{minipage}\captionsetup{justification=raggedright,singlelinecheck=false}

  \caption{Polycatenane properties in shear flow as a function of Weissenberg number (Wi): (a) fractional extension in gradient direction $(\langle Y \rangle / L)$, (f) fractional extension in vorticity direction $(\langle Z \rangle / L)$, and (c) orientation angle ($\theta$). Linear polymers are shown in shades of gray with darker shades indicating more beads. Colors from cold to hot indicate higher number of rings ($N_\mathrm{rings}$) with fixed beads per ring ($\mathrm{BPR}=8$).}
  \label{fig:PC_combined_shear_plots_app}
\end{figure*}

According to Fig.~B3c, at low Wi polymers remain unaligned. 
At moderate Wi, the orientation angle $\theta$ first rises then falls, with polycatenanes of 8 beads per ring exhibiting a secondary increase previously noted in other systems \cite{dutta2024brownian}. 
A terminal decrease, reflecting alignment with very strong flow, could not be verified here because chains broke at high Wi. 
Having more or smaller rings shifts the primary peak to higher Wi and brings the secondary rise to lower Wi. 
Having more or larger rings make the orientation angle approach that of a linear polymer.
\clearpage
\begin{figure*}[h!]
  \centering
  % Top row: two subfigures
  \begin{minipage}[t]{0.48\textwidth}
    \centering
    \includegraphics[width=0.95\textwidth]{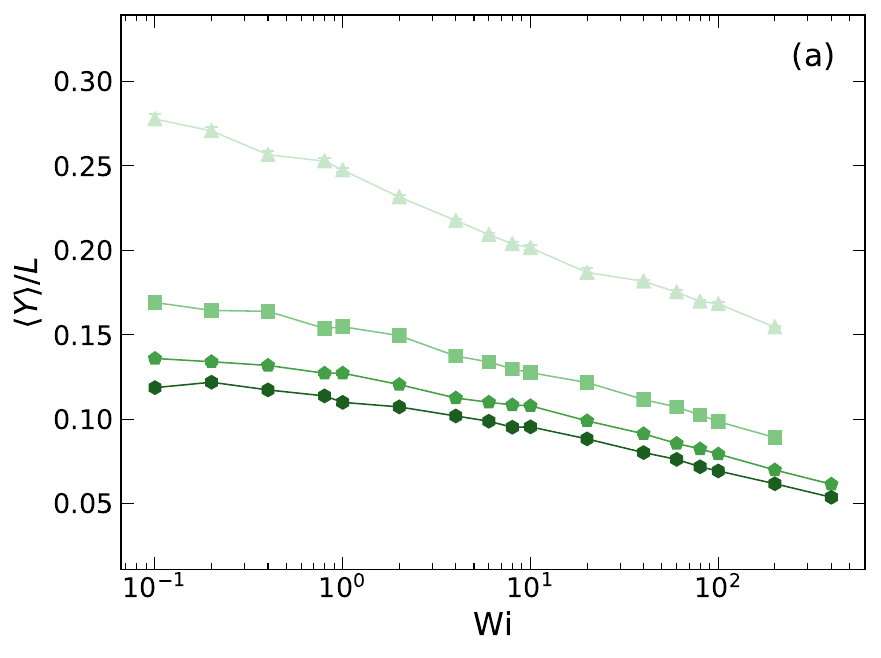}
  \end{minipage}
  \hfill
  \begin{minipage}[t]{0.48\textwidth}
    \centering
    \includegraphics[width=0.97\textwidth]{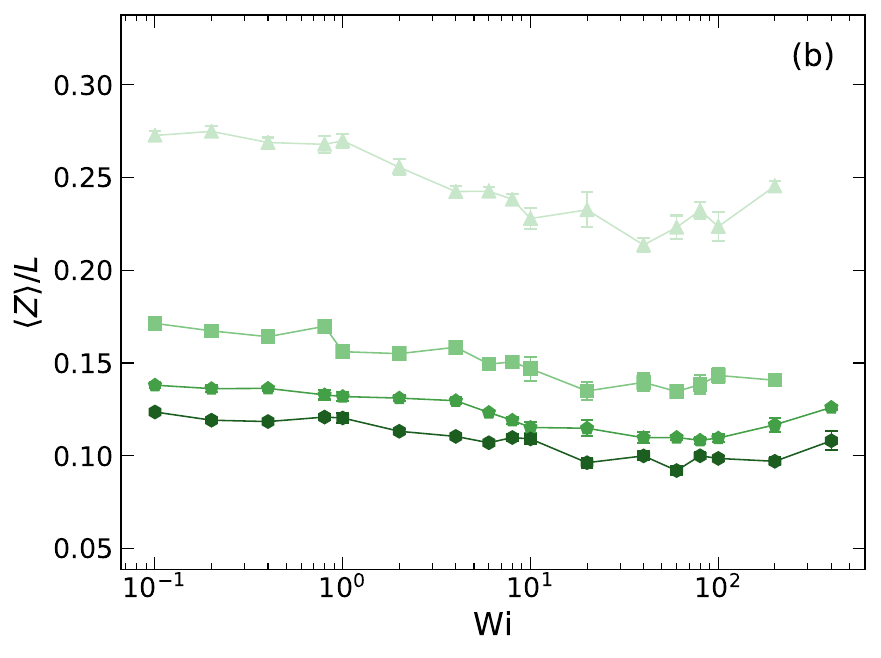}
  \end{minipage}
  
  \vspace{0.2em} % space between rows

  % Bottom row: single centered subfigure
  \begin{minipage}[t]{0.48\textwidth}
    \centering
    \includegraphics[width=0.95\textwidth]{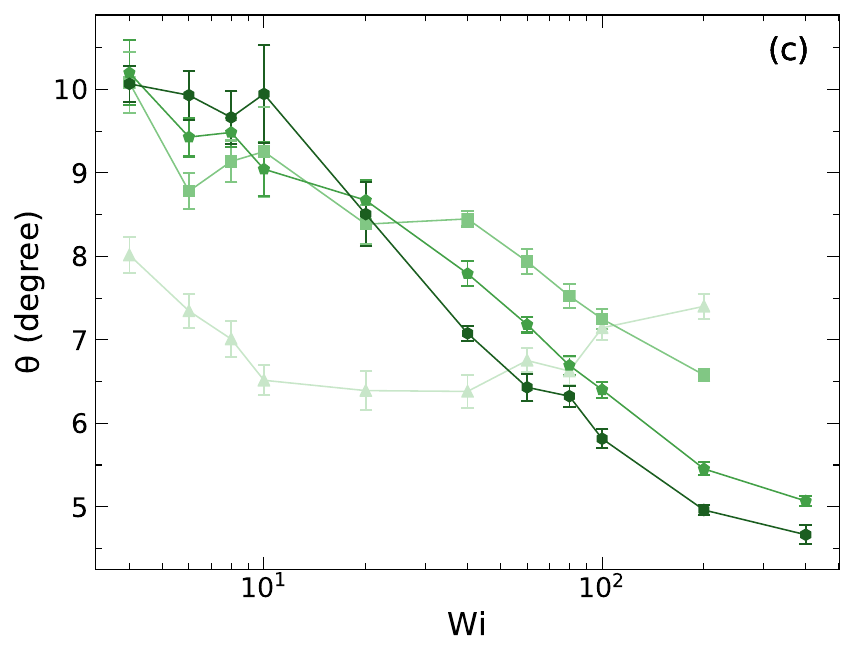}
  \end{minipage}  \captionsetup{justification=raggedright,singlelinecheck=false}

  \caption{Polycatenane properties in shear flow as a function of Weissenberg number (Wi): (a) fractional extension in gradient direction $(\langle Y \rangle / L)$, (b) fractional extension in vorticity direction $(\langle Z \rangle / L)$, and (c) orientation angle ($\theta$). Darker shades indicate higher beads per ring (BPR) in polycatenanes with fixed number of rings ($N_\mathrm{rings}=10)$.}
  \label{fig:PC_inner_combined_shear_plots_app}
\end{figure*}

\clearpage
\begin{figure*}[h!]
  \centering
  % Row 1: fractional extension and viscosity in extension
  \begin{subfigure}[b]{0.48\textwidth}
    \centering
    \includegraphics[width=\textwidth]{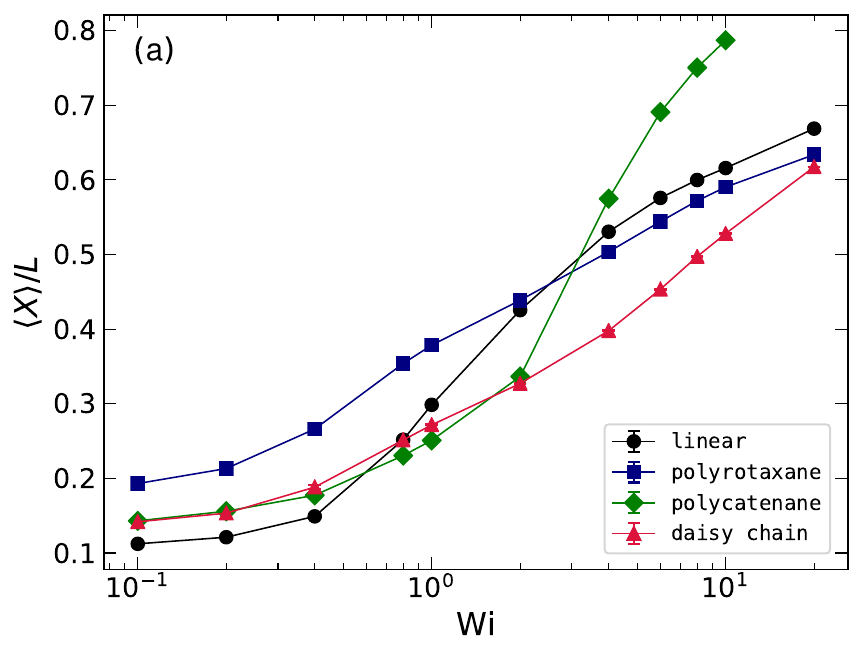}
  \end{subfigure}
  \hfill
  \begin{subfigure}[b]{0.48\textwidth}
    \centering
    \includegraphics[width=\textwidth]{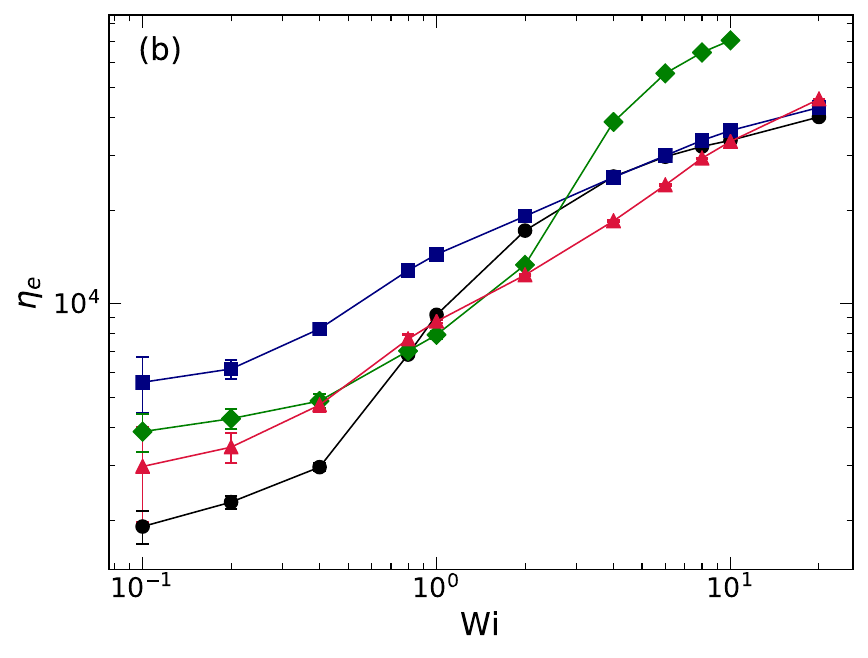}
  \end{subfigure}

  \vspace{0.5em}

  % Row 2: fractional extension and viscosity in shear
  \begin{subfigure}[b]{0.48\textwidth}
    \centering
    \includegraphics[width=\textwidth]{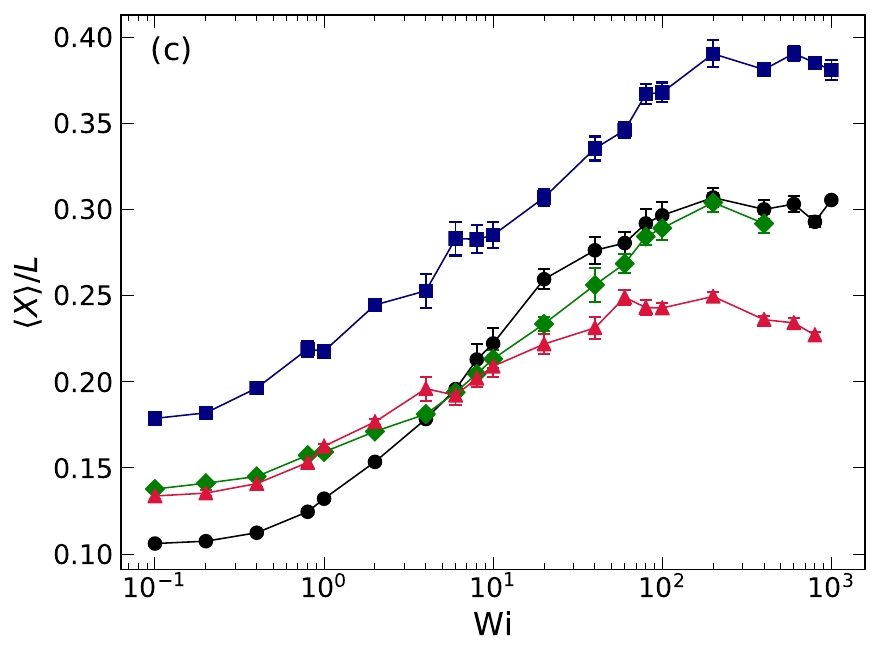}
  \end{subfigure}
  \hfill
  \begin{subfigure}[b]{0.48\textwidth}
    \centering
    \includegraphics[width=\textwidth]{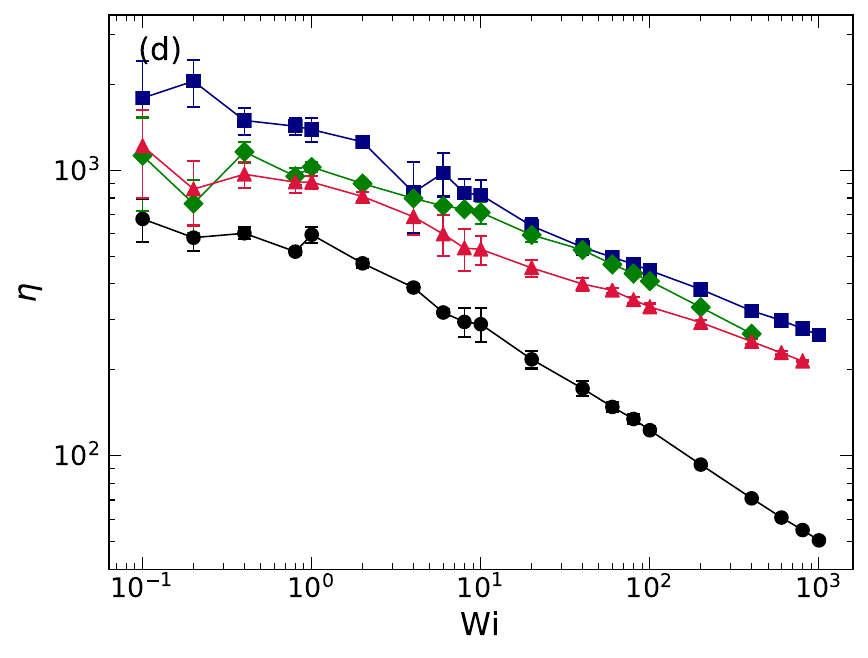}
  \end{subfigure}

  \vspace{0.5em}

   \captionsetup{justification=raggedright,singlelinecheck=false}
  \caption{Comparison of linear and MIP polymers of similar maximum contour length (L) in extensional (a, b) and shear (c, d) flow. Fractional extension in the flow direction ($\langle X_{bb} \rangle / L$) for extension (a) and shear (c) along with extensional viscosity ($\eta_e$) (b) and shear viscosity ($\eta$) are given as a function of Weissenberg number (Wi). The linear polymer (black) has 80 backbone beads. The polyrotaxane (blue) has 80 backbone beads including 2 capping beads and 16 rings. The daisy chain (red) has 10 linear segments of 8 beads each interlocked with 18 8 bead rings and includes 2 capping beads. The polycatenane (green) has 10 interconnected rings of 24 beads each.}
  \label{fig:MIP_comparison_mw}
\end{figure*}

\clearpage

\clearpage
\begin{figure*}[h!]
  \centering
  % Row 1: fractional extension and viscosity in extension
  \begin{subfigure}[b]{0.48\textwidth}
    \centering
    \includegraphics[width=\textwidth]{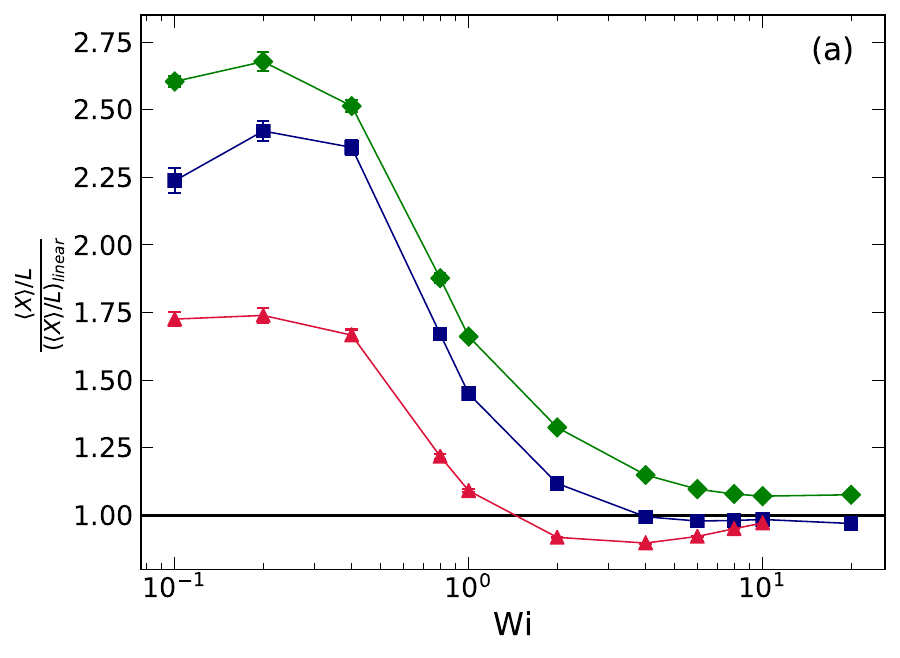}
  \end{subfigure}
  \hfill
  \begin{subfigure}[b]{0.48\textwidth}
    \centering
    \includegraphics[width=\textwidth]{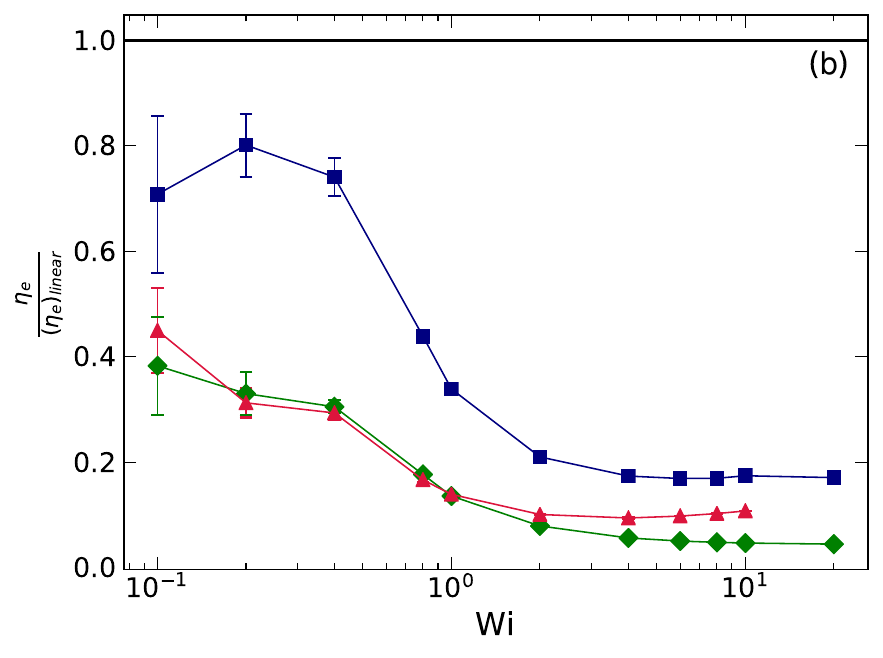}
  \end{subfigure}

  \vspace{0.5em}

  % Row 2: fractional extension and viscosity in shear
  \begin{subfigure}[b]{0.48\textwidth}
    \centering
    \includegraphics[width=\textwidth]{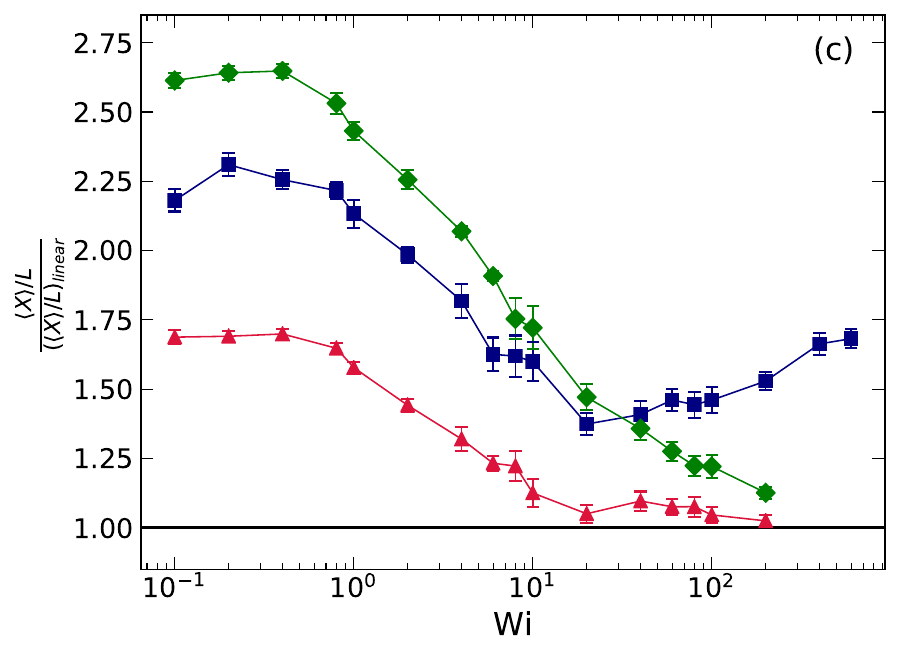}
  \end{subfigure}
  \hfill
  \begin{subfigure}[b]{0.48\textwidth}
    \centering
    \includegraphics[width=\textwidth]{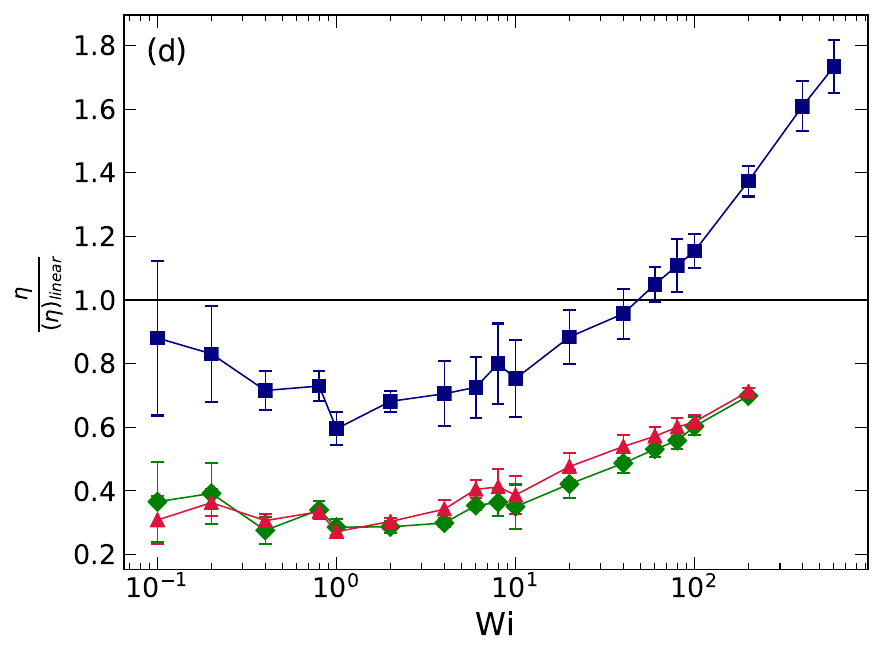}
  \end{subfigure}

  \vspace{0.5em}

   \captionsetup{justification=raggedright,singlelinecheck=false}
    \caption{Comparison of MIPs of similar molecular weight in extensional (a, b) and shear (c, d) flow. All properties are normalized by linear polymer (solid horizontal line).
    Fractional extension in the flow direction $( \langle X_{bb} \rangle / L)$ for extension (a) and shear (c) along with extensional viscosity $(\eta_{e})$ (b) and shear viscosity $(\eta)$ are given as a function of Weissenberg number (Wi).  The linear polymer (black line) has 80 backbone beads ($M_w=80$).  The polyrotaxane (blue) has 40 backbone beads including 2 capping beads and 6 rings ($M_w=88$).  The daisy chain (red) has 4 linear segments of 8 beads each interlocked with 6 8 bead rings ($M_w=74$) and includes 2 capping beads. The polycatenane (green) has 10 interconnected rings of 8 beads each ($M_w=80$).}
  \label{fig:MIP_comparison_mw_norm}
\end{figure*}

\clearpage

\clearpage
\begin{figure*}[h!]
  \centering
  % Row 1: fractional extension and viscosity in extension
  \begin{subfigure}[b]{0.48\textwidth}
    \centering
    \includegraphics[width=\textwidth]{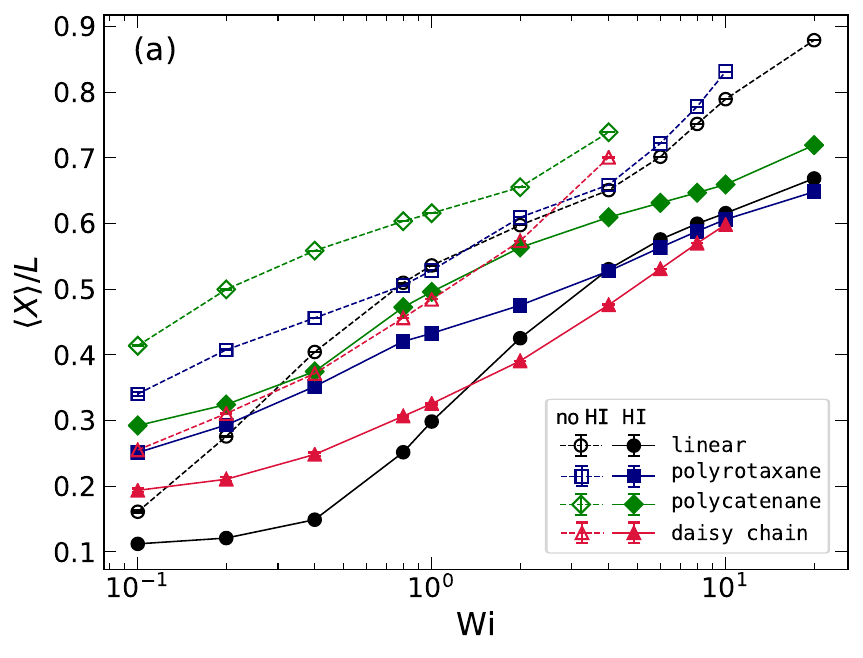}
  \end{subfigure}
  \hfill
  \begin{subfigure}[b]{0.48\textwidth}
    \centering
    \includegraphics[width=\textwidth]{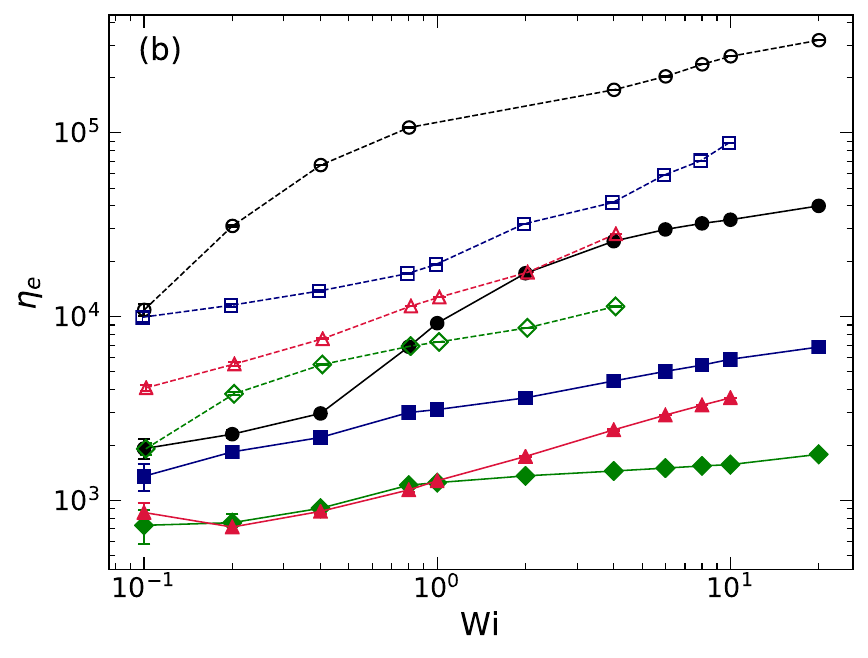}
  \end{subfigure}

  \vspace{0.5em}

  % Row 2: fractional extension and viscosity in shear
  \begin{subfigure}[b]{0.48\textwidth}
    \centering
    \includegraphics[width=\textwidth]{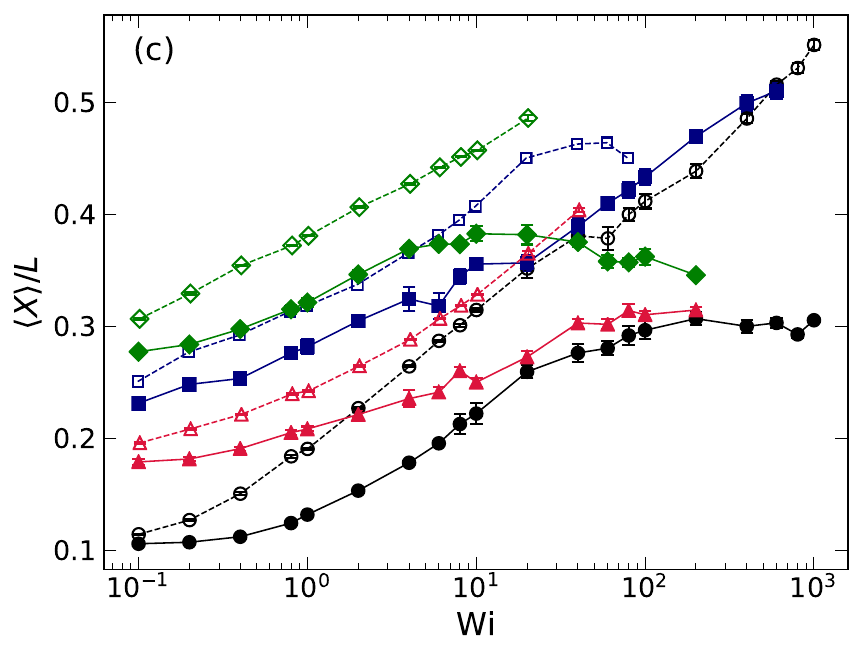}
  \end{subfigure}
  \hfill
  \begin{subfigure}[b]{0.48\textwidth}
    \centering
    \includegraphics[width=\textwidth]{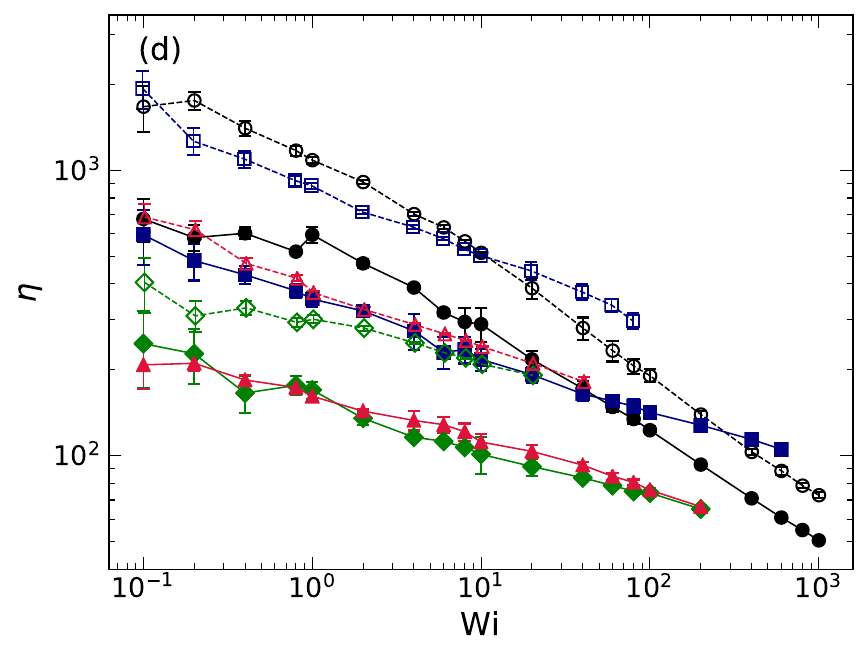}
  \end{subfigure}

  \vspace{0.5em}

   \captionsetup{justification=raggedright,singlelinecheck=false}
    \caption{Comparison of MIPs of similar molecular weight in extensional (a, b) and shear (c, d) flow with (solid lines and filled markers) and without hydrodynamic interaction (dashed lines and hollow markers). 
    Fractional extension in the flow direction $( \langle X_{bb} \rangle / L)$ for extension (a) and shear (c) along with extensional viscosity $(\eta_{e})$ (b) and shear viscosity $(\eta)$ are given as a function of Weissenberg number (Wi).  The linear polymer (black line) has 80 backbone beads ($M_w=80$).  The polyrotaxane (blue) has 40 backbone beads including 2 capping beads and 6 rings ($M_w=88$).  The daisy chain (red) has 4 linear segments of 8 beads each interlocked with 6 8 bead rings ($M_w=74$) and includes 2 capping beads. The polycatenane (green) has 10 interconnected rings of 8 beads each ($M_w=80$).}
  \label{fig:MIP_comparison_Mw_NoHI}
\end{figure*}

\clearpage

\clearpage
\begin{figure*}[h!]
\setcounter{figure}{0}
\section{trajectory snapshots}
    \centering
    % ===== EQ row =====
    \begin{subfigure}[b]{0.22\textwidth}
        \stackon{\includegraphics[width=\textwidth]{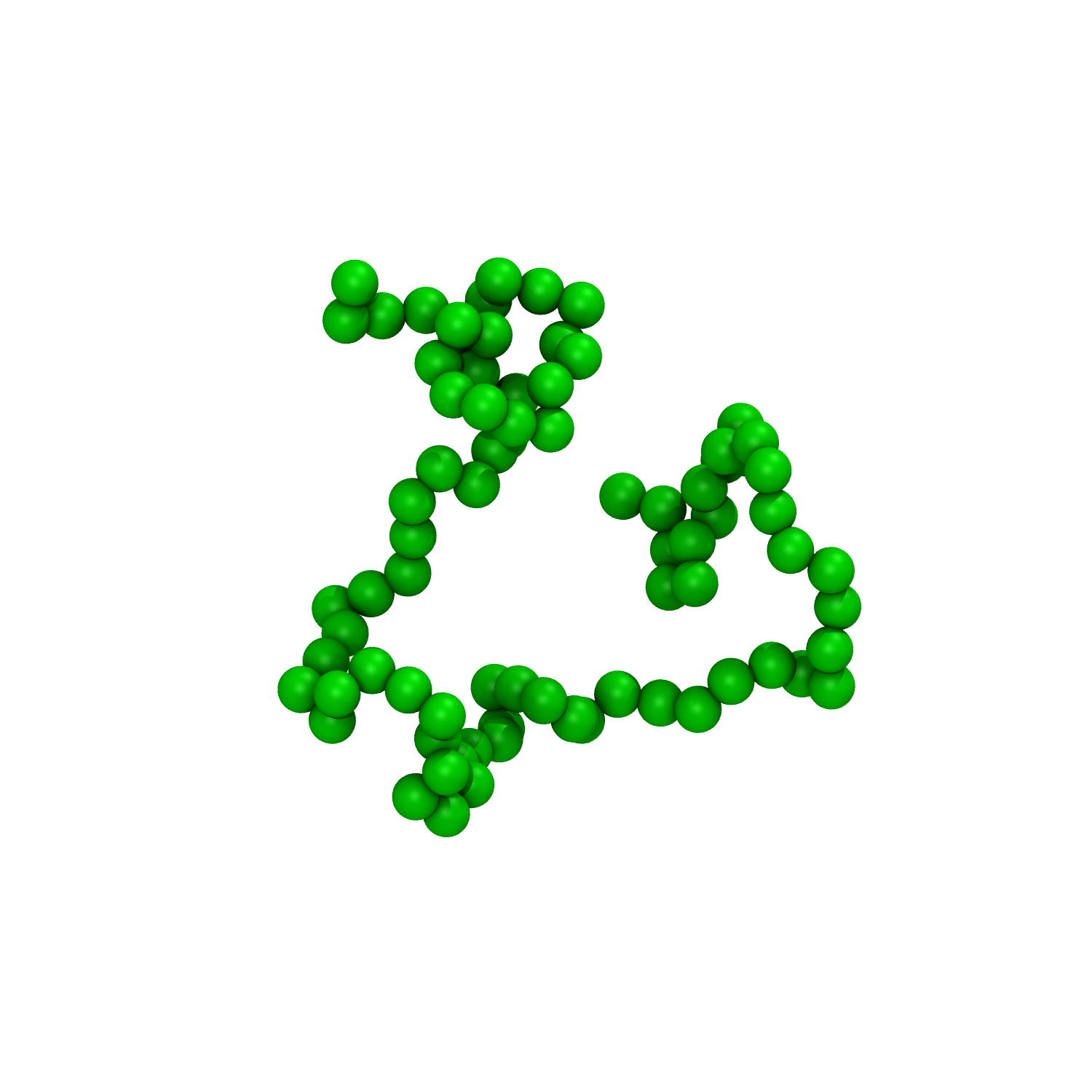}}{\scriptsize Linear (equilibrium)}
    \end{subfigure}
    \begin{subfigure}[b]{0.22\textwidth}
        \stackon{\includegraphics[width=\textwidth]{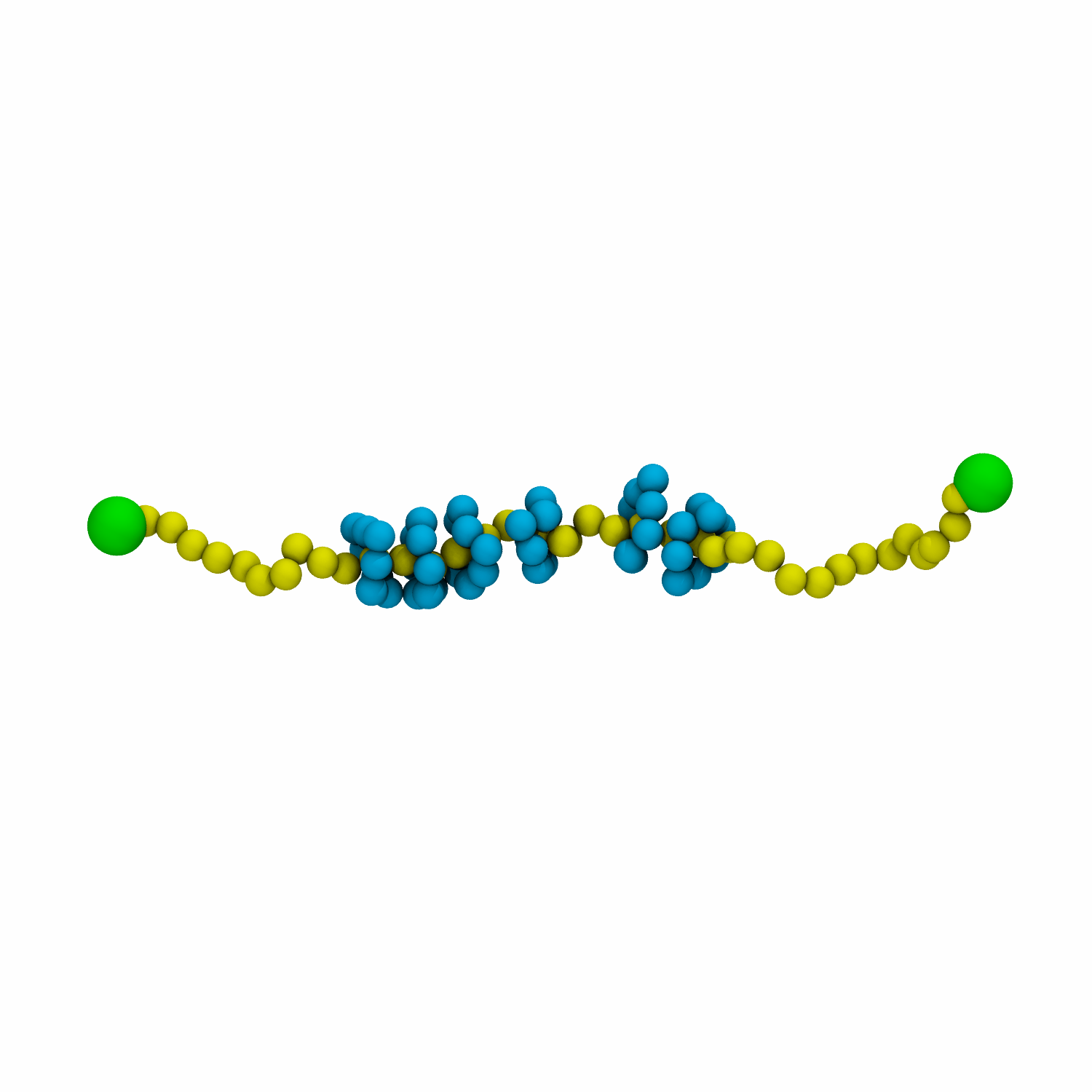}}{\scriptsize Polyrotaxane (equilibrium)}
    \end{subfigure}
    \begin{subfigure}[b]{0.22\textwidth}
        \stackon{\includegraphics[width=\textwidth]{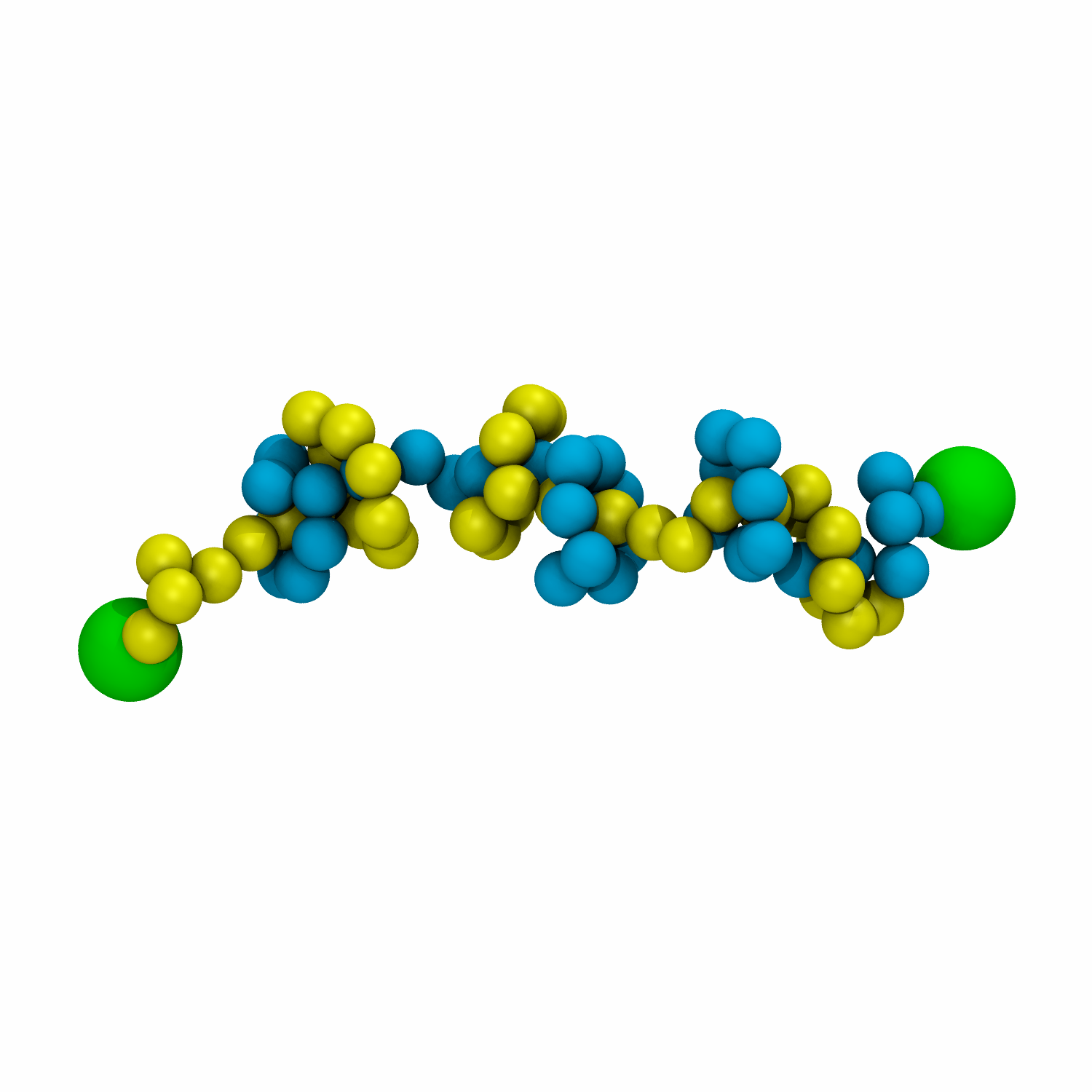}}{\scriptsize Daisy Chain (equilibrium)}
    \end{subfigure}
    \begin{subfigure}[b]{0.22\textwidth}
        \stackon{\includegraphics[width=\textwidth]{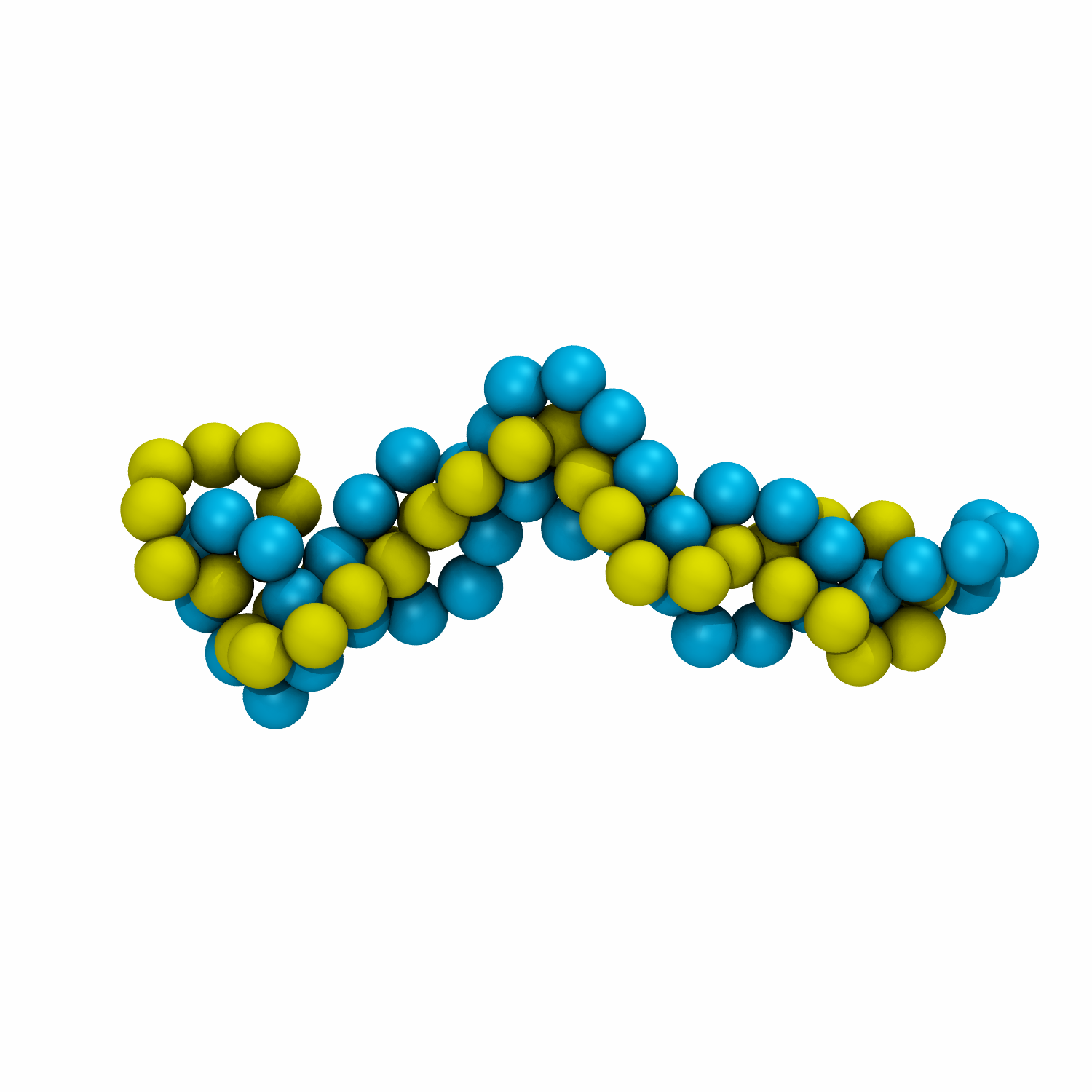}}{\scriptsize Polycatenane (equilibrium)}
    \end{subfigure}
    \\
    % ===== EXT row =====
    \begin{subfigure}[b]{0.22\textwidth}
        \stackon{\includegraphics[width=\textwidth]{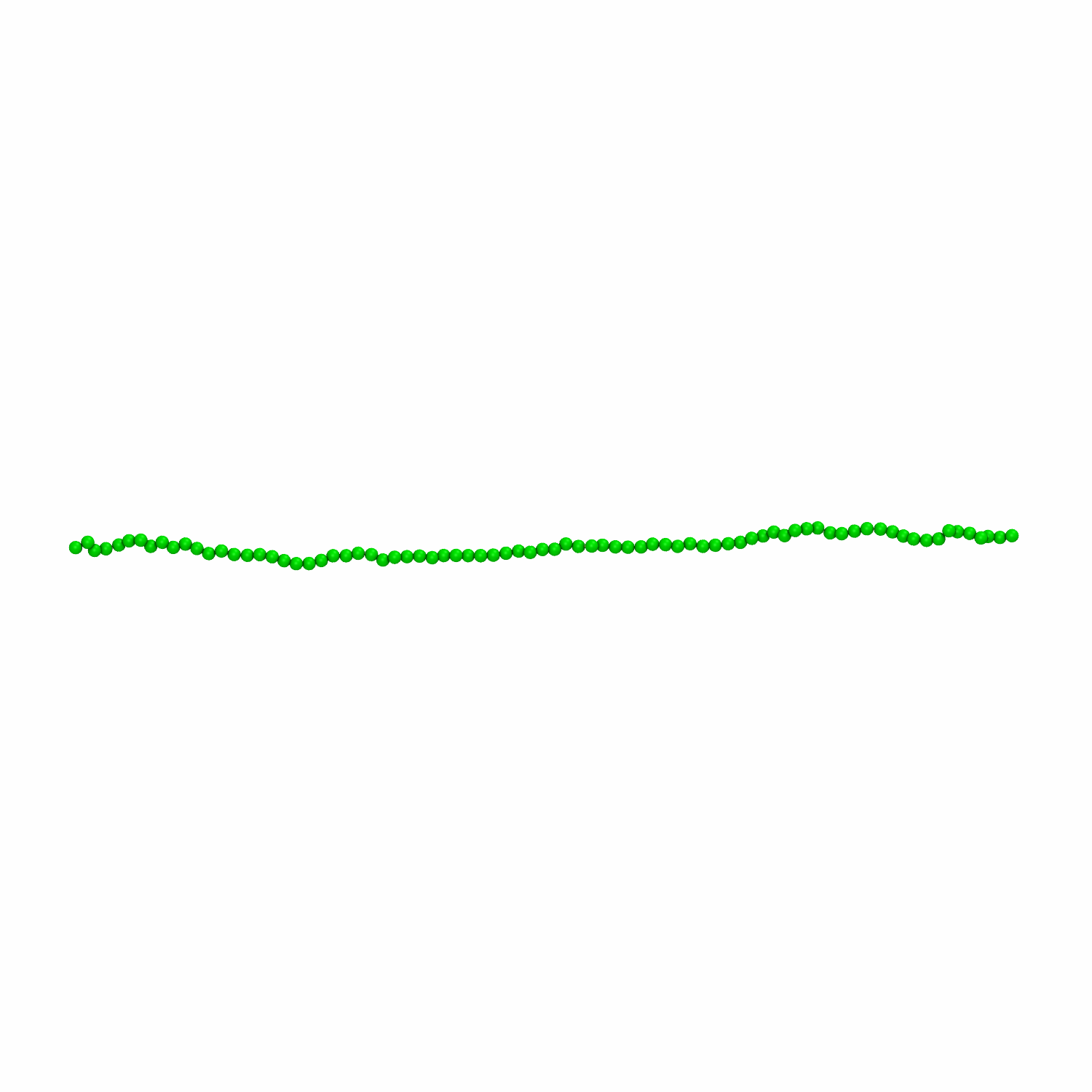}}{\scriptsize Linear (extension)}
    \end{subfigure}
    \begin{subfigure}[b]{0.22\textwidth}
        \stackon{\includegraphics[width=\textwidth]{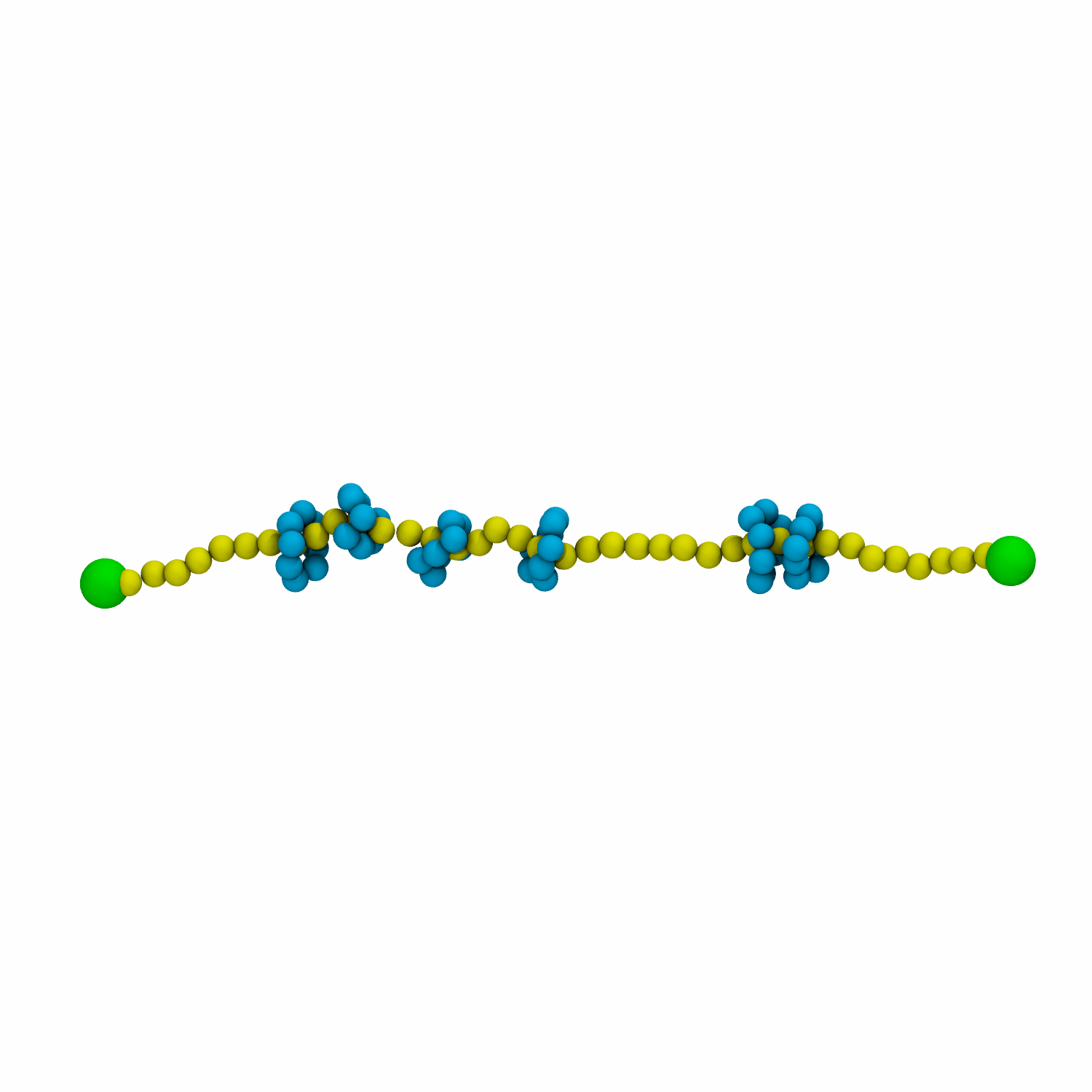}}{\scriptsize Polyrotaxane (extension)}
    \end{subfigure}
    \begin{subfigure}[b]{0.22\textwidth}
        \stackon{\includegraphics[width=\textwidth]{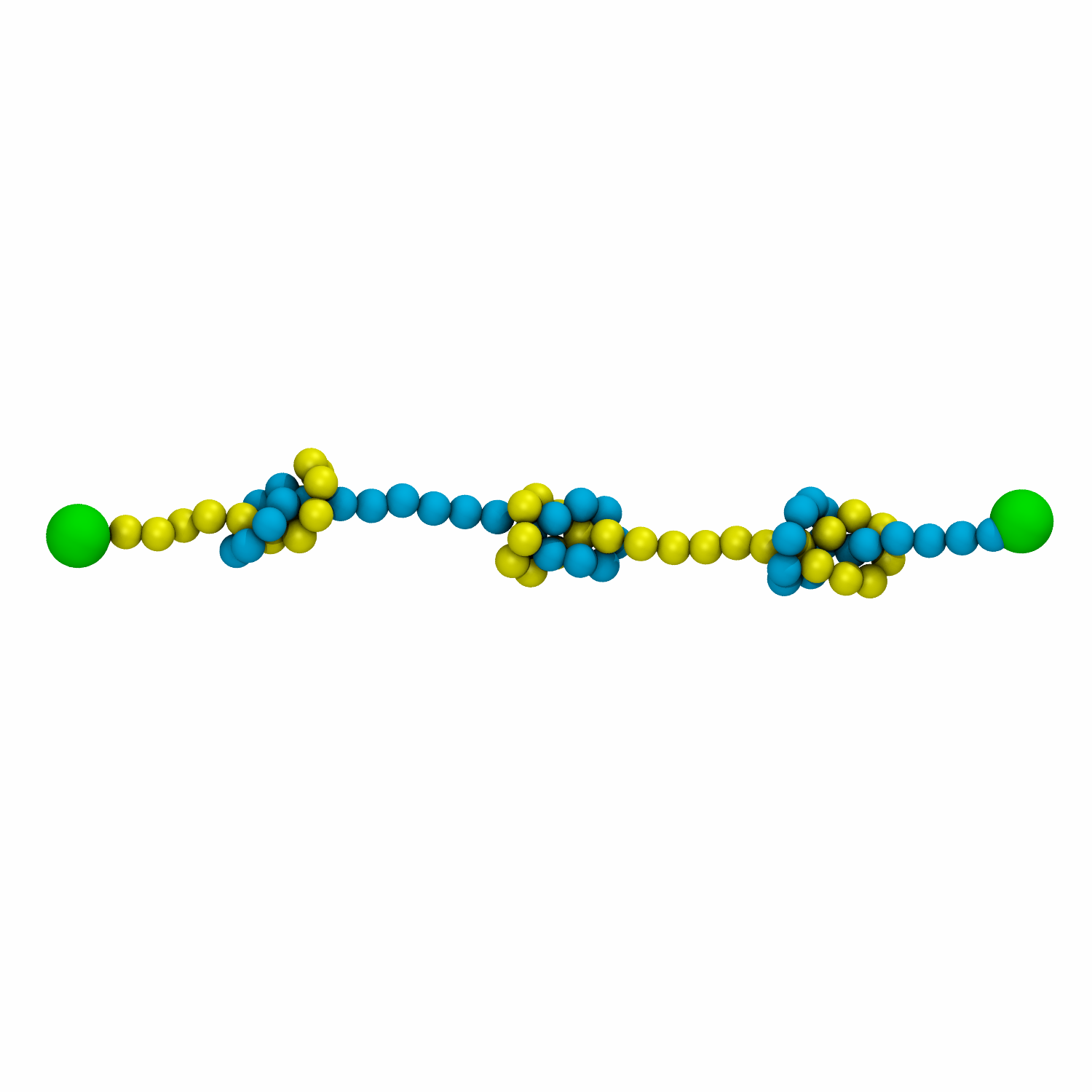}}{\scriptsize Daisy Chain (extension)}
    \end{subfigure}
    \begin{subfigure}[b]{0.22\textwidth}
        \stackon{\includegraphics[width=\textwidth]{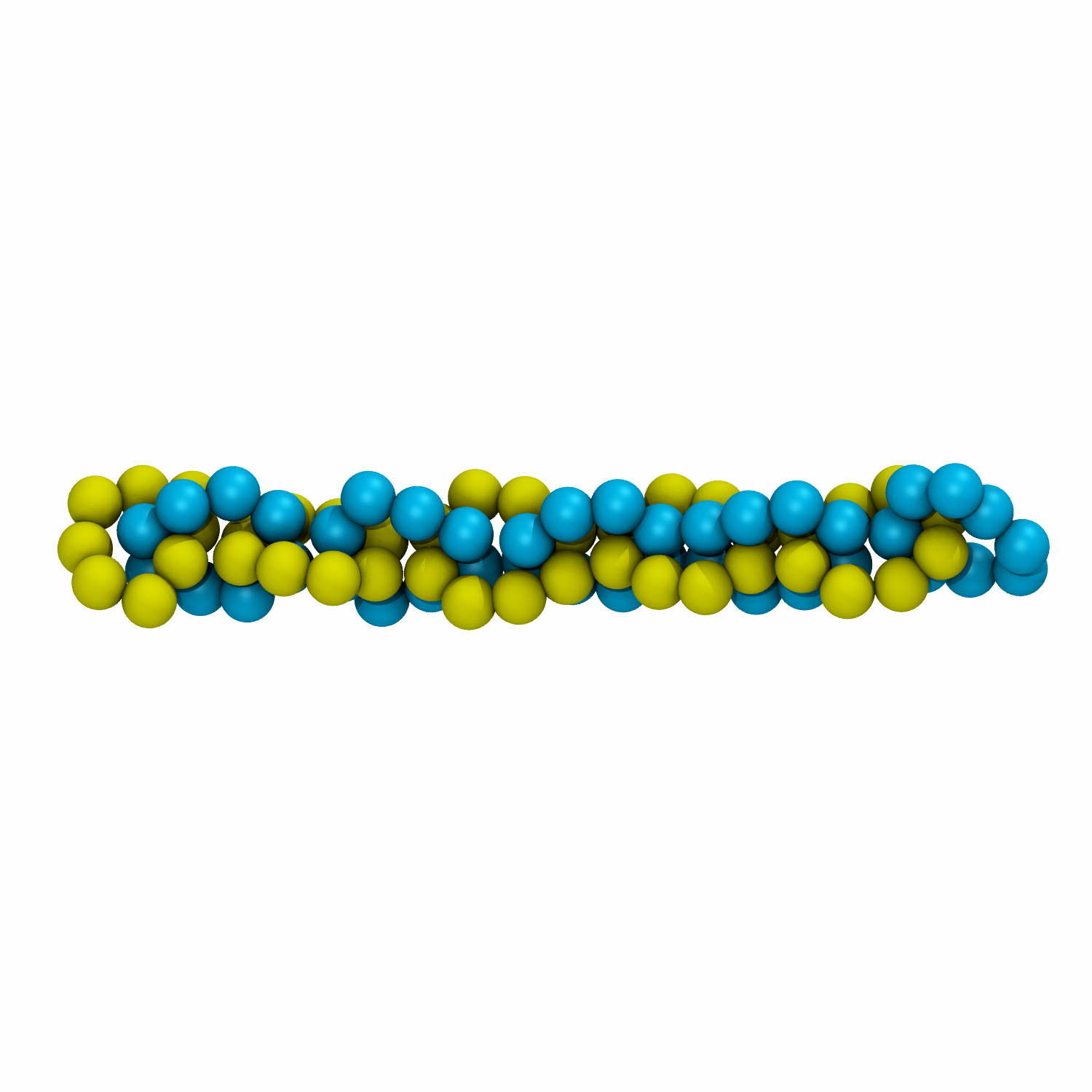}}{\scriptsize Polycatenane (extension)}
    \end{subfigure}
    \\
    % ===== lin-shear =====
    \begin{subfigure}[b]{0.18\textwidth}
        \stackon{\includegraphics[width=\textwidth]{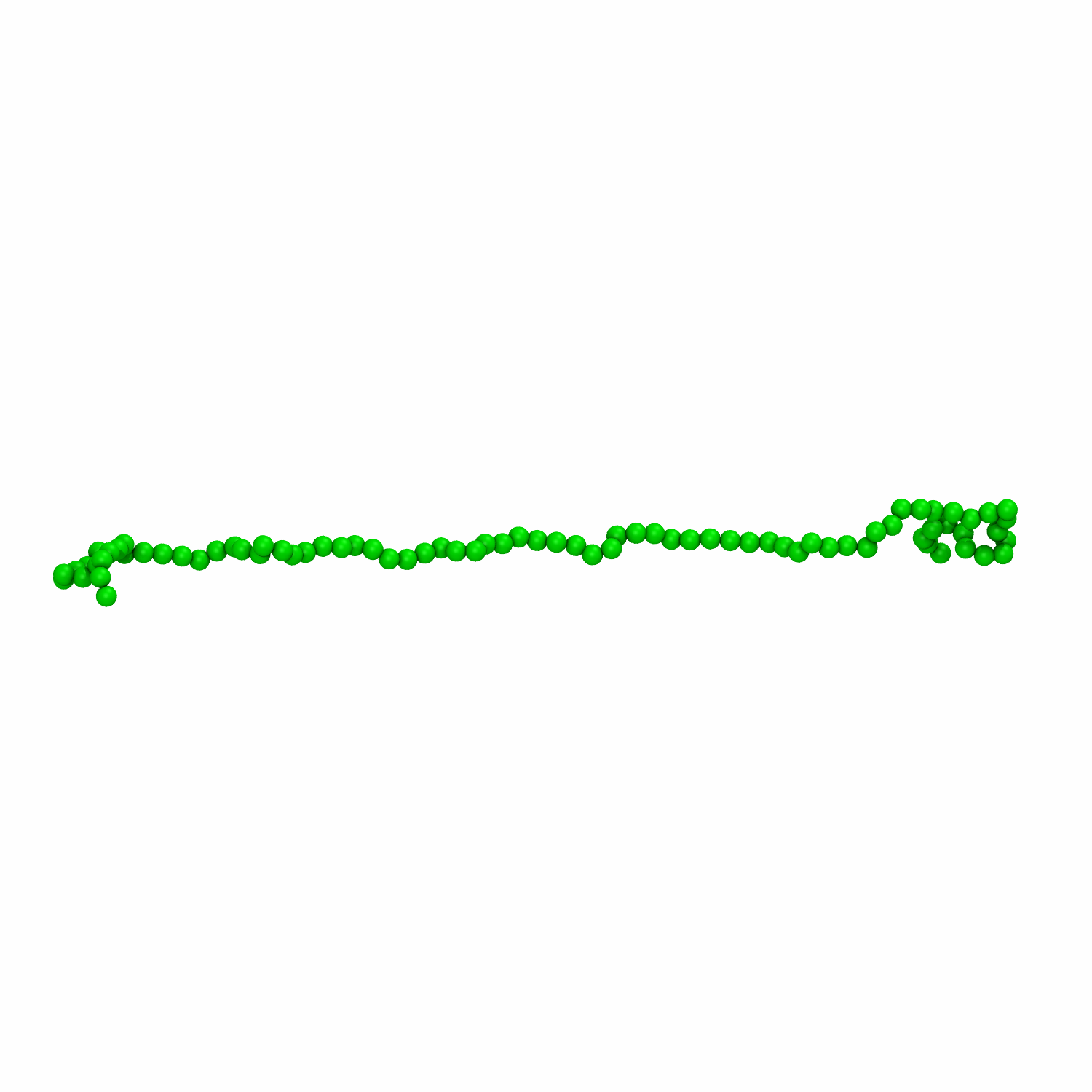}}{\scriptsize Linear (Shear 1)}
    \end{subfigure}
    \begin{subfigure}[b]{0.18\textwidth}
        \stackon{\includegraphics[width=\textwidth]{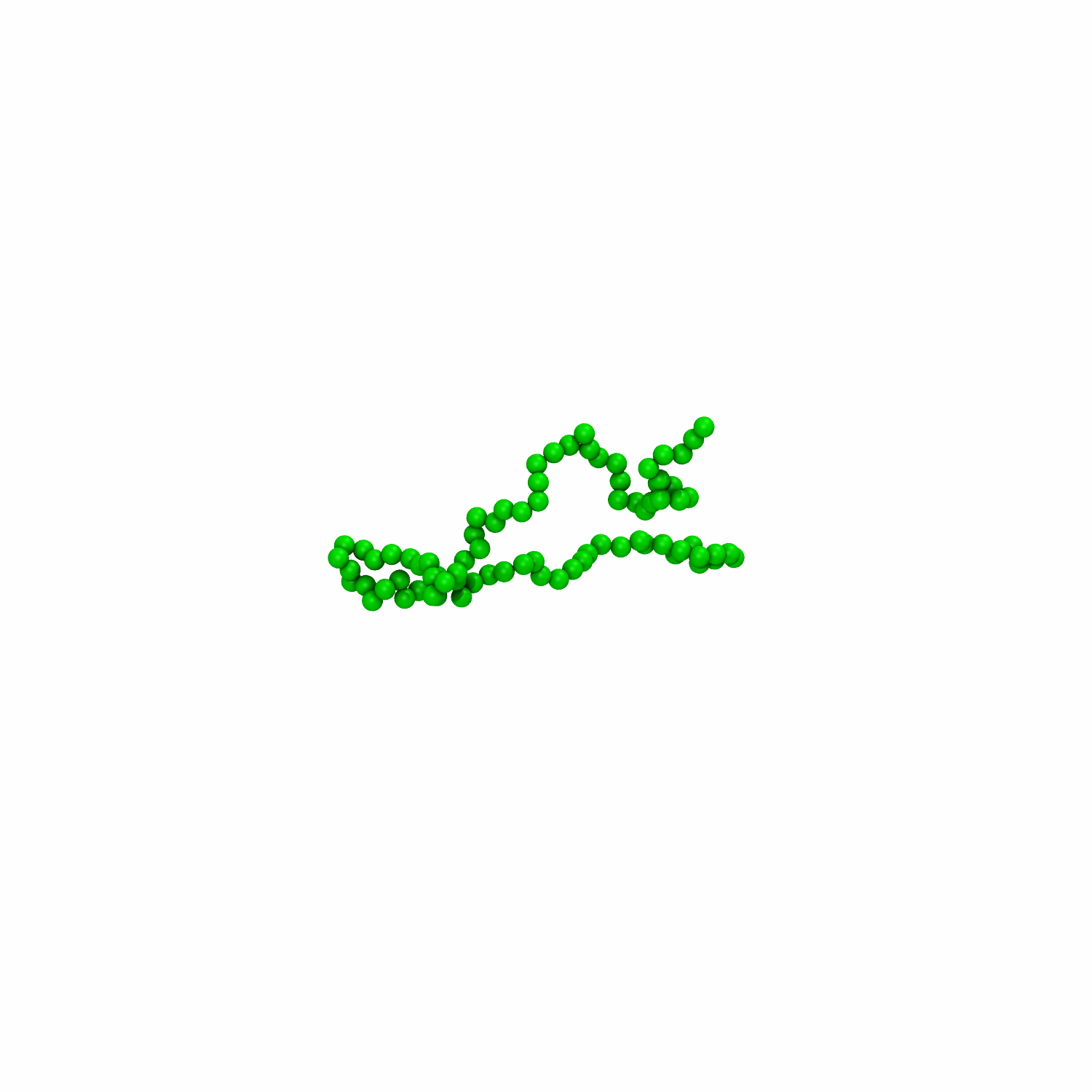}}{\scriptsize Linear (Shear 2)}
    \end{subfigure}
    \begin{subfigure}[b]{0.18\textwidth}
        \stackon{\includegraphics[width=\textwidth]{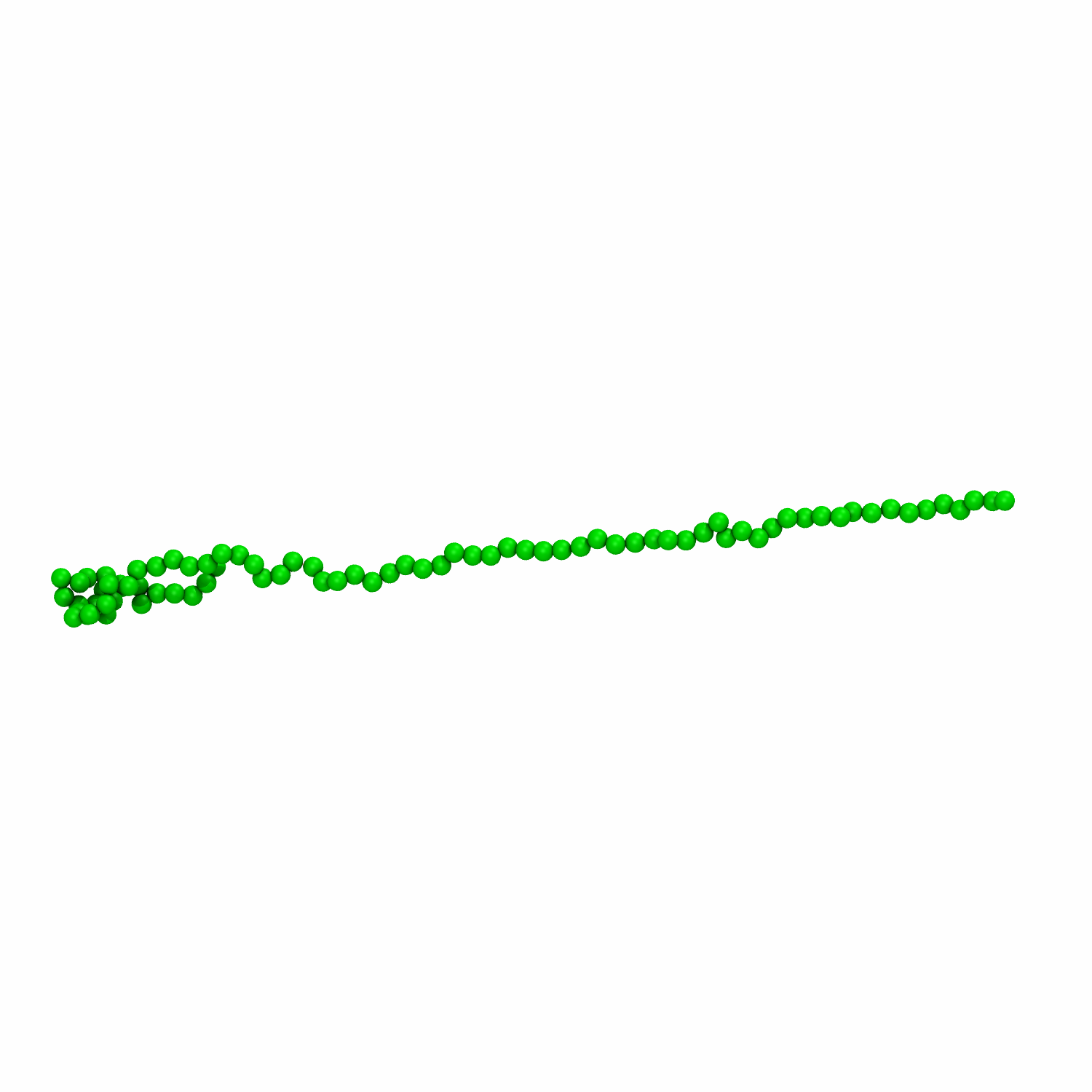}}{\scriptsize Linear (Shear 3)}
    \end{subfigure}
    \begin{subfigure}[b]{0.18\textwidth}
        \stackon{\includegraphics[width=\textwidth]{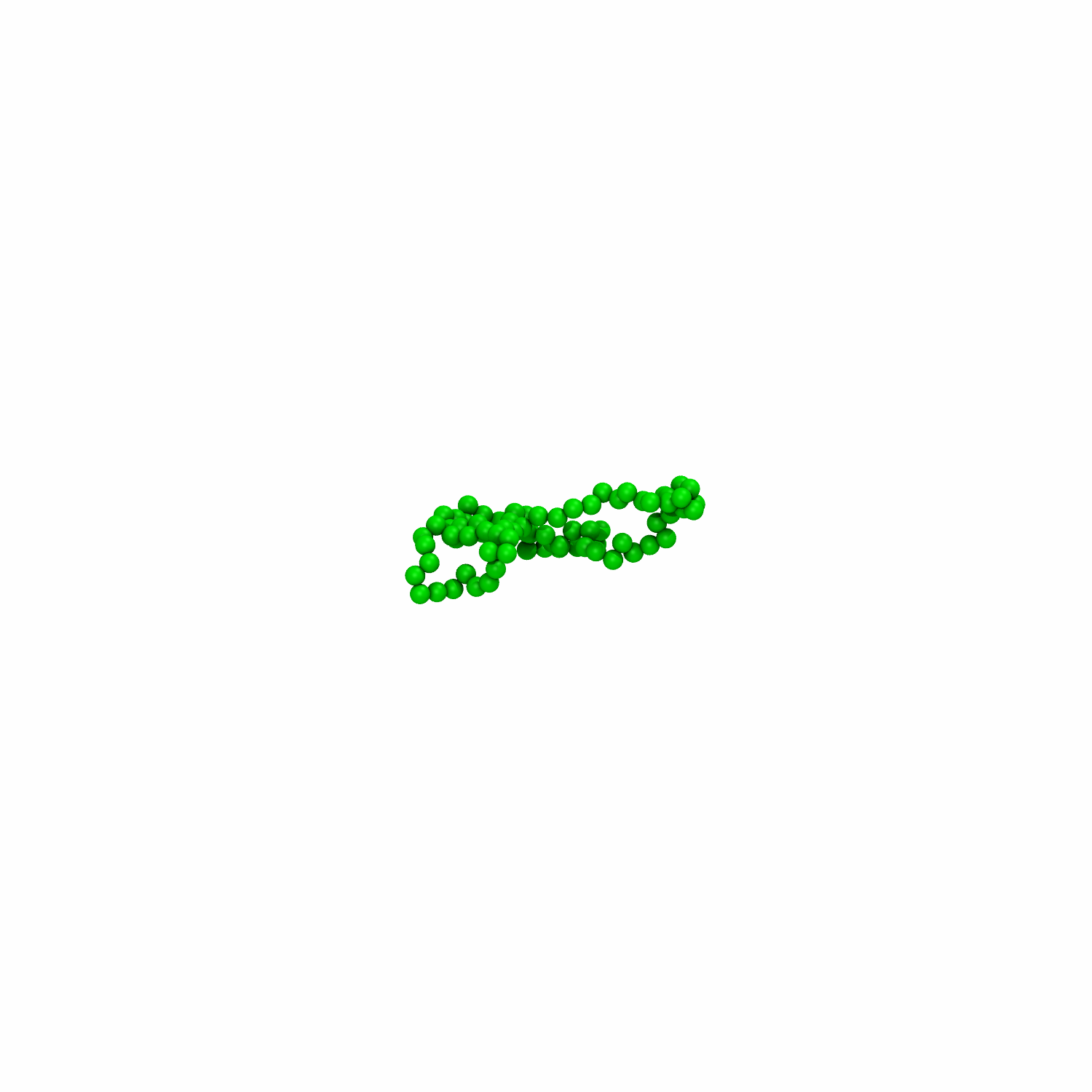}}{\scriptsize Linear (Shear 4)}
    \end{subfigure}
    \begin{subfigure}[b]{0.18\textwidth}
        \stackon{\includegraphics[width=\textwidth]{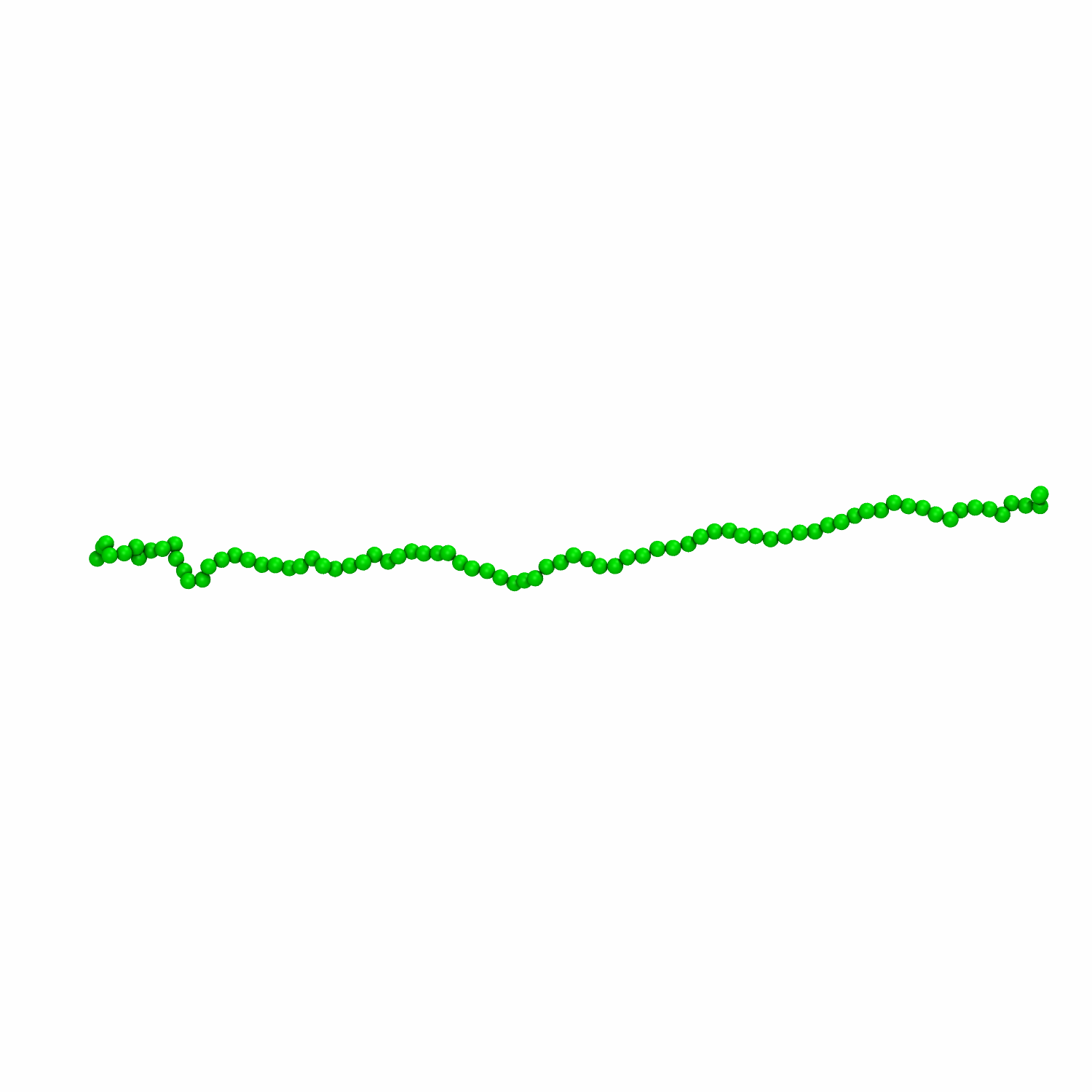}}{\scriptsize Linear (Shear 5)}
    \end{subfigure}
    \\
    % ===== pr-shear =====
    \begin{subfigure}[b]{0.18\textwidth}
        \stackon{\includegraphics[width=\textwidth]{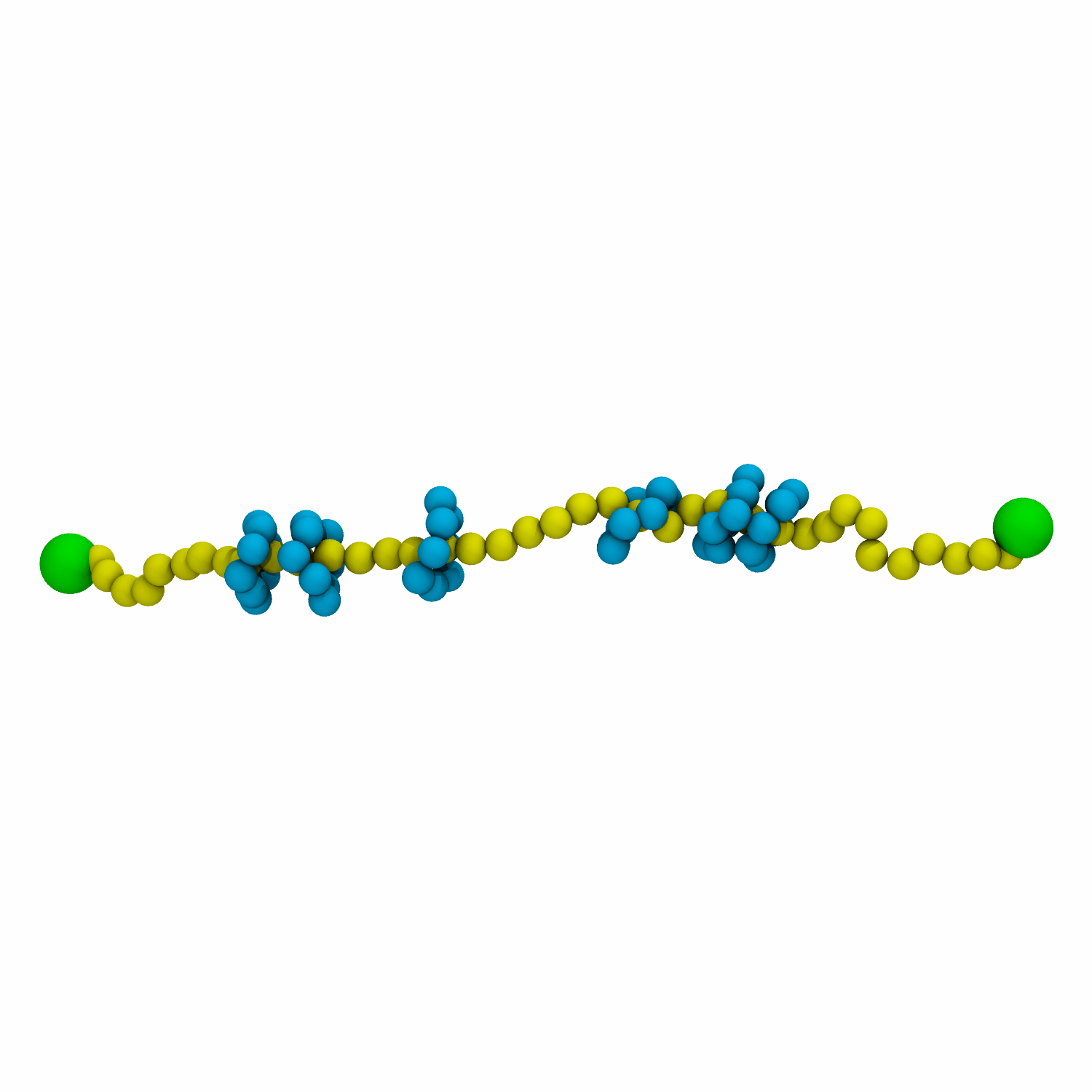}}{\scriptsize Polyrotaxane (Shear 1)}
    \end{subfigure}
    \begin{subfigure}[b]{0.18\textwidth}
        \stackon{\includegraphics[width=\textwidth]{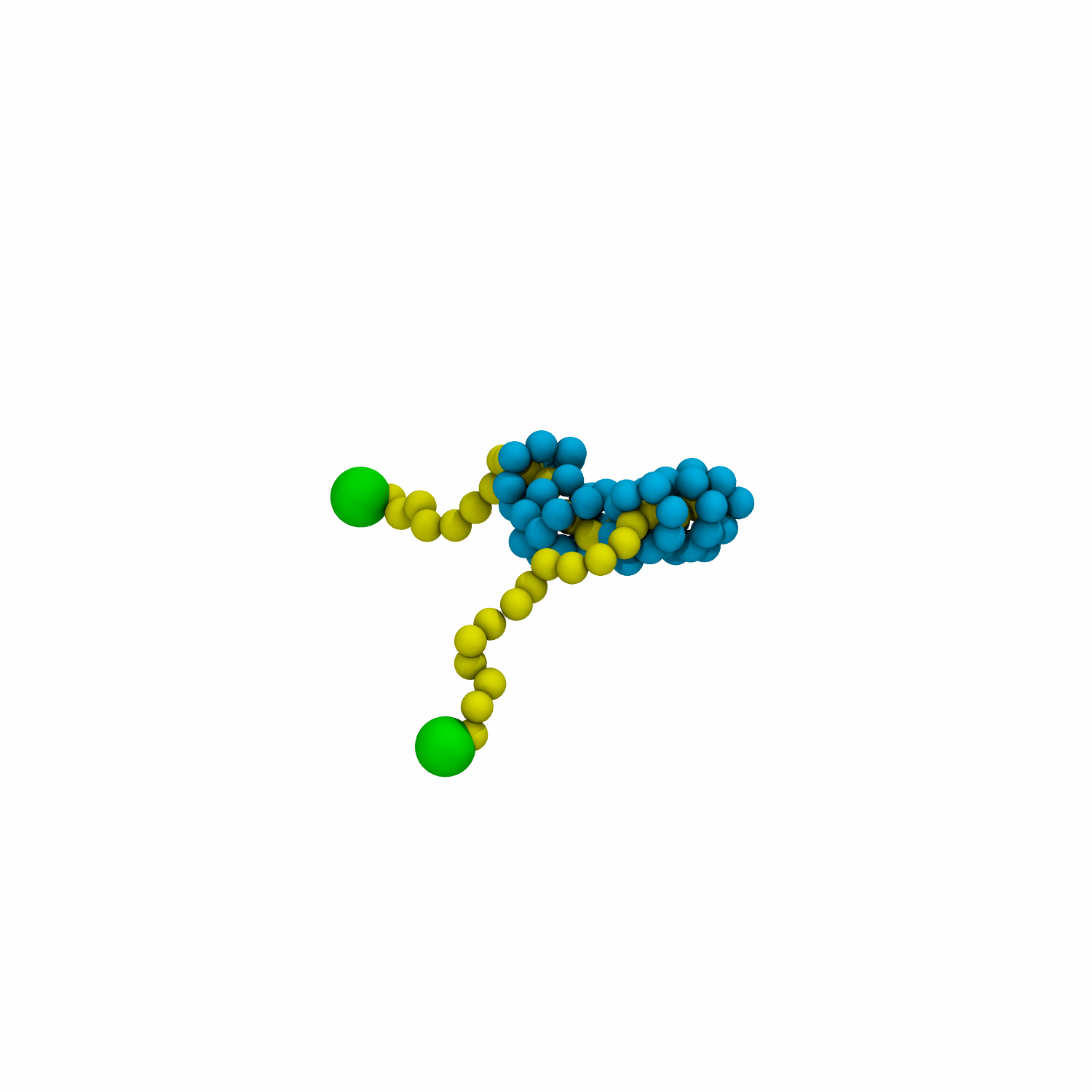}}{\scriptsize Polyrotaxane (Shear 2)}
    \end{subfigure}
    \begin{subfigure}[b]{0.18\textwidth}
        \stackon{\includegraphics[width=\textwidth]{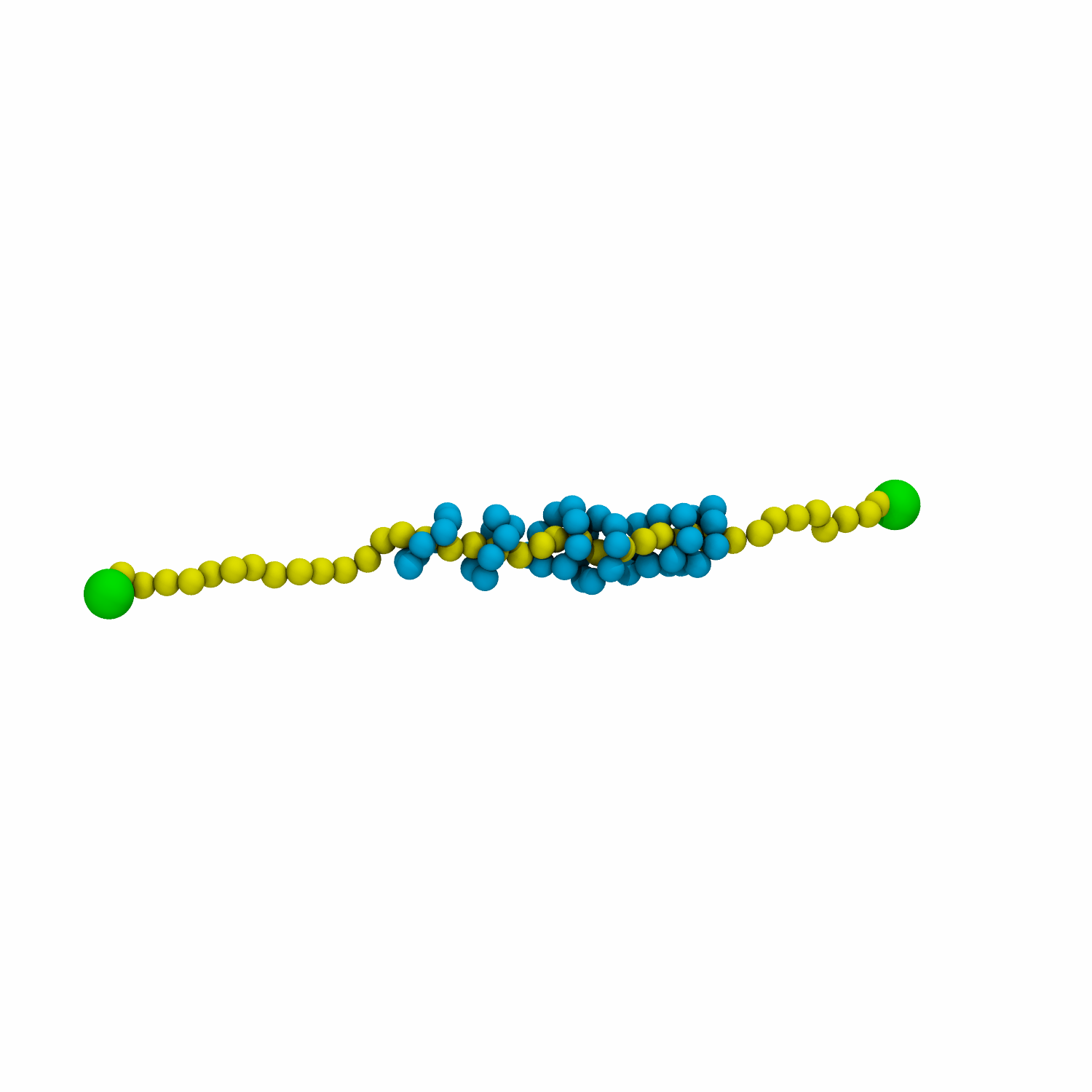}}{\scriptsize Polyrotaxane (Shear 3)}
    \end{subfigure}
    \begin{subfigure}[b]{0.18\textwidth}
        \stackon{\includegraphics[width=\textwidth]{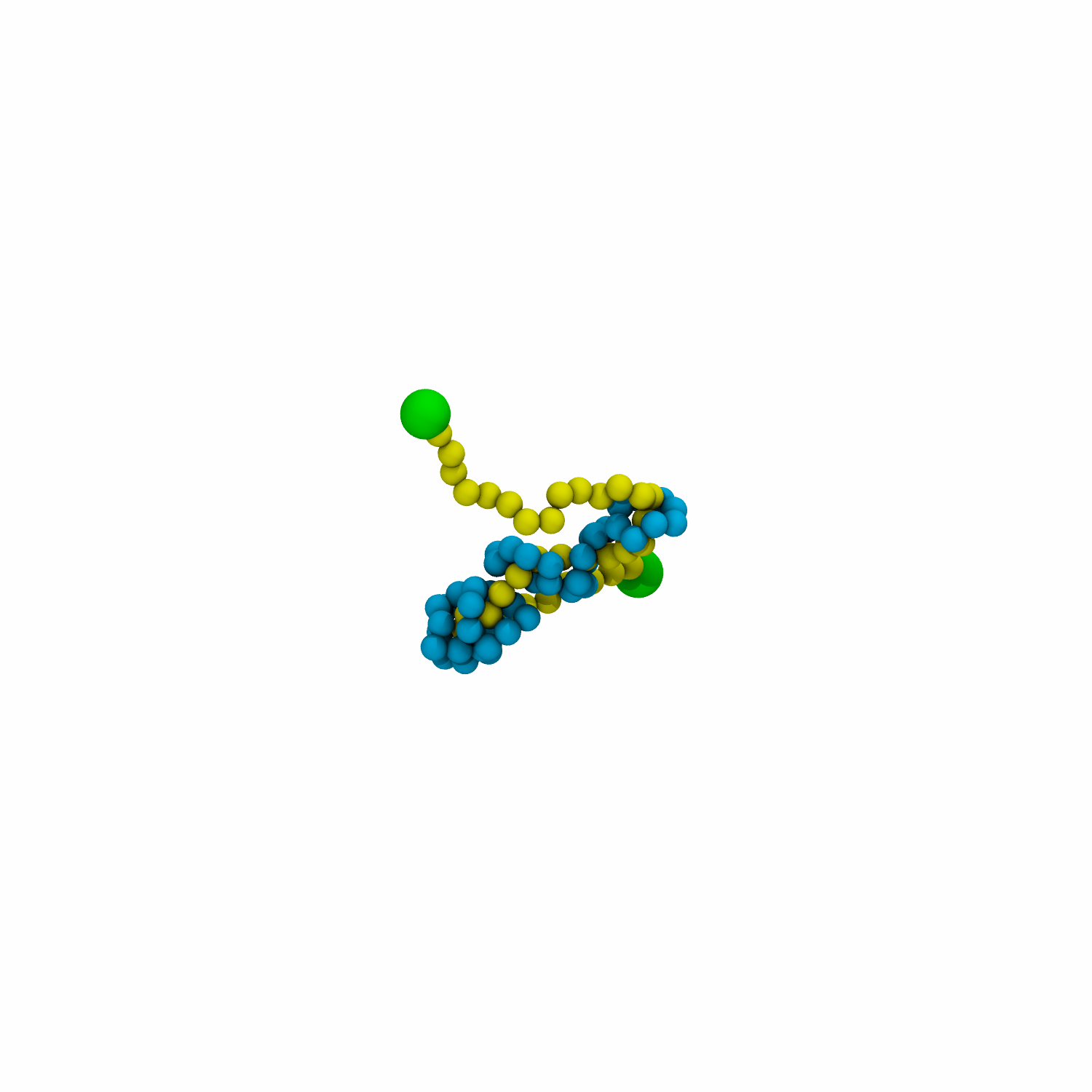}}{\scriptsize Polyrotaxane (Shear 4)}
    \end{subfigure}
    \begin{subfigure}[b]{0.18\textwidth}
        \stackon{\includegraphics[width=\textwidth]{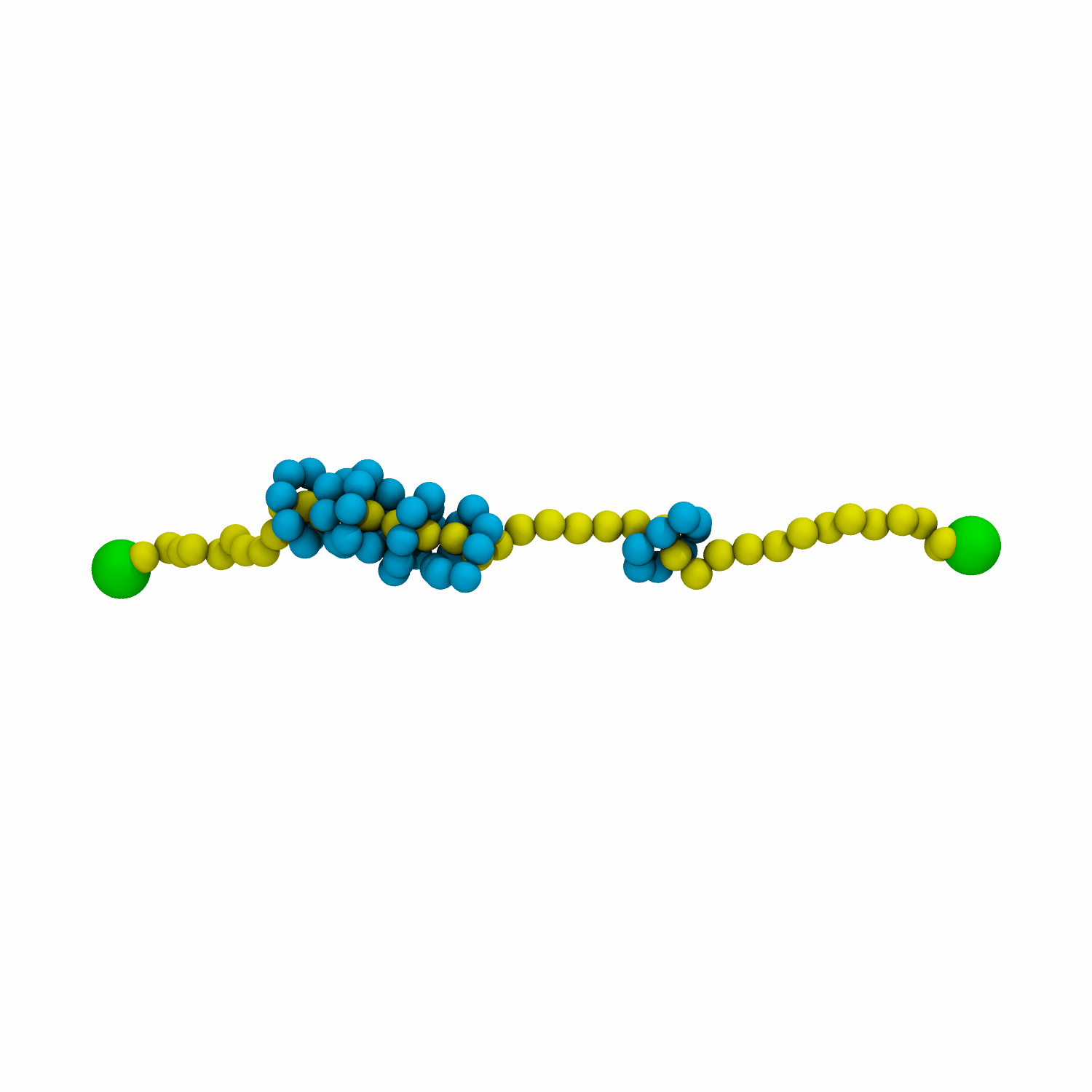}}{\scriptsize Polyrotaxane (Shear 5)}
    \end{subfigure}
    \\
    % ===== dc-shear =====
    \begin{subfigure}[b]{0.18\textwidth}
        \stackon{\includegraphics[width=\textwidth]{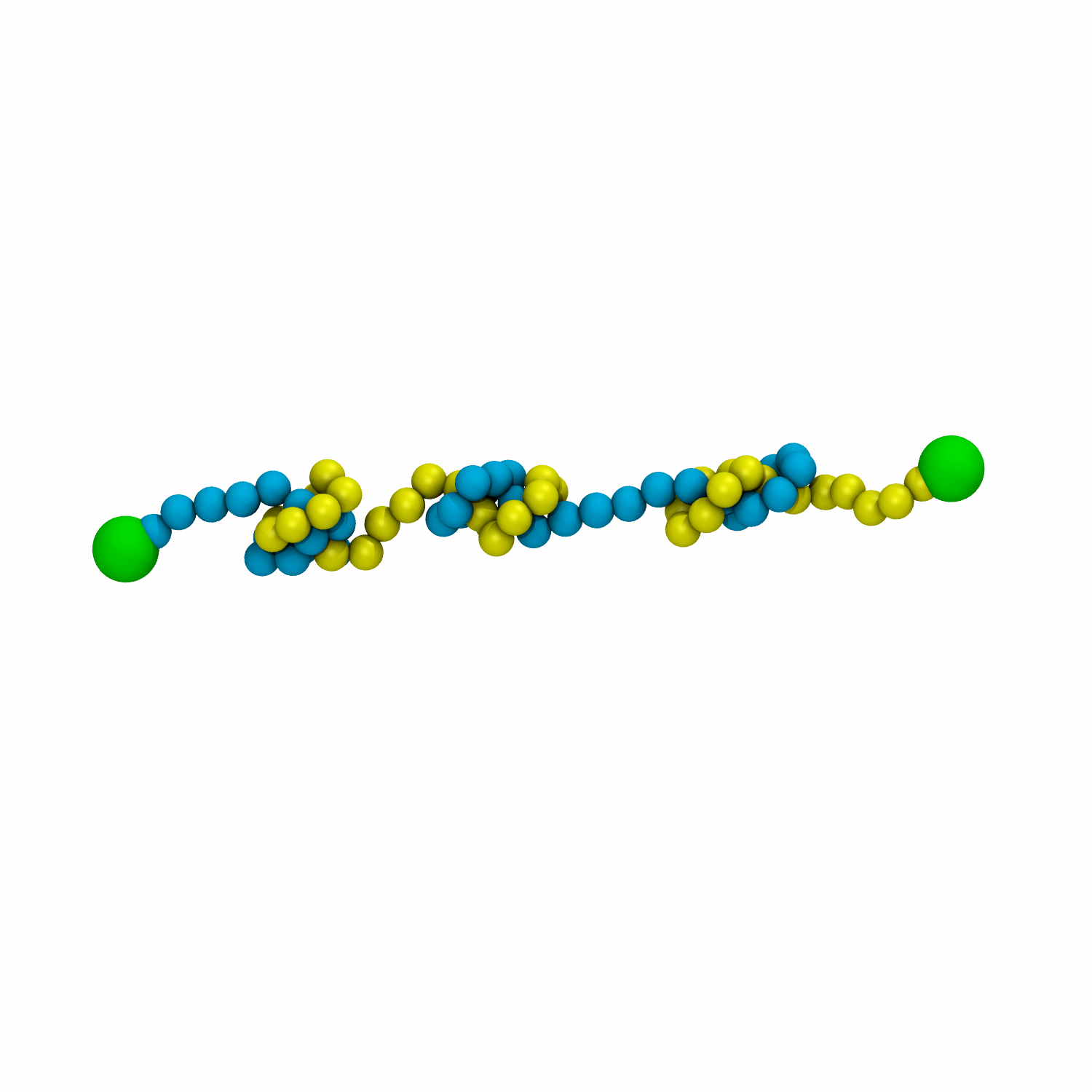}}{\scriptsize Daisy Chain (Shear 1)}
    \end{subfigure}
    \begin{subfigure}[b]{0.18\textwidth}
        \stackon{\includegraphics[width=\textwidth]{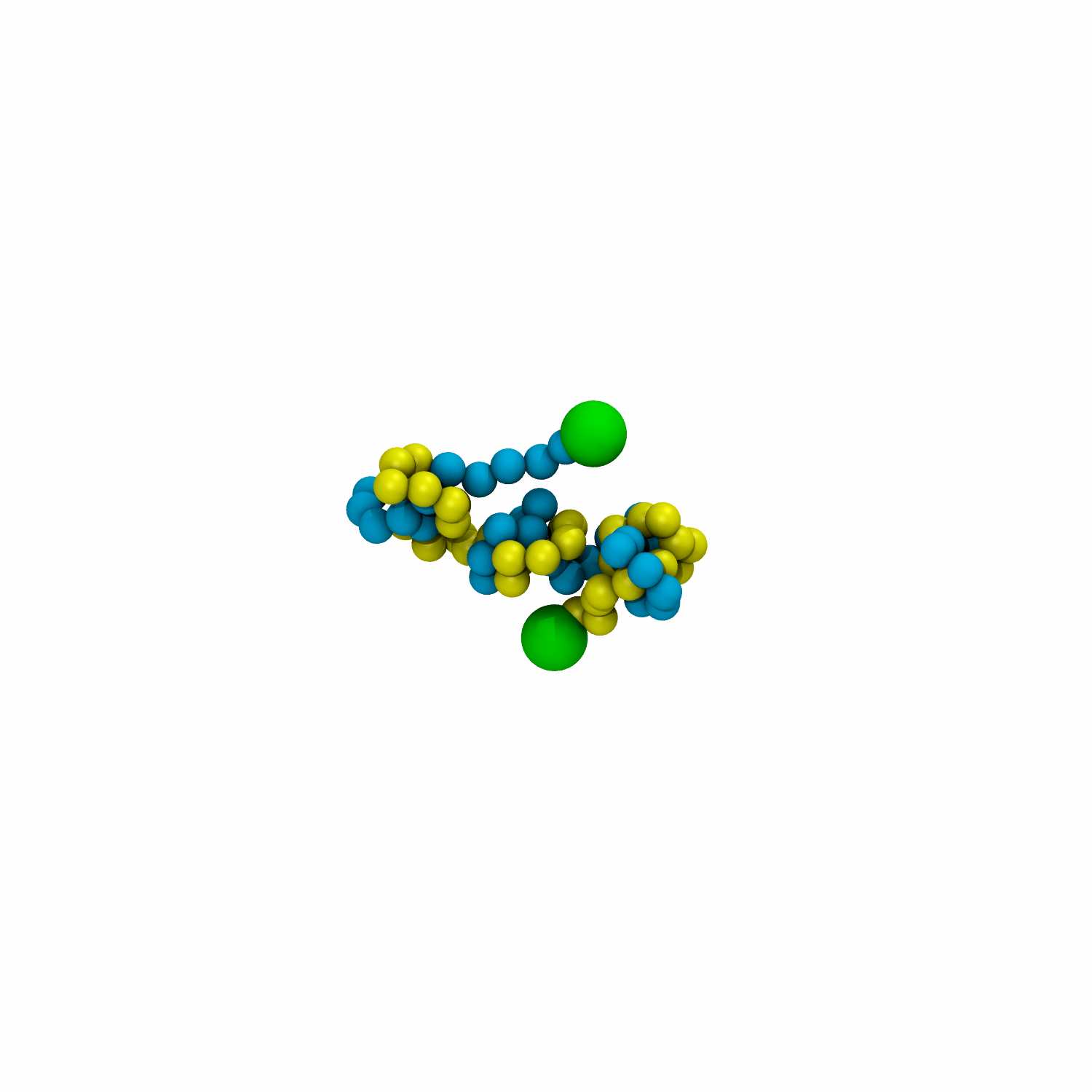}}{\scriptsize Daisy Chain (Shear 2)}
    \end{subfigure}
    \begin{subfigure}[b]{0.18\textwidth}
        \stackon{\includegraphics[width=\textwidth]{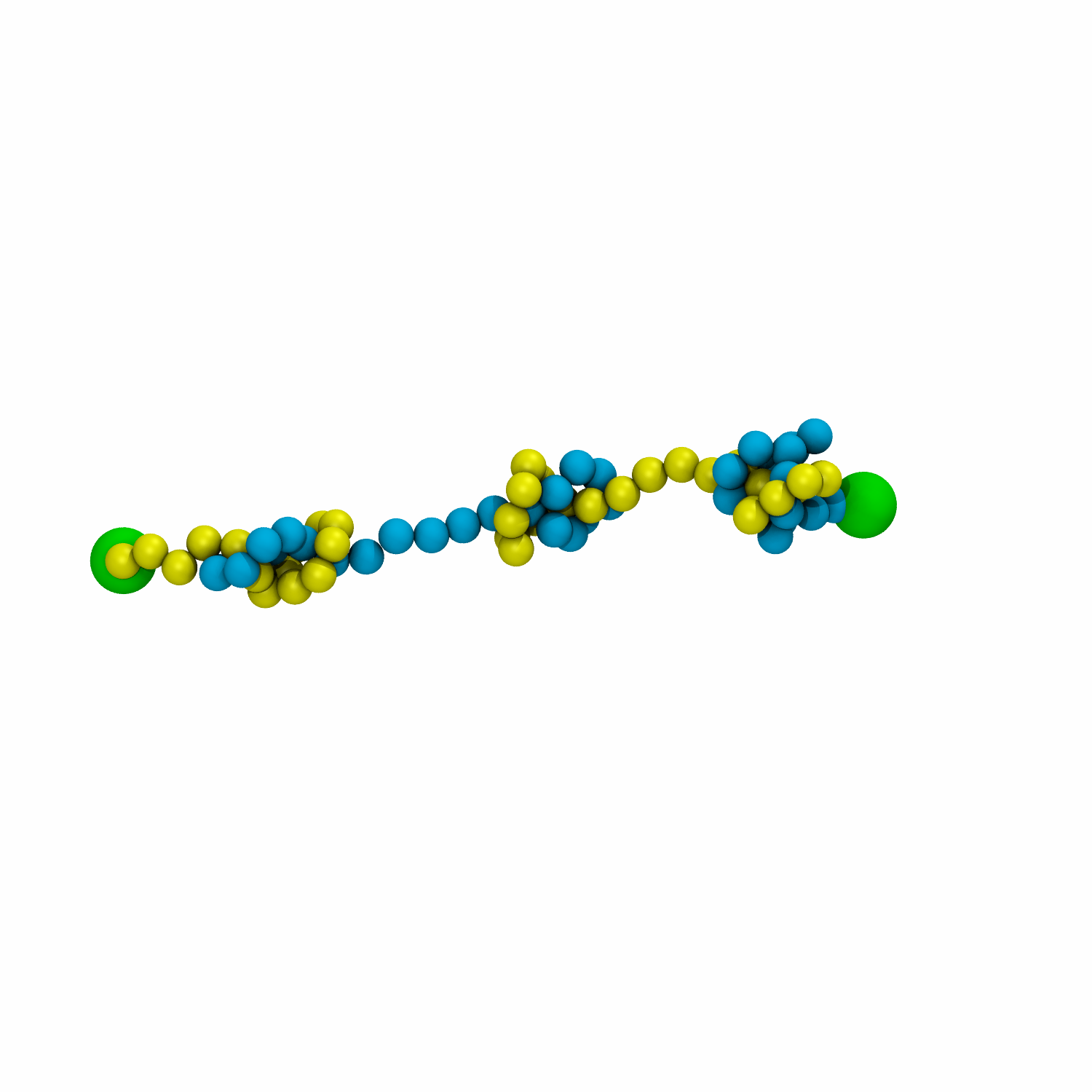}}{\scriptsize Daisy Chain (Shear 3)}
    \end{subfigure}
    \begin{subfigure}[b]{0.18\textwidth}
        \stackon{\includegraphics[width=\textwidth]{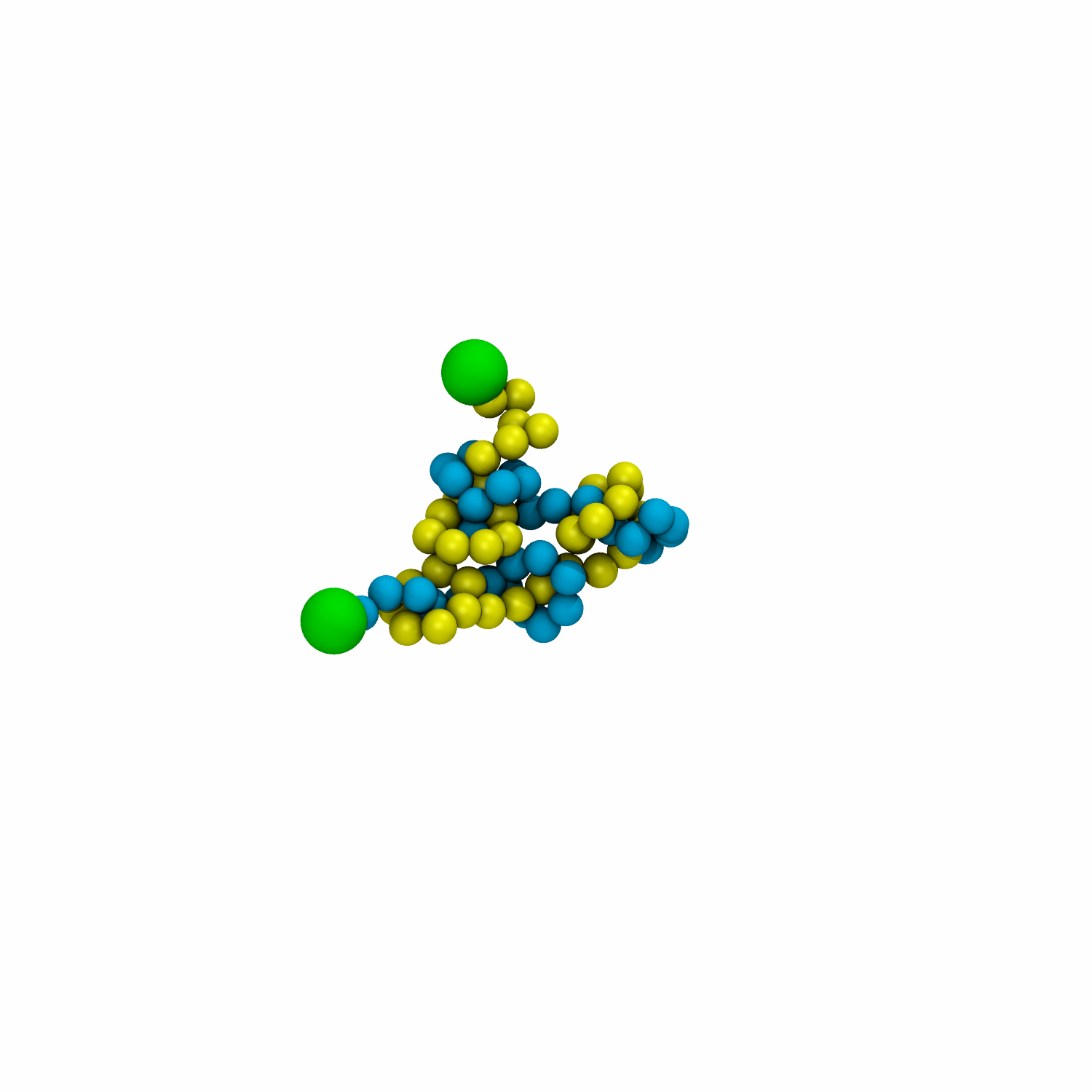}}{\scriptsize Daisy Chain (Shear 4)}
    \end{subfigure}
    \begin{subfigure}[b]{0.18\textwidth}
        \stackon{\includegraphics[width=\textwidth]{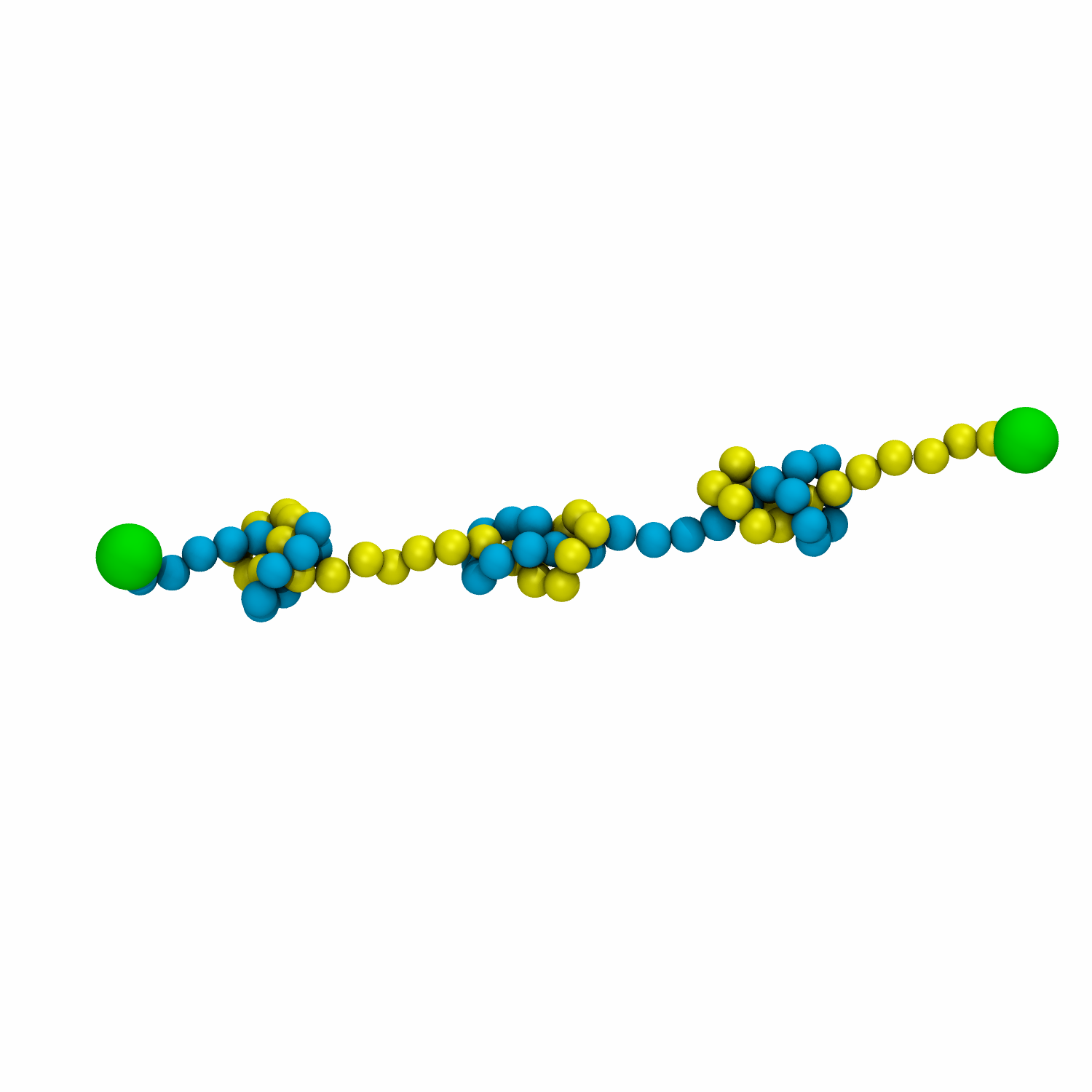}}{\scriptsize Daisy Chain (Shear 5)}
    \end{subfigure}
    \\
    % ===== pc-shear =====
    \begin{subfigure}[b]{0.18\textwidth}
        \stackon{\includegraphics[width=\textwidth]{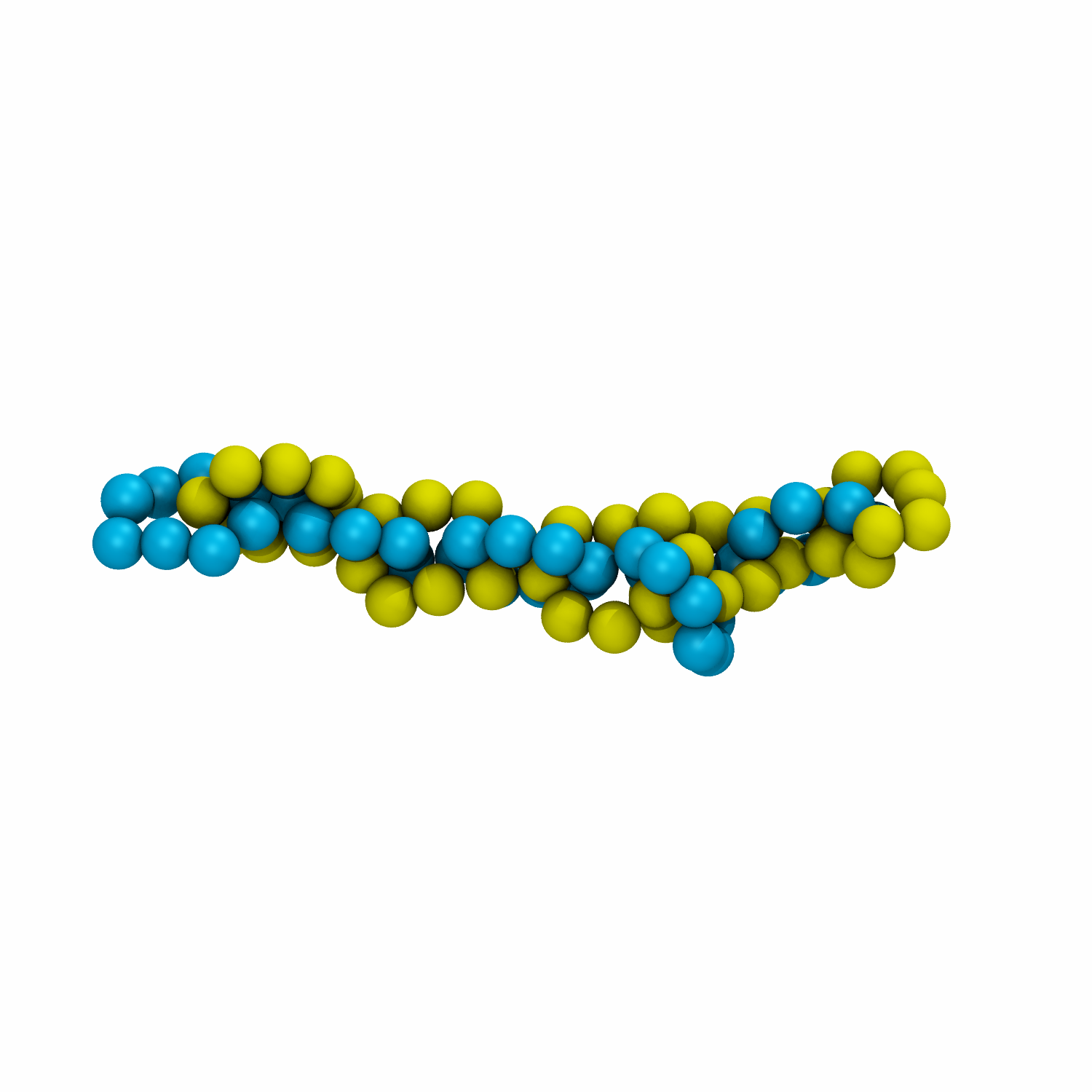}}{\scriptsize Polycatenane (Shear 1)}
    \end{subfigure}
    \begin{subfigure}[b]{0.18\textwidth}
        \stackon{\includegraphics[width=\textwidth]{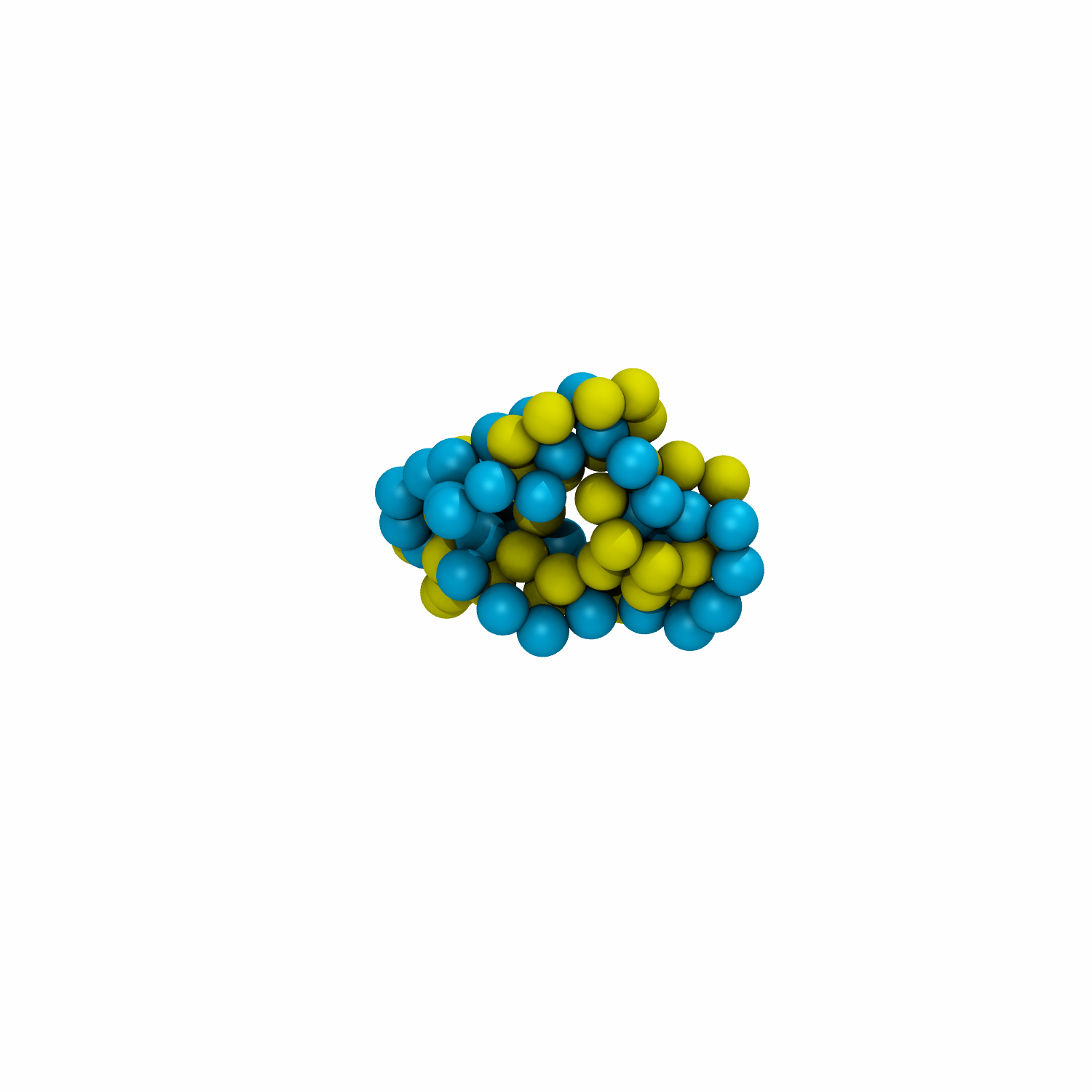}}{\scriptsize Polycatenane (Shear 2)}
    \end{subfigure}
    \begin{subfigure}[b]{0.18\textwidth}
        \stackon{\includegraphics[width=\textwidth]{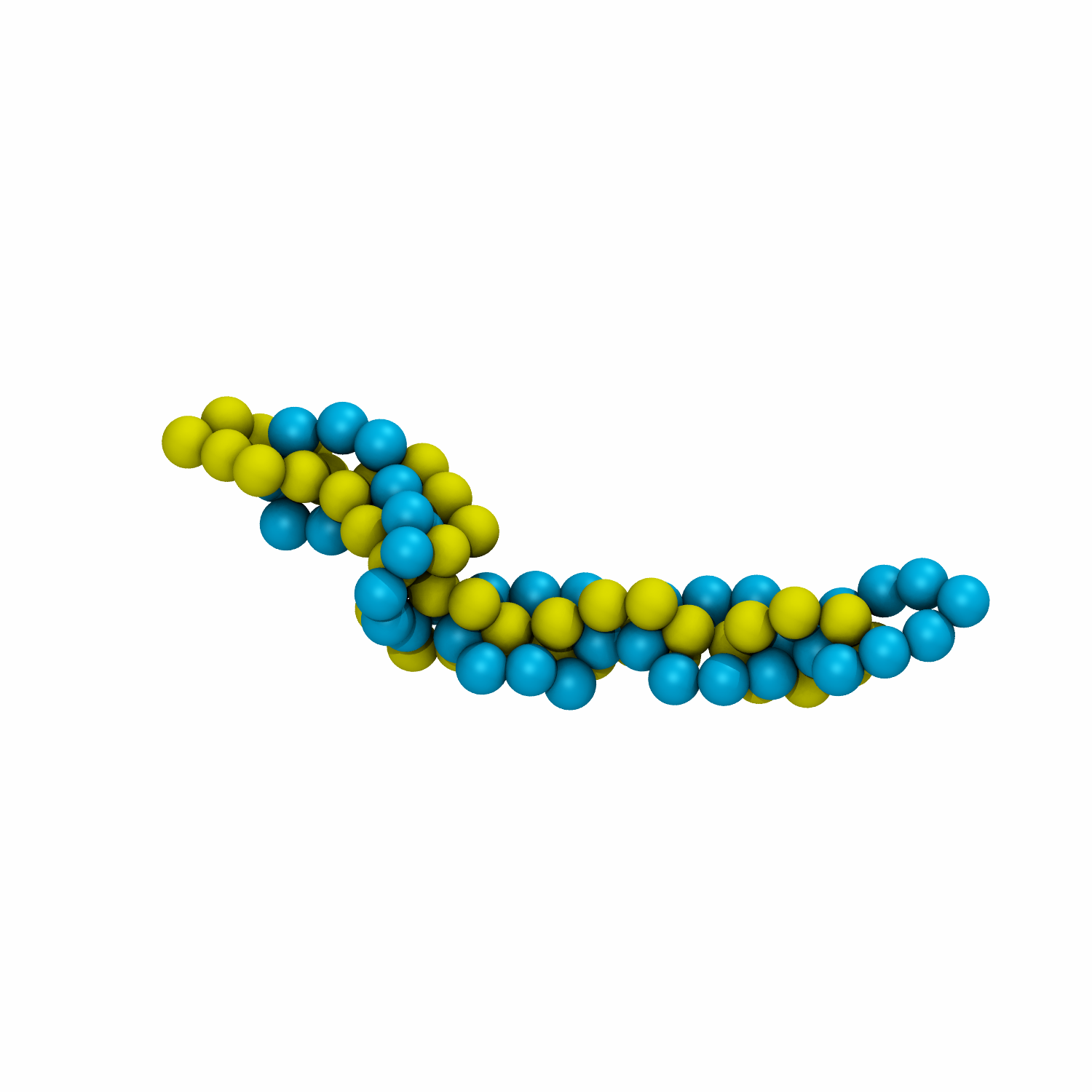}}{\scriptsize Polycatenane (Shear 3)}
    \end{subfigure}
    \begin{subfigure}[b]{0.18\textwidth}
        \stackon{\includegraphics[width=\textwidth]{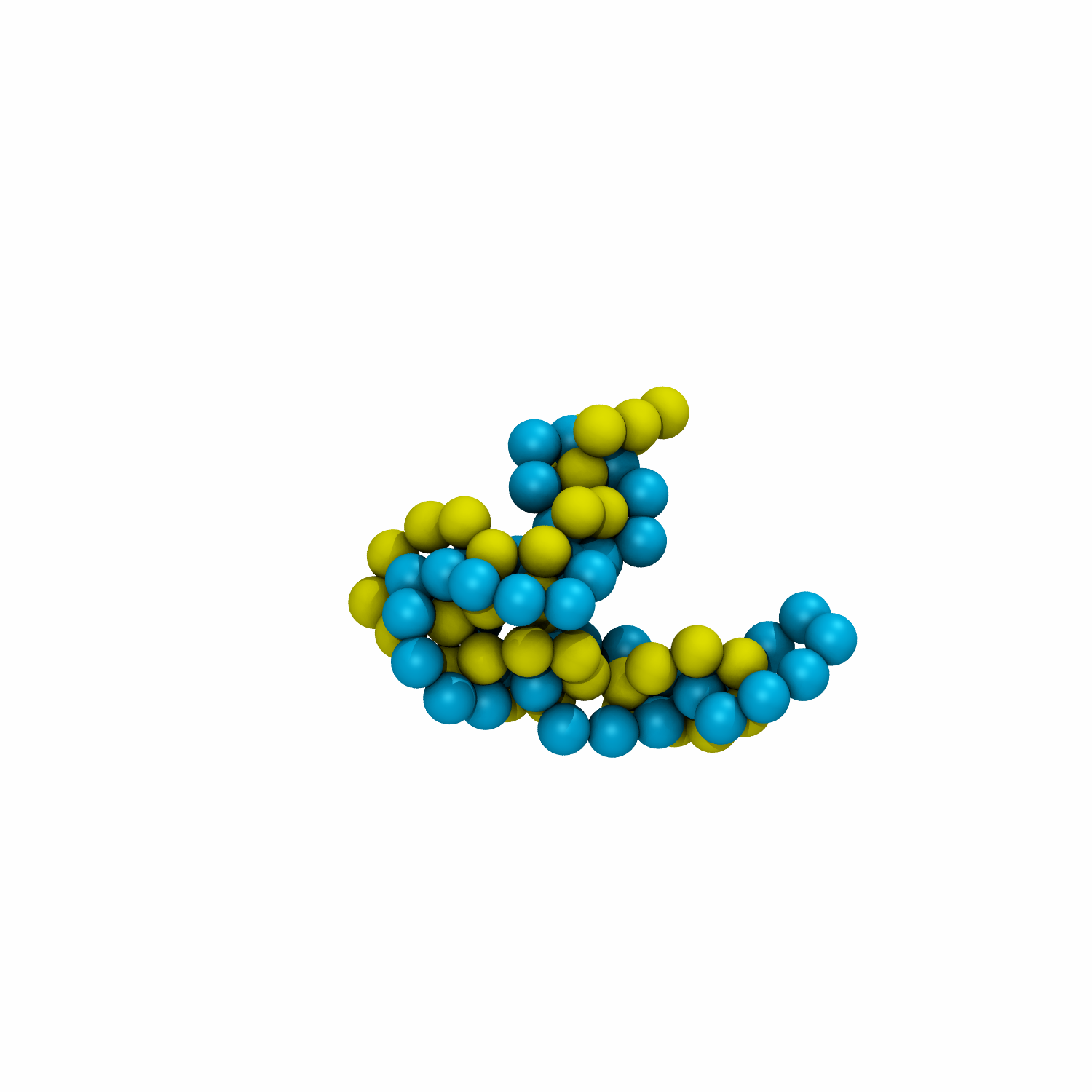}}{\scriptsize Polycatenane (Shear 4)}
    \end{subfigure}
    \begin{subfigure}[b]{0.18\textwidth}
        \stackon{\includegraphics[width=\textwidth]{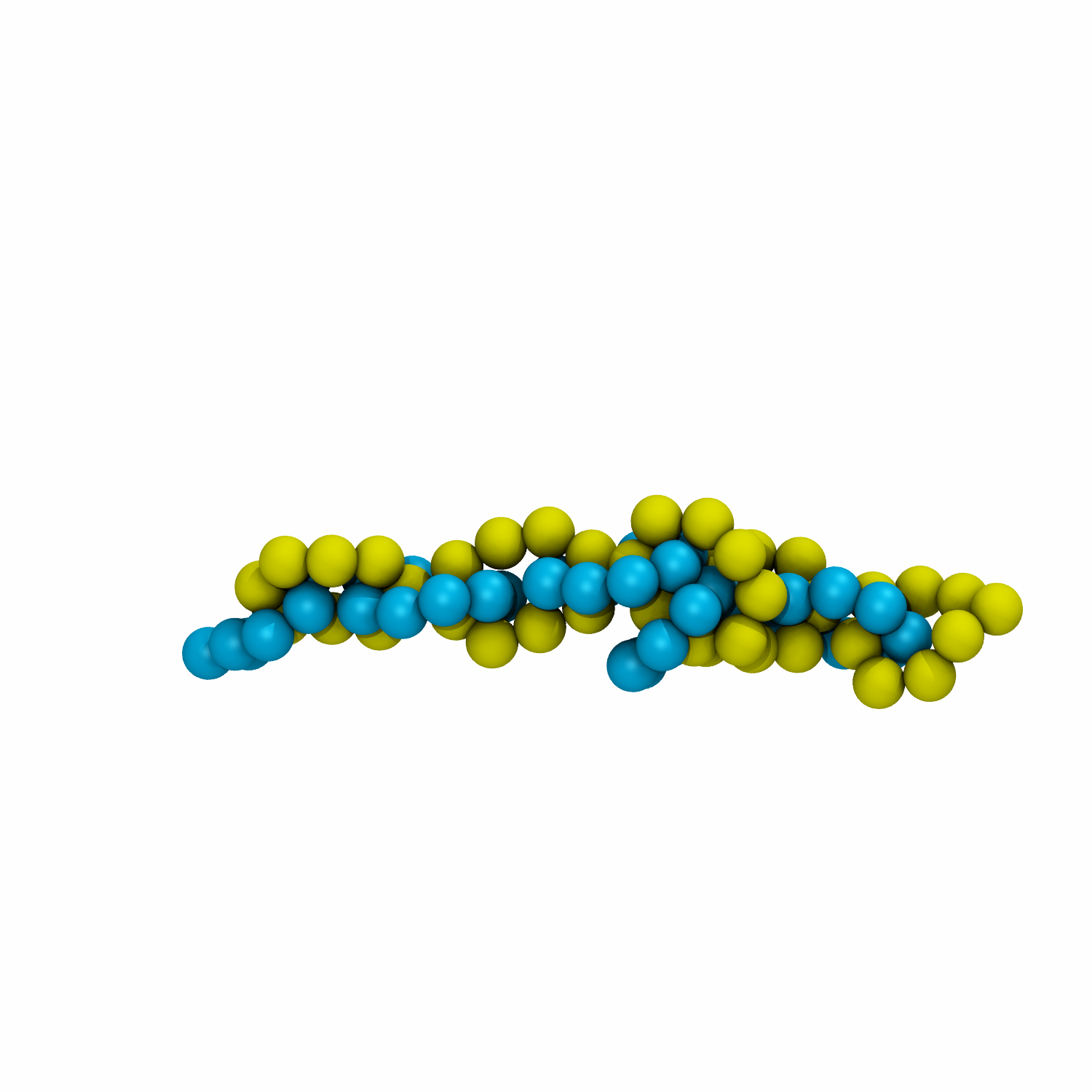}}{\scriptsize Polycatenane (Shear 5)}
    \end{subfigure}
    \\
    \caption{Snapshots of polymer trajectories: equilibrium, extensional flow at Wi=10, and shear flow at Wi=100. Shear flow snapshots show a complete tumbling cycle }
    \label{fig:trajectory_snapshots}
\end{figure*}
\clearpage
\begin{figure*}[h!]
\setcounter{figure}{0}
\section{Probability density distributions of ring positions in polyrotaxanes}
  \centering

    \begin{subfigure}[t]{\textwidth}
      \includegraphics[width=0.55\textwidth]{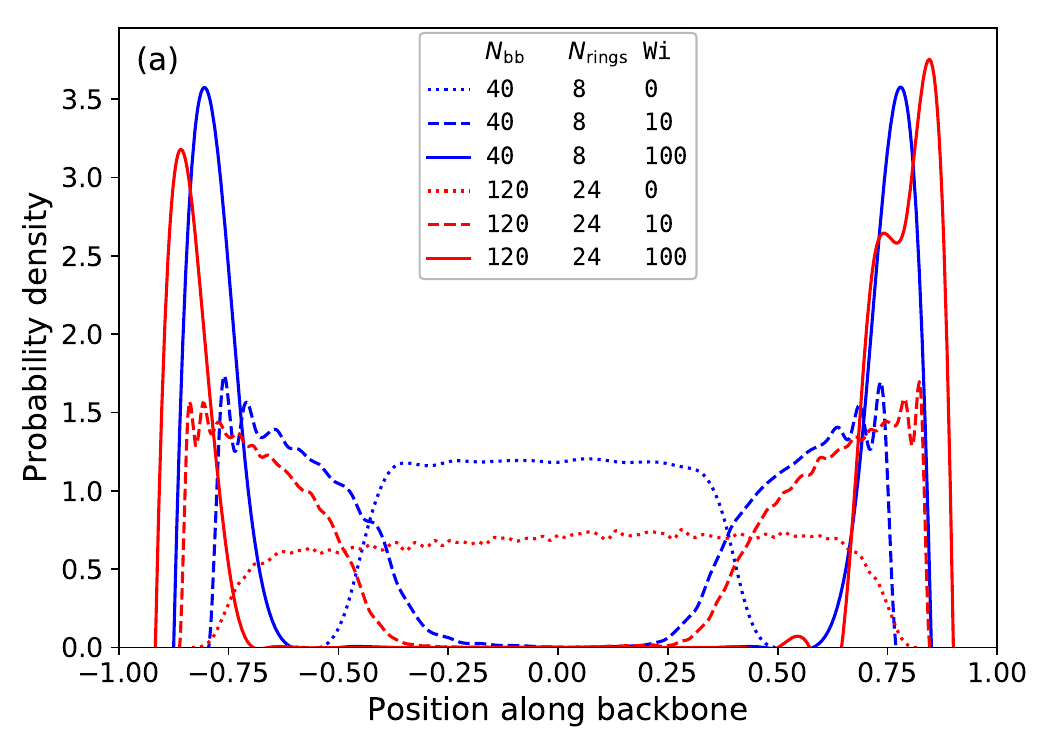}
    \end{subfigure}
    \vspace{0.1em}
    \begin{subfigure}[t]{\textwidth}
      \includegraphics[width=0.55\textwidth]{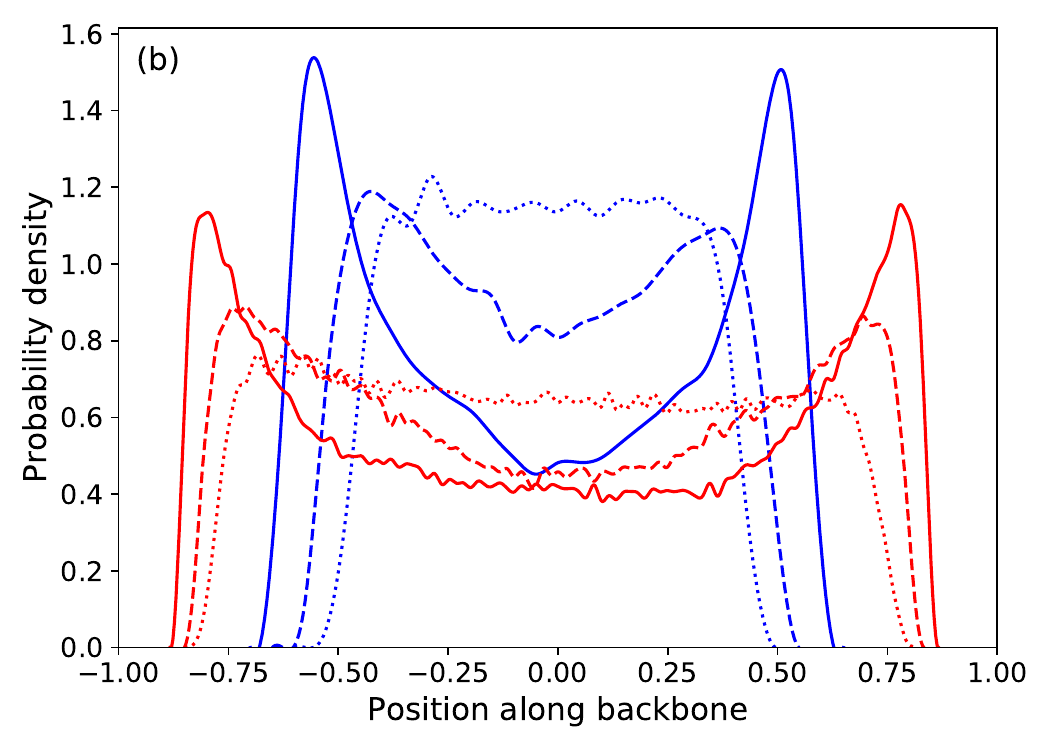}
    \end{subfigure}

  \captionsetup{justification=raggedright,singlelinecheck=false}

  \caption{Probability density distributions of ring positions in polyrotaxanes subjected to (a) extensional and (b) shear flow as functions of the Weissenberg number (Wi). The normalized coordinate ranges from $-1$ to $+1$, corresponding to the two ends of the polymer backbone. The graphs at $\mathrm{Wi}=0$ in (a) indicate equilibrium distributions.}
  \label{fig:ring_dist_app}
\end{figure*}

\clearpage

\begin{table}[h!]
\section{Power-law exponents}
\setcounter{table}{0}
\centering
\caption{Power-law exponents ($\beta$) of shear viscosity for different polymer architectures}
\setlength{\tabcolsep}{6pt}
\begin{tabular}{c c c c c c c c c}
\hline
Architecture & $M_{w}$ & $N_{bb}$ & BPS & BPR & $N_\mathrm{rings}$ & $N_\mathrm{seg}$ & $\beta$ \\
\hline
Linear        & 40   & - & - & - & -  & -  & -0.348 $\pm$ 0.012 \\
Linear        & 80   & - & - & - & -  & -  & -0.380 $\pm$ 0.009 \\
Linear        & 120  & - & - & - & -  & -  & -0.390 $\pm$ 0.019 \\
Polyrotaxane  & 56   & 40 & - & - & 2   & -  & -0.268 $\pm$ 0.028 \\
Polyrotaxane  & 72   & 40 & - & - & 4   & -  & -0.198 $\pm$ 0.028 \\
Polyrotaxane  & 104  & 40 & - & - & 8   & -  & -0.140 $\pm$ 0.023 \\
Polyrotaxane  & 112  & 80 & - & - & 4   & -  & -0.290 $\pm$ 0.019 \\
Polyrotaxane  & 144  & 80 & - & - & 8   & -  & -0.238 $\pm$ 0.024 \\
Polyrotaxane  & 208  & 80 & - & - & 16  & -  & -0.225 $\pm$ 0.028 \\
Polyrotaxane  & 168  & 120 & - & - & 6   & -  & -0.294 $\pm$ 0.018 \\
Polyrotaxane  & 216  & 120 & - & - & 12  & -  & -0.283 $\pm$ 0.023 \\
Polyrotaxane  & 312  & 120 & - & - & 24  & -  & -0.239 $\pm$ 0.018 \\
Daisy chain   & 30   & - & 8  & - & -  & 2  & -0.152 $\pm$ 0.042 \\
Daisy chain   & 74   & - & 8  & - & -  & 4  & -0.207 $\pm$ 0.018 \\
Daisy chain   & 118  & - & 8  & - & -  & 6  & -0.219 $\pm$ 0.009 \\
Daisy chain   & 206  & - & 8  & - & -  & 10 & -0.212 $\pm$ 0.013 \\
Daisy chain   & 46   & - & 16 & - & -  & 2  & -0.245 $\pm$ 0.020 \\
Daisy chain   & 106  & - & 16 & - & -  & 4  & -0.252 $\pm$ 0.011 \\
Daisy chain   & 166  & - & 16 & - & -  & 6  & -0.267 $\pm$ 0.030 \\
Daisy chain   & 226  & - & 16 & - & -  & 8  & -0.275 $\pm$ 0.017 \\
Daisy chain   & 78   & - & 32 & - & -  & 2  & -0.326 $\pm$ 0.021 \\
Daisy chain   & 170  & - & 32 & - & -  & 4  & -0.317 $\pm$ 0.017 \\
Daisy chain   & 262  & - & 32 & - & -  & 6  & -0.325 $\pm$ 0.016 \\
Polycatenane  & 40   & - & -  & 8  & 5   & -  & -0.158 $\pm$ 0.024 \\
Polycatenane  & 80   & - & -  & 8  & 10  & -  & -0.154 $\pm$ 0.016 \\
Polycatenane  & 168  & - & -  & 8  & 21  & -  & -0.163 $\pm$ 0.046 \\
Polycatenane  & 344  & - & -  & 8  & 43  & -  & -0.172 $\pm$ 0.021 \\
Polycatenane  & 520  & - & -  & 8  & 65  & -  & -0.203 $\pm$ 0.039 \\
Polycatenane  & 160  & - & -  & 16 & 10  & -  & -0.238 $\pm$ 0.019 \\
Polycatenane  & 240  & - & -  & 24 & 10  & -  & -0.298 $\pm$ 0.009 \\
Polycatenane  & 320  & - & -  & 32 & 10  & -  & -0.306 $\pm$ 0.014 \\
\hline
\end{tabular}
\label{table:all_combined_separated}
\end{table}

\clearpage

\begin{table}[h!]
\centering
\caption{Power-law exponents ($\beta$) of $\psi_1$ for different polymer architectures}
\setlength{\tabcolsep}{6pt}
\begin{tabular}{c c c c c c c c c}
\hline
Architecture & $M_{w}$ & $N_{bb}$ & BPS & BPR & $N_\mathrm{rings}$ & $N_\mathrm{seg}$ & $\beta$ \\
\hline
Linear        & 40   & - & - & - & -  & -  & -0.976 $\pm$ 0.013 \\
Linear        & 80   & - & - & - & -  & -  & -1.239 $\pm$ 0.018 \\
Linear        & 120  & - & - & - & -  & -  & -1.317 $\pm$ 0.009 \\
Polyrotaxane  & 56   & 40  & - & - & 2   & -  & -1.052 $\pm$ 0.016 \\
Polyrotaxane  & 72   & 40  & - & - & 4   & -  & -0.959 $\pm$ 0.016 \\
Polyrotaxane  & 104  & 40  & - & - & 8   & -  & -0.912 $\pm$ 0.013 \\
Polyrotaxane  & 112  & 80  & - & - & 4   & -  & -1.178 $\pm$ 0.023 \\
Polyrotaxane  & 144  & 80  & - & - & 8   & -  & -1.125 $\pm$ 0.023 \\
Polyrotaxane  & 208  & 80  & - & - & 16  & -  & -1.183 $\pm$ 0.025 \\
Polyrotaxane  & 168  & 120 & - & - & 6   & -  & -1.205 $\pm$ 0.030 \\
Polyrotaxane  & 216  & 120 & - & - & 12  & -  & -1.283 $\pm$ 0.025 \\
Polyrotaxane  & 312  & 120 & - & - & 24  & -  & -1.240 $\pm$ 0.023 \\
Daisy chain   & 30   & - & 8  & - & -  & 2  & -0.947 $\pm$ 0.032 \\
Daisy chain   & 74   & - & 8  & - & -  & 4  & -1.105 $\pm$ 0.018 \\
Daisy chain   & 118  & - & 8  & - & -  & 6  & -1.199 $\pm$ 0.011 \\
Daisy chain   & 206  & - & 8  & - & -  & 10 & -1.255 $\pm$ 0.015 \\
Daisy chain   & 46   & - & 16 & - & -  & 2  & -1.225 $\pm$ 0.017 \\
Daisy chain   & 106  & - & 16 & - & -  & 4  & -1.320 $\pm$ 0.014 \\
Daisy chain   & 226  & - & 16 & - & -  & 8  & -1.423 $\pm$ 0.019 \\
Daisy chain   & 78   & - & 32 & - & -  & 2  & -1.349 $\pm$ 0.033 \\
Daisy chain   & 170  & - & 32 & - & -  & 4  & -1.415 $\pm$ 0.023 \\
Daisy chain   & 262  & - & 32 & - & -  & 6  & -1.497 $\pm$ 0.033 \\
Polycatenane  & 40   & - & -  & 8  & 5   & -  & -0.918 $\pm$ 0.024 \\
Polycatenane  & 80   & - & -  & 8  & 10  & -  & -1.067 $\pm$ 0.023 \\
Polycatenane  & 168  & - & -  & 8  & 21  & -  & -1.342 $\pm$ 0.037 \\
Polycatenane  & 344  & - & -  & 8  & 43  & -  & -1.499 $\pm$ 0.025 \\
Polycatenane  & 520  & - & -  & 8  & 65  & -  & -1.602 $\pm$ 0.027 \\
Polycatenane  & 160  & - & -  & 16 & 10  & -  & -1.064 $\pm$ 0.027 \\
Polycatenane  & 240  & - & -  & 24 & 10  & -  & -1.172 $\pm$ 0.020 \\
Polycatenane  & 320  & - & -  & 32 & 10  & -  & -1.197 $\pm$ 0.027 \\
\hline
\end{tabular}
\label{table:all_combined_separated}
\end{table}

\clearpage

\begin{table}[h!]
\centering
\caption{Power-law exponents ($\beta$) of $\omega\tau_R$ for different polymer architectures}
\setlength{\tabcolsep}{6pt}
\begin{tabular}{c c c c c c c c c}
\hline
Architecture & $M_{w}$ & $N_{bb}$ & BPS & BPR & $N_\mathrm{rings}$ & $N_\mathrm{seg}$ & $\beta$ \\
\hline
Linear        & 40   & - & - & - & -  & -  & 0.741 $\pm$ 0.008 \\
Linear        & 80   & - & - & - & -  & -  & 0.752 $\pm$ 0.016 \\
Linear        & 120  & - & - & - & -  & -  & 0.732 $\pm$ 0.006 \\
Polyrotaxane  & 56   & 40  & - & - & 2   & -  & 0.740 $\pm$ 0.011 \\
Polyrotaxane  & 72   & 40  & - & - & 4   & -  & 0.751 $\pm$ 0.014 \\
Polyrotaxane  & 104  & 40  & - & - & 8   & -  & 0.826 $\pm$ 0.011 \\
Polyrotaxane  & 112  & 80  & - & - & 4   & -  & 0.748 $\pm$ 0.020 \\
Polyrotaxane  & 144  & 80  & - & - & 8   & -  & 0.763 $\pm$ 0.017 \\
Polyrotaxane  & 208  & 80  & - & - & 16  & -  & 0.794 $\pm$ 0.012 \\
Polyrotaxane  & 168  & 120 & - & - & 6   & -  & 0.766 $\pm$ 0.013 \\
Polyrotaxane  & 216  & 120 & - & - & 12  & -  & 0.770 $\pm$ 0.015 \\
Polyrotaxane  & 312  & 120 & - & - & 24  & -  & 0.787 $\pm$ 0.018 \\
Daisy chain   & 30   & - & 8  & - & -  & 2  & 0.838 $\pm$ 0.006 \\
Daisy chain   & 74   & - & 8  & - & -  & 4  & 0.822 $\pm$ 0.009 \\
Daisy chain   & 118  & - & 8  & - & -  & 6  & 0.838 $\pm$ 0.011 \\
Daisy chain   & 206  & - & 8  & - & -  & 10 & 0.867 $\pm$ 0.016 \\
Daisy chain   & 46   & - & 16 & - & -  & 2  & 0.875 $\pm$ 0.021 \\
Daisy chain   & 106  & - & 16 & - & -  & 4  & 0.864 $\pm$ 0.018 \\
Daisy chain   & 226  & - & 16 & - & -  & 8  & 0.882 $\pm$ 0.022 \\
Daisy chain   & 78   & - & 32 & - & -  & 2  & 0.834 $\pm$ 0.022 \\
Daisy chain   & 170  & - & 32 & - & -  & 4  & 0.822 $\pm$ 0.021 \\
Daisy chain   & 262  & - & 32 & - & -  & 6  & 0.840 $\pm$ 0.021 \\
Polycatenane  & 40   & - & -  & 8  & 5   & -  & 0.917 $\pm$ 0.007 \\
Polycatenane  & 80   & - & -  & 8  & 10  & -  & 0.935 $\pm$ 0.010 \\
Polycatenane  & 168  & - & -  & 8  & 21  & -  & 0.912 $\pm$ 0.030 \\
Polycatenane  & 344  & - & -  & 8  & 43  & -  & 0.989 $\pm$ 0.013 \\
Polycatenane  & 520  & - & -  & 8  & 65  & -  & 1.029 $\pm$ 0.015 \\
Polycatenane  & 160  & - & -  & 16 & 10  & -  & 0.790 $\pm$ 0.029 \\
Polycatenane  & 240  & - & -  & 24 & 10  & -  & 0.747 $\pm$ 0.031 \\
Polycatenane  & 320  & - & -  & 32 & 10  & -  & 0.753 $\pm$ 0.021 \\
\hline
\end{tabular}
\label{table:all_combined_separated}
\end{table}

\end{document}